\documentclass[epsf,useAMS,usenatbib,usegraphicx]{mn2e}
\usepackage{epsfig}
\usepackage{aas_macros} % required to get journal abbreviations
\usepackage{amsmath}
\usepackage{amssymb}
\usepackage{appendix}
\usepackage{booktabs}
\usepackage{upgreek}
\usepackage{url}
\usepackage[usenames]{color}

\newcommand{\ion}[2]{{#1}\,{\sc #2}}
\newcommand{\specline}[3]{{#1}\,{\sc #2}\:{#3}}

\usepackage{graphicx}
\usepackage{rotating}
\usepackage{color}
\usepackage{pifont}
\usepackage{longtable}
\usepackage{pdflscape,lscape}
\usepackage{natbib}

\title[Spectroscopic survey of \textit{Kepler} stars]{Spectroscopic survey of \textit{Kepler} stars.\thanks{Based on observations made with the Mercator Telescope, operated on the 
island of La Palma by the Flemish Community, at the Spanish Observatorio del Roque de los Muchachos of the Instituto de Astrofsica de Canarias. Based on observations obtained with the HERMES spectrograph, which is supported by the Fund for Scientific Research of Flanders (FWO), Belgium , 
the Research Council of K.U.Leuven, Belgium, the Fonds National Recherches Scientific (FNRS), Belgium, the Royal Observatory of Belgium, the Observatoire de Genve, Switzerland and the Thringer Landessternwarte Tautenburg, Germany.} \\
I. HERMES/Mercator observations of A- and F-type stars}
\author[E.\,Niemczura et al.]{E.\,Niemczura$^1$\thanks{E-mail: eniem@astro.uni.wroc.pl},
S.\,J.\,Murphy$^{2,3}$,
B.\,Smalley$^4$,
K.\,Uytterhoeven$^{5,6}$,
A.\,Pigulski$^1$, \and
H.\,Lehmann$^7$,
D.M.\,Bowman$^8$,
G.\,Catanzaro$^9$,
E.\,van Aarle$^{10}$, 
S.\,Bloemen$^{11}$, 
M.\,Briquet$^{12}$, \and
P.\,De Cat$^{13}$, 
D.\,Drobek$^1$, 
L.\,Eyer$^{14}$,
J.\,F.\,S.\,Gameiro$^{15}$,
N.\,Gorlova$^{10}$, 
K.\,Kami{\'n}ski$^{16}$,\and
P.\,Lampens$^{13}$, 
P.\,Marcos-Arenal$^{10}$, 
P.\,I.\,P\'apics$^{10}$, 
B.\,Vandenbussche$^{10}$,
H.\,Van Winckel$^{10}$, \and
M.\,St\c e\'slicki$^{17}$,
M.\,Fagas$^{16}$
\\
$^1$ Instytut Astronomiczny, Uniwersytet Wroc{\l}awski, Kopernika 11, 51-622 Wroc{\l}aw, Poland\\
$^2$ Sydney Institute for Astronomy (SIfA), School of Physics, University of Sydney NSW 2006, Australia\\
$^3$ Stellar Astrophysics Centre, Department of Physics and Astronomy, Aarhus University, 8000 Aarhus C, Denmark\\
$^4$ Astrophysics Group, Keele University, Staffordshire, ST5 5BG, United Kingdom\\
$^5$ Instituto de Astrofisica de Canarias, E-38205 La Laguna, Tenerife, Spain\\
$^6$ Universidad de La Laguna, Departamento de Astrofisica, E-38206 La Laguna, Tenerife, Spain\\
$^7$ Th\"{u}ringer Landessternwarte Tautenburg (TLS), Sternwarte 5, 07778 Tautenburg, Germany\\
$^8$ Jeremiah Horrocks Institute, University of Central Lancashire, Preston PR1 2HE, UK\\
$^9$ INAF-Osservatorio Astrofisico di Catania, Via S. Sofia 78, I-95123, Catania, Italy\\
$^{10}$ Instituut voor Sterrenkunde, KU Leuven, Celestijnenlaan 200D, 3001, Leuven, Belgium\\
$^{11}$ Department of Astrophysics, IMAPP, Radboud University Nijmegen, PO Box 9010, NL-6500 GL Nijmegen, The Netherlands\\
$^{12}$ Institut d'Astrophysique et de G\'eophysique, Universit\'e de Li\`ege, All\'ee du 6 Ao\^ut 17, B-4000 Li\`ege, Belgium\\
$^{13}$ Royal observatory of Belgium, Ringlaan 3, B-1180 Brussel, Belgium\\
$^{14}$ University of Geneva, Department of Astronomy, Chemin des Maillettes, 51, Sauverny, CH-1290, Switzerland\\
$^{15}$ Instituto de Astrof\'isica e Ci\^{e}ncias Espaciais and Faculdade de Ci\^{e}ncias, Universidade do Porto, Rua das Estrelas, \\PT4150-762 Porto, Portugal\\
$^{16}$ Astronomical Observatory, Adam Mickiewicz University, ul. S\l{}oneczna 36, PL-60-286 Pozna\'{n}, Poland \\
$^{17}$ Centrum Bada\'{n} Kosmicznych, Polska Akademia Nauk, Kopernika 11, 51-622 Wroc{\l}aw. }

\begin{document}
\def\teff{${T}_{\rm eff}$}
\def\kms{{km\,s}$^{-1}$}
\def\logg{$\log g$}
\def\turb{$\xi_{\rm t}$}
\def\rad{$v_{\rm r}$}
\def\vsini{$v\sin i$}
\def\ebv{$E(B-V)$}
\date{Accepted ... Received ...; in original form ...}

\pagerange{\pageref{firstpage}--\pageref{lastpage}} \pubyear{2002}

\maketitle

\label{firstpage}

\begin{abstract}
The \textit{Kepler} space mission provided near-continuous and high-precision photometry of about 207,000 stars, which can be used for asteroseismology. 
However, for successful seismic modelling it is equally important to have accurate stellar physical parameters. 
Therefore, supplementary ground-based data are needed. We report the results of the analysis of high-resolution spectroscopic data of A- and F-type stars 
from the \textit{Kepler} field, which were obtained with the HERMES spectrograph on the Mercator telescope. We determined spectral types, atmospheric 
parameters and chemical abundances for a sample of 117 stars. Hydrogen Balmer, \ion{Fe}{i}, and \ion{Fe}{ii} lines were used to derive effective temperatures, surface gravities, 
and microturbulent velocities. We determined chemical abundances and projected rotational velocities using a spectrum synthesis technique.
The atmospheric parameters obtained were compared with those from the \textit{Kepler Input Catalogue} (KIC), confirming that the KIC effective temperatures 
are underestimated for A stars. Effective temperatures calculated by spectral energy distribution fitting are in good agreement with those determined from the spectral line analysis.
The analysed sample comprises stars with approximately solar chemical abundances, as well as chemically peculiar stars of the Am, Ap, and $\lambda$\,Boo types. 
The distribution of the projected rotational velocity, \vsini, is typical for A and F stars and ranges from $8$ to about $280$\,\kms, with a mean of $134$\,\kms.
\end{abstract}


\section{Introduction}
\label{intro}
The NASA space mission \textit{Kepler} was successfully launched on 7th March, 2009. The main scientific goal of the mission was to discover Earth-sized planets 
around Sun-like stars \citep{borucki}. During May 2013 the satellite lost the second of its four reaction wheels (the first one failed in July 2012), 
preventing further observation of the original field. Nevertheless, during the four years of near-continuous data collection, \textit{Kepler} produced 
photometric time series of an exceptional precision of a few ppm (parts per million), 
which provide a unique and excellent database for studying variability due to pulsations in thousands of stars across the Hertzsprung-Russell (H-R) diagram \citep{kjeldsen}.

Asteroseismology is the only way to probe the internal structure of stars and significantly improve our understanding of stellar evolution. 
It provides direct tests of the modelling of processes taking place in stellar interiors, such as diffusion, transport of angular momentum \citep{mosser, marques}, 
convective overshooting, mechanisms driving pulsations \citep[e.g.][]{moravveji, 2014ApJ...796..118A}, and the value of the mixing-length 
parameter in stars of different effective temperatures \citep[][]{trampedach, bonaca}. The key ingredients for an asteroseismic study are precise pulsation 
frequencies, mode identification, and strong constraints on atmospheric parameters, such as effective temperature $T_{\rm eff}$, surface gravity $\log g$, 
chemical abundances, and rotational velocity $v \sin i$. Accurate values of the pulsation frequencies are provided by the ultra-precise \textit{Kepler} photometry. 
The accuracy of atmospheric parameters available from the \textit{Kepler Input Catalogue} (KIC) \citep{brown} is generally too low for asteroseismic modelling, 
particularly for stars of spectral types earlier than F0 \citep{huber2014}. Also, information on the stellar chemical composition and rotational velocity is lacking. 
Therefore, high-resolution spectra which offer a wide range of information are needed to fully characterise \textit{Kepler} targets \citep{uytterhoeven2014}. 

For more than a thousand \textit{Kepler} stars high- and medium-resolution spectra have been collected \citep{uytterhoeven2010}. 
These observations are summarized in the {\it Kepler Asteroseismic Science Operations Center} (KASOC) database\footnote{\url{http://kasoc.phys.au.dk/}}. 
In this paper, we focus on the analysis of high-resolution spectra of main-sequence A- and F-type stars. The main groups of pulsators among these objects are $\delta$\,Scuti, 
$\gamma$\,Doradus, and $\gamma$\,Dor/$\delta$\,Sct hybrid stars. The $\delta$\,Sct stars are main-sequence or evolved stars situated in the classical instability strip. 
They pulsate in low-order radial and non-radial pressure modes (p-modes) with typical frequencies of 5 to 50\,d$^{-1}$ \citep[see reviews by][]{breger2000, handler2009, balona2014}. 
The $\gamma$\,Dor stars are pulsating variables with spectral types A7--F5 that are on or close to the main-sequence and 
pulsate in high-order gravity modes (g-modes) with typical frequencies of 0.3 to 3\,d$^{-1}$ \citep{kaye}. The instability strip for $\gamma$\,Dor stars partly 
overlaps with the $\delta$\,Sct instability strip. The $\gamma$\,Dor variables are of particular interest in asteroseismology, as their g-mode 
pulsations probe the deep interiors of these stars, reaching their convective cores. Additionally, these pulsators occupy a region in the H-R 
diagram where $\gamma$\,Dor-like g-mode pulsations, $\delta$\,Sct-like p-mode pulsations and possibly solar-like oscillations can co-exist in a single star. 
This overlap means that some individual objects show g-mode as well as p-mode pulsations. These hybrid variables permit simultaneous probing of two distinct regions of the interior: 
the deep layers between the convective core and the envelope, where the g-modes propagate, and the outer layers of the envelope where the p-modes 
propagate \citep[][]{kurtz2014, 2015MNRAS.447.3264S}. Therefore, hybrid stars are key objects for testing stellar models with seismic techniques. 

\citet{grigahcene} found that the hybrid stars in the \textit{Kepler} field are spread over the entire $\delta$\,Sct and $\gamma$\,Dor instability domains.
The analysis of \textit{Kepler} data and KIC parameters for 750 A and F stars presented by \citet{uytterhoeven} confirms this result.
They found that 23\% of their sample of A and F objects shows both $\delta$\,Sct pulsations and low-frequency variations. 
\citet{balonadziembowski} analysed observations of 1568 $\delta$\,Sct pulsators taken from the \textit{Kepler} archive and identified 
a group of $\delta$\,Sct stars with low-frequency peaks similar to those characteristic for $\gamma$\,Dor stars. Contrary to the results 
of \citet{uytterhoeven} and \citet{grigahcene}, these variables are located in the same region of the H-R diagram as the ground-based $\gamma$\,Dor 
pulsators and none is found in the $\delta$\,Sct instability domain. The low frequencies are observed in the hotter stars, but their origin is not clear \citep{balona2014}.
A high-resolution spectroscopic analysis by \citet{tkachenko2013a} found each of their 38 $\gamma$\,Dor stars to lie within the $\gamma$\,Dor instability strip. 
Accurate stellar atmospheric parameters are clearly required to place $\gamma$\,Dor/$\delta$\,Sct hybrids unambiguously in the H-R diagram.

This paper is organised as follows: in Sect.\,\ref{sec:observations}, we describe observational data used in the analysis. Details on the spectral classification 
methods are given in Sect.\,\ref{sec:classification}. In Sect.\,\ref{sec:methods}, we present all the methods applied for determining the atmospheric parameters of 
the investigated objects, the sources of possible errors, and discuss consistency of the results obtained from different approaches. The distribution of 
rotational velocities is presented and discussed in Sect.\,\ref{sec:rotational}. The derived abundances of elements, their correlation with the other parameters and a sample 
of chemically peculiar (CP) stars are discussed in Sect.\,\ref{sec:abundances}. Positions of the analysed stars in the $\log$\,\teff\ -- \logg\ diagram, conclusions 
and future prospects are given in Sect.\,\ref{sec:conclusions}.

\section{Spectroscopic surveys of \textit{Kepler} stars}
\label{sec:observations}
High-resolution spectra were collected for several hundred stars located in the \textit{Kepler} field-of-view. Most of these
spectra were obtained by \'{e}chelle spectrographs attached to the Mercator Telescope\footnote{\url{http://www.mercator.iac.es}}, Nordic Optical 
Telescope\footnote{\url{http://www.not.iac.es/}} (NOT), Tautenburg 2-m Alfred-Jensch Telescope\footnote{\url{http://www.tls-tautenburg.de/TLS}} (TLS), 
Bernard Lyot Telescope\footnote{\url{http://www.tbl.omp.eu/}} and the Canada-France-Hawaii Telescope\footnote{\url{http://www.cfht.hawaii.edu/}} (CFHT).

In this paper we analyse high-resolution spectra taken with the cross-dispersed, fibre-fed \'{e}chelle spectrograph HERMES 
\citep[High Efficiency and Resolution Mercator \'{E}chelle Spectrograph;][]{raskin} attached to the 1.2-m Mercator Telescope located on La Palma (Canary Islands, Spain). 
The spectra have a resolving power $\rm{R}\sim85\,000$ and cover the spectral range from 3770 to 9000\,{\AA}. The typical signal-to-noise (S/N) ratio for an individual spectrum 
at 5500\,{\AA} is 80--100. The spectra have been reduced with a dedicated pipeline\footnote{\url{hermes-as.oma.be/doxygen/html/index.html}}, which includes bias subtraction, extraction of scattered light produced by 
the optical system, cosmic ray filtering, division by a normalized flat-field, wavelength calibration by a ThArNe lamp and order merging. The normalisation to 
the continuum was performed manually by using the standard {\sc iraf}\footnote{Image Reduction and Analysis Facility, \url{http://iraf.noao.edu/}} procedure {\it continuum}. 
Most of the stars in our sample have \vsini\,\,$> 40$\,\kms. The spectra of rapid rotators are very difficult to normalize correctly in spectral regions rich in blended lines, e.g. for $\lambda < 4100$\,{\AA}. 
For this reason, we checked the quality of our normalisation by fitting synthetic spectra to all Balmer lines in the spectrum of each star. 
If the spectrum has been normalized correctly, all hydrogen lines can be fitted satisfactorily with the same \teff. The process of normalization is otherwise repeated.
For the few stars for which we have more than one spectrum, we analysed the averaged spectrum. The averaging process is applied after normalization 
and forms an additional test of the quality of the normalization process.

The data analysed in this paper were collected in spectroscopic surveys carried out in the years 2010 to 2012 (Table\,\ref{surveys}). 
Table\,\ref{journal1} lists the analysed stars and gives their KIC\ numbers, dates of the observations, number of 
spectra, visual ($V$) magnitudes, and spectral types from the literature and determined in this paper. We also provide notes on chemical peculiarity, binarity and pulsation characteristics. 
Chemically peculiar stars in our sample were identified in the process of spectroscopic classification and confirmed by the detailed
spectroscopic analysis afterwards. Twenty stars were found to be double-lined spectroscopic binaries (SB2) and the atmospheric parameters
of these stars will be presented in a separate paper (Catanzaro\ et. al, in preparation). 

The results of the spectroscopic analysis of stars observed with the other telescopes will be presented in a forthcoming paper (Niemczura et al., in preparation).

\begin{table}
\centering
\caption{HERMES spectroscopic surveys of \textit{Kepler} stars.}
\label{surveys}
\begin{tabular}{lll}
\toprule
Observation       &Primary            & Observers \\
dates             &Investigator       &  \\ 
\midrule
2010 May-Sept.    & M. Briquet        & KU Leuven observers$^{\star}$\\
2011 July         & E. Niemczura      & D. Drobek \\
2011 Sept.        & E. Niemczura      & J.\,F.\,S. Gameiro \\
2012 July         & K. Uytterhoeven   & E. Niemczura \\
2012 August       & P. De Cat         & P.\,I.\,P\'apics \\
2012 Sept.        & P. De Cat         & P.\,M. Arenal \\
\bottomrule
\end{tabular}
 \begin{description}
 \item[ ]{$^{\star}$ } E. Van Aarle, S. Bloemen, N. Gorlova, P. Lampens, H. Van Winckel, B. Vandenbussche
 \end{description}

\end{table}

\begin{table*}
\centering
\caption{Journal of spectroscopic observations and the derived spectral classes. $N$ is the number of available spectra; SpT1 indicates the spectral
classification from the literature, whereas SpT2 means spectral classifications obtained in this work. Superscript $\star$ in the first column 
denotes the stars with additional information below the table. Additional notations for luminosity class: ``b'' -- lower luminosity main-sequence star; 
``a'' or ``a+'' -- higher luminosity main-sequence star; ``IV-V'' -- between IV and V; ``IV/V'' -- either IV or V; ``n'' -- nebulous (i.e. broad-lined); ``s'' -- sharp lines;
``nn'' -- very rapid rotators; ``wk'' met -- weak metal lines; ``met str'' -- strong metal lines. In the last column additional information taken from spectra and 
\textit{Kepler} data are presented. ``SB2'' means double-lined spectroscopic binary; ``rot.'' -- rotational frequency detected; 
``bin.'' -- orbital frequency detected; ``mult.'' -- orbital frequencies of both components detected; ``hybrid'' -- frequencies of $\gamma$\,Dor and $\delta$\,Sct visible in the spectra; 
``contamination'' -- \textit{Kepler} data contaminated by a nearby star; ``?'' -- our classification of variability type is uncertain. 
The full table is available in the electronic form.} 
\label{journal1}
\begin{tabular}{llcrlll}
\toprule
KIC          & HERMES        & N   & $V$   & SpT1     & SpT2        & Notes \\
Number       & observations  &     & [mag]& & [this paper]&       \\
\midrule
 1294756               & 2010 May-September  & 1     & 8.96&  A2$^{1}$ & A3\,IV              & mult. + rot. + $\delta$\,Sct \\ 
 2571868               & 2011 July           & 1     & 8.67&  A0$^{1}$ & A3\,IVn             & hybrid (low. ampl. $\gamma$\,Dor) \\ 
 2694337               & 2010 May-September  & 1     &10.35&           & F0.5\,IVs           & hybrid\\   
 2695344               & 2012 July           & 1     & 9.62&  F2$^{1}$ & F2\,V               & SB2; $\gamma$\,Dor\\
                       & 2012 September      & 1     &     &           &                     & \\
 3230227$^{\star}$     & 2010 May-September  & 1     & 8.94&  A5$^{1}$ & A5\,IV:             & SB2; ecl. bin. + $\delta$\,Sct\\  
 3231985$^{\star}$     & 2011 September      & 1     & 9.22&  A2$^{1}$ & A4\,IV Ca weak (A3) & rot.\\  
 3347643               & 2010 May-September  & 1     & 7.69&  A2$^{1}$ & A3\,Van             & hybrid (low. ampl. $\gamma$\,Dor)\\  
 3441230               & 2012 July           & 1     & 9.79&  A2$^{1}$ & kA2hA5mF0\,V Am     & rot.\\     
 3656913               & 2012 July           & 1     & 9.81&  A0$^{1}$ & kA2hA6mF0\,IV Am    & SB2; bin.\\
 3850810$^{\star}$     & 2010 May-September  & 1     &10.10&           & F1\,Vs wk met (F0)  & hybrid + rot.\\         
\bottomrule
\end{tabular}
\begin{description}
\item[ ]{KIC\,3230227:} Incorrectly normalised H lines; wings are unusable but core matches A5\,IV. Metal lines indicate A5\,V.
\item[ ]{KIC\,3231985:} Weak Ca\,K and Ca\,$\lambda$4226 line.
\item[ ]{KIC\,3850810:} Considerably metal weak or poorly normalised.
\end{description}
 \begin{description}
 \item[ ]{$^1$ } \citet{1975AGK3..C......0H}
 \end{description}
\end{table*}

\section{Spectral Classification}
\label{sec:classification}
Stellar spectral classification in the Morgan-Keenan (MK) system involves determining the luminosity class and spectral type for a given star 
by comparing its spectrum with spectra of standards \citep{gray&corbally2009}. The spectral type in A stars is determined mainly from three spectral features: 
(i) hydrogen lines, (ii) the \ion{Ca}{ii}\,K line, and (iii) lines of metals. In a non-peculiar star, all three features should consistently provide the same spectral type, 
if the correct luminosity class has been adopted. Determination of luminosity class in A stars is difficult. While the hydrogen 
lines are very sensitive to luminosity in early A stars, the A5 and later spectral types lose much of this sensitivity and at F0 the hydrogen 
lines become almost insensitive to the luminosity class. For F0, ionised lines of Fe and Ti become useful in determination of the luminosity class.

At late A spectral type, metallic (Am) stars are common. For these stars spectral types determined from the \ion{Ca}{ii} K-line 
and metal lines differ by five or more subclasses. The \ion{Ca}{ii}\,K line is weak in their spectra and thus yields an earlier spectral type than 
hydrogen lines, while the metal lines are enhanced and typical for a later subtype. They are denoted as, for example, kA2hA5mF0 (IV), 
where k, h and m precedes the spectral type obtained from the \ion{Ca}{ii}\,K line, hydrogen lines and metal lines, respectively. These symbols are 
followed by the luminosity class. If the luminosity class is in parentheses, as it is in this example, then the star exhibits the anomalous luminosity 
effect (ALE, see \citealt{gray&corbally2009}) where the violet part of the spectrum indicates a more luminous class than the blue part. 
Stars showing mild Am features, where there are fewer than five spectral subtypes between the \ion{Ca}{ii} K line and metal line types, 
are called marginal Am stars, denoted as ``Am:''. 

Most of the objects were classified as A stars, including slow, fast and very fast rotators (``s'', ``n'' and ``nn'' added to the spectral type, respectively). 
The classified CP stars are of Am (17), Ap CrSrEu (1) and $\lambda$\,Bootis types (2). Additional, mild peculiarities were discovered in some of the stars, 
such as weak Ca or metal lines. Only three stars were classified as late B, with one of them appearing to be He strong (He-s).
The assigned spectral classes are presented in Table~\ref{journal1} and some notes are provided in the comments below the table.

\section{Atmospheric parameters}
\label{sec:methods}
To perform an abundance analysis, one needs an appropriate atmospheric model of the star, which requires the effective temperature, surface gravity and metallicity to be known. To determine these parameters, we applied both photometric and spectroscopic methods. 

\begin{table*}
\centering
\caption{Atmospheric parameters of the investigated \textit{Kepler} A and F stars. We present parameters derived from the KIC photometry (KIC), 
those obtained from spectral energy distributions (Na\,D \& SED) and from analysis of metal and Balmer lines. The $1\sigma$ uncertainties are given for effective temperatures 
derived from SED fitting, and for \vsini\ and $\log\epsilon{\rm (Fe)}$ values derived from metal lines. The full table is available in electronic form.}
\begin{scriptsize}
\label{parameters1}
\begin{tabular}{c|cccc|cc|rccrc}
\toprule
KIC          &\multicolumn{4}{c}{\hrulefill \,KIC \,\hrulefill}
             &\multicolumn{2}{c}{\hrulefill \,Na\,D \& SED\,\hrulefill}
             &\multicolumn{5}{c}{\hrulefill \,Metal \& Balmer lines\,\hrulefill}\\
Number       & \teff      & \logg     & $E(B-V)$   & $\log\epsilon{\rm (Fe)}$   & $E(B-V)$   &  \teff    &  \multicolumn{1}{c}{\teff}   & \logg    &  \turb    &   \multicolumn{1}{c}{\vsini} & $\log\epsilon{\rm (Fe)}$\\
             & [K]        &           & [mag]      &                            & [mag]      &     [K]   & \multicolumn{1}{c}{[K]}      &          & [\kms]    &    \multicolumn{1}{c}{[\kms]}&  \\
\midrule

10263800  &$  8573 $&$   3.89 $&$   0.09 $&$   7.44 $&$   0.01 $&$    8590 \pm   210 $&$  8900 \pm   200 $&$   4.0 \pm    0.2 $&$   2.0 \pm    0.4 $&$   230 \pm     7 $&$   6.91 \pm   0.15 $\\
10264728  &$  7794 $&$   3.85 $&$   0.09 $&$   7.42 $&$   0.03 $&$    7790 \pm   200 $&$  7900 \pm   200 $&$   3.8 \pm    0.2 $&$   1.3 \pm    0.4 $&$   257 \pm     8 $&$   7.59 \pm   0.12 $\\
10355055  &$  8110 $&$   3.74 $&$   0.10 $&$   7.43 $&$   0.05 $&$    8420 \pm   270 $&$  8300 \pm   200 $&$   3.8 \pm    0.2 $&$   2.9 \pm    0.4 $&$   223 \pm     7 $&$   7.29 \pm   0.11 $\\
10533616  &$  8317 $&$   3.77 $&$   0.10 $&$   7.48 $&$   0.02 $&$    8260 \pm   250 $&$  8200 \pm   100 $&$   3.8 \pm    0.2 $&$   2.0 \pm    0.2 $&$   128 \pm     3 $&$   7.68 \pm   0.11 $\\
10549371  &$  6973 $&$   3.95 $&$   0.06 $&$   7.12 $&$   0.06 $&$    7300 \pm   170 $&$  7200 \pm   100 $&$   3.8 \pm    0.1 $&$   3.9 \pm    0.1 $&$    71 \pm     2 $&$   7.32 \pm   0.09 $\\
10555142  &$  6998 $&$   3.51 $&$   0.08 $&$   7.32 $&$   0.04 $&$    7180 \pm   160 $&$  7000 \pm   200 $&$   3.8 \pm    0.2 $&$   2.5 \pm    0.4 $&$   210 \pm     5 $&$   7.45 \pm   0.14 $\\
10590857  &$  7564 $&$   3.76 $&$   0.09 $&$   7.41 $&$   0.02 $&$    7470 \pm   180 $&$  7300 \pm   100 $&$   4.0 \pm    0.2 $&$   2.9 \pm    0.2 $&$   108 \pm     4 $&$   7.53 \pm   0.09 $\\
10721930  &$  8311 $&$   3.73 $&$   0.10 $&$   7.47 $&$   0.06 $&$    8480 \pm   210 $&$  8200 \pm   100 $&$   3.7 \pm    0.1 $&$   1.7 \pm    0.1 $&$    16 \pm     1 $&$   7.48 \pm   0.17 $\\
10977859  &$  8052 $&$   3.93 $&$   0.06 $&$   7.65 $&$   0.01 $&$    8400 \pm   250 $&$  8200 \pm   100 $&$   3.7 \pm    0.1 $&$   1.6 \pm    0.1 $&$    65 \pm     2 $&$   7.55 \pm   0.16 $\\
11013201  &$  7779 $&$   3.82 $&$   0.07 $&$   7.44 $&$   0.01 $&$    7740 \pm   260 $&$  7800 \pm   100 $&$   3.8 \pm    0.2 $&$   1.7 \pm    0.2 $&$   117 \pm     2 $&$   7.71 \pm   0.08 $\\

\bottomrule
\end{tabular}
\end{scriptsize}
\end{table*}

\subsection{Effective temperatures from Spectral Energy Distributions}
Effective temperature can be determined from the stellar spectral energy distribution (SED). For our target stars these were constructed from the available 
photometry, using 2MASS \citep{2006AJ....131.1163S}, Tycho $B$ and $V$ magnitudes \citep{1997A&A...323L..57H}, USNO-B1 $R$ magnitudes
\citep{2003AJ....125..984M}, and TASS $I$ magnitudes \citep{2006PASP..118.1666D}, supplemented, if available, by Geneva photometry \citep{1999yCat.2169....0R}, 
CMC14 $r'$ magnitudes \citep{2002A&A...395..347E} and TD-1 ultraviolet flux measurements \citep{1979BICDS..17...78C}.   

SEDs can be significantly affected by interstellar reddening. We have, therefore, estimated this parameter from the interstellar Na\,D$_{2}$ (5889.95\,{\AA}) lines. 
The equivalent widths of the D$_2$ lines were measured and $E(B-V)$ values were obtained using the relation given by \citet{1997A&A...318..269M}.
For resolved multi-component interstellar Na\,D$_{2}$ lines, the equivalent widths of the individual components were measured. The total $E(B-V)$ in these cases is
the sum of the reddening per component, since interstellar reddening is additive \citep{1997A&A...318..269M}. The SEDs were de-reddened using the analytical
extinction fits of \citet{1979MNRAS.187P..73S} for the ultraviolet and \citet{1983MNRAS.203..301H} for the optical and infrared.                    

Effective temperatures were determined by fitting \citet{1993KurCD..13.....K} model fluxes to the de-reddened SEDs. The model
fluxes were convolved with photometric filter response functions. A weighted Levenberg-Marquardt non-linear least-squares fitting procedure was used 
to find the solution that minimized the difference between the observed and model fluxes. Since \logg\ and metallicity [M/H] are poorly constrained by our SEDs, 
we fixed $\log g = 4.0$ and [M/H]\,$=0.0$\,dex for all of the fits. The results are given in Table\,\ref{parameters1}. 
The uncertainties in \teff\  include the formal least-squares error and
adopted uncertainties in $E(B-V)$ of $\pm$0.02\,mag, $\log g$ of $\pm$0.5\,dex and [M/H] of $\pm0.5$\,dex added in quadrature.

\subsection{Atmospheric parameters from spectroscopy}
We used atmospheric models and synthetic spectra to obtain atmospheric parameters, abundances of chemical elements and rotational velocities from the 
high-resolution HERMES spectra. All the necessary atmospheric models were computed with the line-blanketed, local thermodynamical equilibrium 
(LTE) {\small\sc ATLAS9} code \citep{1993KurCD..13.....K}. 
The physics of the models includes plane-parallel geometry, and hydrostatic and radiative equilibrium. 
Models were pre-calculated in a grid for effective temperatures between $5000$ and $12000$\,K with a step of $100$\,K, surface gravities 
from $2.0$ to $4.6$\,dex with a step of $0.1$\,dex, microturbulence velocities between $0.0$ and $6.0$\,km\,s$^{-1}$ with a step of $0.1$\,km\,s$^{-1}$, 
and for solar metallicity. Additional models of non-solar metallicity were calculated for smaller ranges of effective temperatures and surface gravities.
The synthetic spectra were computed with the {\small\sc SYNTHE} code \citep{1993KurCD..18.....K}. Both codes, {\small\sc ATLAS9} and {\small\sc SYNTHE}, 
were ported to GNU/Linux by \citet{sbordone} and are available online\footnote{wwwuser.oats.inaf.it/castelli/}. The stellar line identification and the abundance analysis 
in the entire observed spectral range were performed on the basis of the line list from \citet{castelli}. 

We analysed the metal lines following the spectrum synthesis methodology established by \citet{niemczurapolubek}
and \citet{niemczuramorel}. It relies on an efficient least-squares optimisation algorithm \citep[see][]{bevington, takeda}, and is appropriate over a wide range of \vsini\ values. 
Our observed $v\sin i$ range is $8$ to $283$\,\kms, which includes stars rotating too rapidly for equivalent width analysis to be possible due to line blending. The spectrum synthesis method allows 
for a simultaneous determination of various parameters influencing stellar spectra and consists of the minimisation of the deviation between the theoretical 
and observed spectra. The synthetic spectrum depends on stellar parameters such as \teff, \logg, \turb, \vsini, 
radial velocity \rad, and the relative abundances of the elements $\log\epsilon{\rm (El)}$, where $\rm {El}$ denotes the individual element. 
All atmospheric parameters are correlated. 
In our method, the \teff, {\logg} and \turb\ parameters were obtained prior to the determination of abundances of chemical elements and were considered as input 
parameters. The remaining parameters ($\log\epsilon{\rm (El)}$, {\vsini} and \rad) were determined simultaneously because they produce detectable and different spectral signatures. 
The \vsini\ values were determined by comparing the shapes of observed metal line profiles with the computed profiles, as shown by \citet{gray}. 

The starting values of \teff\ were taken from the SED fits, and values \logg\ came from the KIC. These parameters were then improved by using the sensitivity of Balmer 
lines to effective temperature and gravity, following the method proposed by \citet{catanzaro1}. However, hydrogen lines cease to be good indicators 
of \logg\ for effective temperatures cooler than $8000$\,K. Therefore, for stars with lower effective temperatures, the surface gravity 
was initially assumed to be equal to $4.0$\,dex. We used an iterative approach to minimize the differences between observed and synthetic H$\delta$, 
H${\gamma}$ and H$\beta$ profiles \citep[see][]{catanzaro1}. 

The effective temperatures and surface gravities determined from Balmer lines were improved through the analysis of lines of neutral and ionised iron.
In this method, we adjusted \teff, {\logg} and \turb\ by comparing the abundances determined from \ion{Fe}{i} and \ion{Fe}{ii} lines. 
The analysis was based on iron lines because they are the most numerous in our spectra. In general, we required that 
the abundances measured from \ion{Fe}{i} and \ion{Fe}{ii} lines yield the same result. 
In the spectra of stars with lower effective temperatures, the number of \ion{Fe}{i} lines outnumbers those of the \ion{Fe}{ii} lines. 
The strength of \ion{Fe}{i} lines depends on \teff, {\turb} and metallicity [M/H], but is practically independent of \logg. 
On the other hand, the lines of \ion{Fe}{ii} are weakly sensitive to effective temperature and metallicity, but highly sensitive 
to gravity \citep{gray}. First, we adjusted microturbulence until we saw no correlation 
between iron abundances and line depths for the \ion{Fe}{i} lines. Next, \teff\ was changed until we saw no trend in the abundance versus 
excitation potential for the \ion{Fe}{i} lines. Determinations of microturbulence and effective temperature are not independent, 
but since the influence of microturbulence on spectral lines is stronger, we adjusted this parameter first. Then, surface gravity is obtained by fitting the 
\ion{Fe}{ii} and \ion{Fe}{i} lines, and by requiring the same abundances from the lines of both ions. 
For hotter stars, whose spectra show more \ion{Fe}{ii} than \ion{Fe}{i} lines, the \ion{Fe}{ii} 
lines were used to determine \teff\ and \turb, whereas \logg\ was derived from Balmer lines \citep{gray}.
This approach is limited by stellar rotation, since line blending grows rapidly with \vsini. In the spectra of rapid rotators all the lines are blended and
it is impossible to use individual iron lines for \teff, \logg\ and \turb\ determination.
In such cases we assumed that the abundances obtained from \ion{Fe}{i} and \ion{Fe}{ii} lines are the same for the correct \teff, {\logg} and \turb.
The analysis relies on the same physics, regardless of rotation. The only difference is that for slow and moderate rotators we can use separate iron lines, 
including weak lines, to calculate the atmospheric parameters. For rapid rotators we are comparing iron abundances derived from broader 
spectral regions. Typically such regions contain many iron lines, both neutral and ionised. Of course, in such cases, stronger lines have more influence on the result.

Having derived \teff, \logg\ and \turb, the determination of abundances could be performed. The analysis of chemical abundances began with the selection 
of spectral regions suitable for abundance analysis. The length of a chosen part depended mainly on \vsini.  
For slowly rotating stars (\vsini\,$<$\,$40$\,\kms), short sections covering only one or a few blended spectral features were used. 
For stars with \vsini\,$>$\,$40$\,\kms\ we used broader spectral ranges re-normalised by comparison with theoretical spectra, if necessary. 
The next step was line identification in the chosen region. In the case of rapidly rotating stars, where all lines are blended, only the most important elements 
producing spectral features were considered. 
In our method we can take into account all elements which have lines in the chosen spectral region. Elements 
that have little or no influence on the spectra are included in the analysis, but no attempt is made to calculate their abundances.
The stellar rotation and signal-to-noise ratio of the spectrum dictate for which elements the abundance analysis is performed.

The line list used is that of \citet{castelli}, which is based on Kurucz's line list\footnote{http://kurucz.harvard.edu/linelists.html} and 
supplemented by additional sources. It was prepared for the analysis of chemically peculiar Ap and HgMn stars, hence is suitable for the analysis of our CP stars.
Each selected spectral region was analysed by the spectral synthesis method described above. We adjusted \rad, \vsini\ and chemical abundances to minimize
the difference between the calculated and observed spectra. The final solution was typically reached after ten iterations. The iterations stopped if the values of the 
determined parameters remained the same within 2\% for three consecutive steps. 
In our method, chemical abundances and \vsini\ values were determined from different parts of the spectrum. In the final step of the analysis, 
we derived the average values of \vsini\ and abundances of all the chemical elements considered for a given star and estimated the uncertainty on the derived abundances.

In our analysis we do not take into account macroturbulence, whose influence on the line profile is degenerate with \vsini. 
The effect of macroturbulence on F stars was discussed in detail by \citet{2011MNRAS.417..495F}, who found that stronger lines required higher macroturbulent velocities 
and interpreted this as the effect of unmodelled depth-dependent velocities in the atmosphere.
Since these are not observable in A stars, the required macroturbulent velocities remain uncertain, but the effect of macroturbulence is mainly limited to \vsini.
The other parameters can be slightly affected, especially in case of slowly rotating stars \citep{2011MNRAS.417..495F}. 
The \vsini\ values of slow rotators may be overestimated due to non-inclusion of macroturbulence. 
However, for \vsini\, $>$ $50$\,\kms\ the effect of modest (e.g. $10$\,\kms) macroturbulence is negligible at our resolution and signal-to-noise.

The atmospheric parameters obtained from our analysis are given in Table\,\ref{parameters1}, and the abundances of analysed elements are given in Table\,\ref{abundance-table}.

\begin{table*}
\centering
\caption{Abundances of chemical elements for a sample of analysed stars (on the scale in which $\log \epsilon$(H) = 12). The median value is given; standard deviations were calculated only if the number of analysed
parts of a spectrum was greater than three. In other cases the average value calculated from standard deviations of other elements are given. 
Number of analysed parts is given in brackets. The full table is available in the electronic version.}
\label{abundance-table}
\begin{tabular}{c|llllllll}
\toprule
   & KIC\,10263800 &     KIC\,10264728 &     KIC\,10355055 &     KIC\,10533616 &     KIC\,10549371 &     KIC\,10555142 &     KIC\,10590857\\ 
\midrule
    C & 8.85$\pm$0.14 (1) &   8.48$\pm$0.12 (2) &   8.90$\pm$0.12 (2) &   8.38$\pm$0.07 (7) &  8.31$\pm$0.14 (14) &   8.78$\pm$0.16 (5) &   8.36$\pm$0.19 (5)  \\
    N               & $-$ &                 $-$ &                 $-$ &                 $-$ &   8.33$\pm$0.14 (1) &                 $-$ &                 $-$  \\
    O & 8.74$\pm$0.14 (2) &   9.04$\pm$0.12 (1) &   8.44$\pm$0.12 (1) &   8.92$\pm$0.14 (2) &   8.83$\pm$0.14 (1) &                 $-$ &   8.91$\pm$0.17 (1)  \\
   Ne               & $-$ &                 $-$ &                 $-$ &                 $-$ &                 $-$ &                 $-$ &                 $-$  \\
   Na               & $-$ &   6.60$\pm$0.12 (1) &   7.10$\pm$0.12 (1) &   6.89$\pm$0.14 (2) &   6.46$\pm$0.14 (2) &   6.63$\pm$0.18 (1) &   6.22$\pm$0.17 (1)  \\
   Mg & 7.68$\pm$0.14 (1) &   8.01$\pm$0.12 (4) &   7.28$\pm$0.19 (3) &   7.98$\pm$0.10 (6) &   7.56$\pm$0.13 (8) &   7.64$\pm$0.18 (2) &   7.57$\pm$0.10 (4)  \\
   Al               & $-$ &                 $-$ &                 $-$ &                 $-$ &                 $-$ &                 $-$ &                 $-$  \\
   Si & 6.80$\pm$0.14 (2) &   7.94$\pm$0.12 (2) &   7.50$\pm$0.12 (2) &   7.60$\pm$0.18 (5) &  7.54$\pm$0.20 (21) &   7.52$\pm$0.29 (5) &   7.56$\pm$0.25 (8)  \\
    P               & $-$ &                 $-$ &                 $-$ &                 $-$ &                 $-$ &                 $-$ &                 $-$  \\
    S               & $-$ &                 $-$ &                 $-$ &                 $-$ &   7.32$\pm$0.01 (3) &                 $-$ &   7.29$\pm$0.17 (1)  \\
   Cl               & $-$ &                 $-$ &                 $-$ &                 $-$ &                 $-$ &                 $-$ &                 $-$  \\
    K               & $-$ &                 $-$ &                 $-$ &                 $-$ &                 $-$ &                 $-$ &                 $-$  \\
   Ca & 6.62$\pm$0.14 (2) &   6.64$\pm$0.12 (4) &   6.61$\pm$0.16 (8) &  6.65$\pm$0.16 (15) &  6.54$\pm$0.16 (25) &   6.30$\pm$0.23 (7) &   6.67$\pm$0.23 (4)  \\
   Sc & 4.03$\pm$0.14 (1) &   3.09$\pm$0.12 (3) &   2.96$\pm$0.11 (3) &   3.10$\pm$0.16 (5) &   3.24$\pm$0.10 (9) &   2.91$\pm$0.21 (4) &   3.28$\pm$0.13 (6)  \\
   Ti & 4.43$\pm$0.14 (3) &   5.07$\pm$0.16 (8) &   5.05$\pm$0.18 (9) &  5.22$\pm$0.17 (16) &  5.06$\pm$0.15 (47) &   5.05$\pm$0.15 (8) &  5.27$\pm$0.20 (14)  \\
    V               & $-$ &   2.79$\pm$0.12 (1) &                 $-$ &                 $-$ &   4.43$\pm$0.11 (9) &   4.67$\pm$0.18 (2) &                 $-$  \\
   Cr & 5.46$\pm$0.14 (2) &   5.69$\pm$0.19 (5) &   5.60$\pm$0.04 (6) &   5.86$\pm$0.09 (8) &  5.53$\pm$0.16 (26) &   5.69$\pm$0.18 (2) &  5.67$\pm$0.18 (16) \\
   Mn               & $-$ &   5.60$\pm$0.12 (2) &   5.29$\pm$0.12 (1) &   5.57$\pm$0.14 (2) &  5.35$\pm$0.16 (10) &   5.43$\pm$0.04 (3) &   5.59$\pm$0.17 (2)  \\
   Fe & 6.93$\pm$0.15 (5) &  7.54$\pm$0.12 (11) &  7.29$\pm$0.11 (17) &  7.70$\pm$0.11 (36) & 7.32$\pm$0.09 (101) &  7.45$\pm$0.14 (20) &  7.54$\pm$0.09 (36)  \\
   Co               & $-$ &                 $-$ &                 $-$ &                 $-$ &   5.88$\pm$0.14 (2) &                 $-$ &                 $-$ \\
   Ni               & $-$ &   6.36$\pm$0.07 (4) &   6.11$\pm$0.11 (6) &  6.35$\pm$0.23 (11) &  6.19$\pm$0.18 (41) &   6.27$\pm$0.13 (6) &  6.41$\pm$0.16 (13)  \\
   Cu               & $-$ &                 $-$ &                 $-$ &                 $-$ &   4.41$\pm$0.14 (1) &                 $-$ &                 $-$  \\
   Zn               & $-$ &                 $-$ &                 $-$ &                 $-$ &   4.78$\pm$0.14 (2) &                 $-$ &                 $-$  \\
   Ga               & $-$ &                 $-$ &                 $-$ &                 $-$ &                 $-$ &                 $-$ &                 $-$  \\
   Sr               & $-$ &   2.55$\pm$0.12 (1) &   1.67$\pm$0.12 (1) &   2.96$\pm$0.14 (2) &   3.23$\pm$0.14 (1) &   1.36$\pm$0.18 (1) &                 $-$  \\
    Y               & $-$ &   3.14$\pm$0.12 (1) &   2.24$\pm$0.12 (1) &   2.82$\pm$0.14 (1) &  2.51$\pm$0.17 (10) &   3.03$\pm$0.29 (3) &   2.85$\pm$0.17 (1)  \\
   Zr               & $-$ &                 $-$ &                 $-$ &                 $-$ &  3.18$\pm$0.20 (10) &   2.98$\pm$0.18 (1) &                 $-$  \\
   Ba               & $-$ &   3.78$\pm$0.12 (2) &   3.11$\pm$0.12 (1) &   2.48$\pm$0.20 (3) &   2.75$\pm$0.13 (3) &   2.69$\pm$0.18 (2) &   2.43$\pm$0.17 (2)  \\
   La               & $-$ &                 $-$ &                 $-$ &                 $-$ &   1.60$\pm$0.14 (2) &                 $-$ &                 $-$  \\
   Ce               & $-$ &                 $-$ &                 $-$ &                 $-$ &   1.69$\pm$0.14 (2) &                 $-$ &                 $-$ \\
   Pr               & $-$ &                 $-$ &                 $-$ &                 $-$ &                 $-$ &                 $-$ &                 $-$  \\
   Nd               & $-$ &                 $-$ &                 $-$ &                 $-$ &   1.72$\pm$0.20 (5) &                 $-$ &                 $-$ \\
\bottomrule
\end{tabular}
\end{table*}

%____________________________________________________________________________

\subsection{Discussion of uncertainties}
The derived atmospheric parameters and chemical abundances are influenced by errors from a number of sources. First of all, the adopted atmosphere models 
and synthetic spectra were calculated taking into account several simplifications, e.g. LTE, 
plane parallel geometry (1-D), and hydrostatic equilibrium \citep{1993KurCD..13.....K}, instead of a 3-D and non-LTE approach 
\citep[e.g.][]{bergemann2012}. 

Recently, the non-LTE line formation for \ion{Fe}{i} and \ion{Fe}{ii} for stars with \teff\ between $6500$\,K and $8500$\,K and \logg\, between $3.0$ and $4.0$\,dex 
was discussed by \citet{2011mast.conf..314M}. In this paper the non-LTE calculations were performed with a complete model atom, as described therein.
The departures from LTE caused the depletion of total absorption in the \ion{Fe}{i} lines and positive abundance corrections, leading to an underestimation 
of Fe abundance in LTE calculations. However, the corrections to LTE abundances were found to be smaller than $0.1$\,dex for main-sequence stars. 
For \ion{Fe}{ii}, the non-LTE effect was again found to be negligible, with corrections of  $-0.01$ to $-0.03$\,dex over the whole range of considered stellar parameters.
Most stars in our sample have effective temperatures lower than $8500$\,K. The non-LTE corrections of Fe abundances for these stars less than $0.1$\,dex. 
Application of this correction would change \logg\ determinations from the ionisation balance of \ion{Fe}{i} and \ion{Fe}{ii} lines. 
If we add non-LTE corrections to the obtained abundances of \ion{Fe}{i}, the determined surface gravity would be higher by about $0.1$\,dex. However, in most of our investigated stars, 
high projected rotational velocities prevents us from obtaining abundances of \ion{Fe}{i} and \ion{Fe}{ii} separately. 
Therefore, we did not include the non-LTE corrections in the present analysis.

Other important factors affecting spectral analysis are atomic data \citep{kurucz-alllines}, quality (resolution, S/N), wavelength range of the observed spectra and their normalisation. 
It is quite easy to normalise the optical spectrum of an A star with \vsini\ $< 40$\,\kms. The difficulty increases for heavily blended spectra of stars with \vsini\ much 
higher than this value, even for high-resolution and high S/N data. 

All the above-mentioned factors influenced our derived atmospheric parameters and chemical abundances. The chemical abundances are also affected by inaccurate  atmospheric parameters.
The influence of errors in atmospheric parameters on the derived chemical abundances can be checked by perturbing the synthetic spectrum by the adopted step size in the grid of atmospheric models. 
For example, $\Delta$\teff\,$ = 100$\,K leads to changes in abundances smaller than $0.1$\,dex 
in most cases. Similarly, $\Delta$\logg\,$ = 0.1$ changes chemical abundance by about $0.05$\,dex, and $\Delta$\turb\,$ = 0.1$\,\kms\ changes abundances by $0.15$\,dex at most. 
It should be noted here that the uncertainty on the microturbulence velocity increases with \vsini, from  $0.1$\,\kms\ for 
small and moderate velocities, to $0.4$\,\kms\ for rapid rotators. Chemical abundances determined for rapidly rotating stars are affected by the greatest uncertainty.
The combined errors of chemical abundances were calculated as:
\begin{displaymath}
\Delta \log \epsilon_{el}^{\rm tot} = \sqrt{(\Delta \log \epsilon_{el}^{T_{\rm eff}})^2 +(\Delta \log \epsilon_{el}^{\log g})^2 + (\Delta \log \epsilon_{el}^{\xi_{\rm t}})^2}
\end{displaymath}
In most cases $\Delta \log \epsilon_{el}^{\rm tot}$ was less than $0.20$\,dex.

%%%%%%%%%%%%%%%%%%%%%%%%%%%%%%%%%%%%%%%%%%%%%%%%%%%%%%%%%%%%%%%%%%%%%%%%%%%%%%%%%%%%%%%%%%%%%%%%%%%%%%%%%%%%%%%%%
\subsection{Comparison of results from different methods}

\begin{figure*}
\centering
\includegraphics[width=18cm,angle=0]{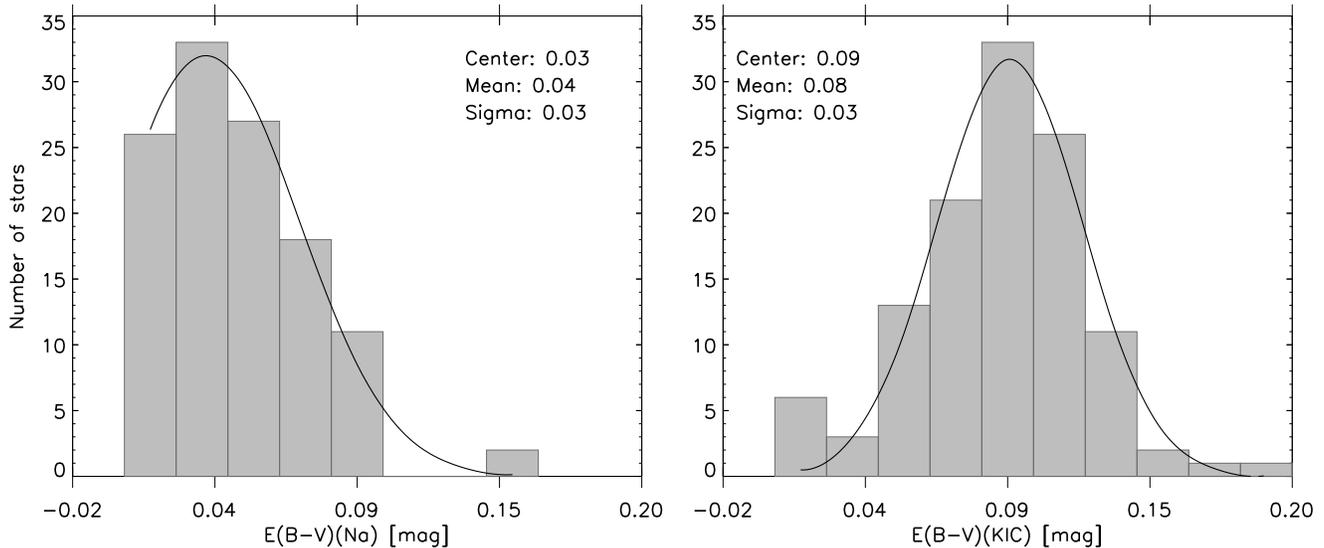}
\caption[]{Distributions of $E(B-V){\rm(Na)}$ (left panel) and $E(B-V){\rm(KIC)}$ (right panel). Gaussian fits are shown by solid lines. 
Central values, arithmetic means and standard deviations of both distributions are indicated in figures.}
\label{ebvcomparison1}
\end{figure*}

\begin{figure*}
\centering
\includegraphics[width=18cm,angle=0]{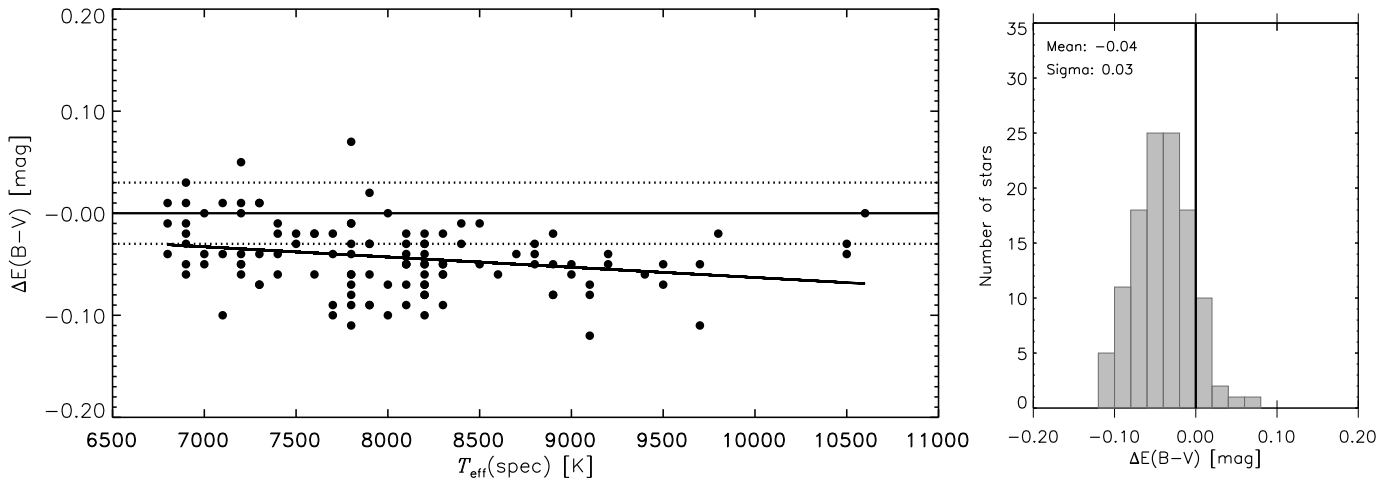}
\caption[]{Left panel: differences $\Delta E(B-V) = E(B-V){\rm(Na)} - E(B-V){\rm(KIC)}$ as a function of \teff({\sc spec}). A linear function is fitted. 
The dotted horizontal line indicates a difference of $\pm 0.03$ mag. 
Right panel: the distribution of differences $\Delta E(B-V)$. Arithmetic mean and standard deviation of the distribution are indicated in the figure.}
\label{ebvcomparison}
\end{figure*}

%--
\subsubsection{\it Interstellar reddening}
The $E(B-V)$ values from the KIC, $E(B-V){\rm (KIC)}$, and those determined here from the equivalent widths of sodium D$_2$ lines, $E(B-V){\rm (Na)}$, 
are given in Table\,\ref{parameters1}. The distributions of these parameters are presented in Fig.\,\ref{ebvcomparison1}. 
It is clear that the photometric interstellar reddening given in the KIC is higher than $E(B-V){\rm (Na)}$. The average value 
of $E(B-V)(\rm KIC)$ is equal to $0.09$\,mag, whereas from sodium lines it is $0.03$\,mag. The reddenings obtained from Na\,D$_{2}$ lines 
are lower than $0.10$\,mag for all stars, except for KIC\,9552758 (HD\,190293) and KIC\,8915335 (HD\,190566), for which they equal $0.14$\,mag. 
We can see from Fig.\,\ref{ebvcomparison} that the differences between interstellar reddenings obtained from Na\,D$_{2}$ lines and 
those listed in KIC slightly increase with effective temperature. As explained by \citet{brown}, the values of $E(B-V)$ given in the KIC were approximated using 
a simple model of dust distribution and with an adopted disk scale-height of $150$\,pc -- larger than suggested by earlier estimates \citep[e.g.][]{koppen, marshall}.
\citet{brown} assumed a wavelength dependence of reddening taken from \citet{cardelli} and used $R_{\rm V} = 3.1$ in all cases.
As described by \citet{brown}, reddening was precomputed for all filters and for a range of stellar parameters. 
Then, the authors calculated the reddening vectors for typical stars in the \textit{Kepler} field (\teff\,$= 5000$\,K, \logg\,$= 4.0$, [M/H]$=0.0$) 
and applied them to all stars. Values of interstellar reddening influenced the other parameters determined 
from KIC photometry, especially the effective temperatures. 
For this reason, it is necessary to check the available KIC $E(B-V)$ values by determining them with an independent method. \\
%--

\begin{figure*}
\centering
\includegraphics[width=18cm,angle=0]{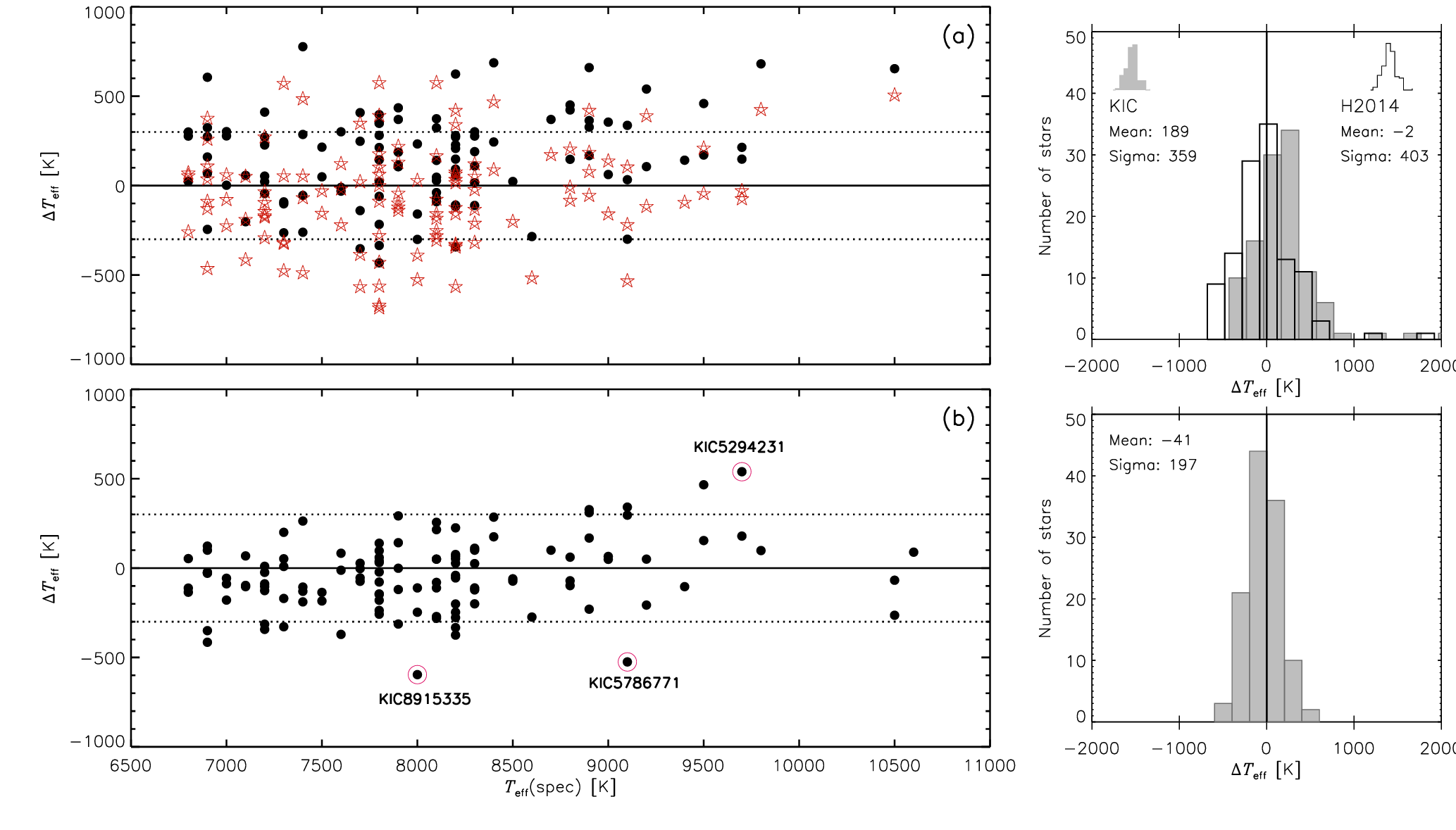}
\caption[]{A comparison of effective temperatures derived by different methods. In the panel (a) the differences $\Delta$\teff\,$=$\teff({\sc spec})$-$\teff({\sc KIC})
are plotted as filled circles, whereas $\Delta$\teff\,$=$\teff({\sc spec})$-$\teff({\sc H2014}) are shown as stars. In the panel (b) 
$\Delta$\teff\,$=$\teff({\sc spec})$-$\teff({\sc SED}) are plotted. The stars described in the text are indicated by names and open circles. 
The distributions of differences $\Delta$\teff\ and their characteristics are given in the right panels.}
\label{tempcomparison}
\end{figure*}

%--
\subsubsection{\it Effective temperatures}
The effective temperatures determined using the three methods described in Sect.\,\ref{sec:methods} are presented in Table\,\ref{parameters1} and compared 
in Fig.\,\ref{tempcomparison}, where the differences between \teff({\sc spec}) obtained from the analysis of the high-resolution spectra and values taken 
from the KIC, \teff({\sc KIC}), and from SED fitting, \teff(SED), are shown as a function of \teff({\sc spec}). 

For most of the stars, the values of \teff({\sc spec})$-$\teff({\sc KIC}) agree to within $\pm 300$\,K.
Differences larger than $500$\,K occur for stars with peculiarities identified in their spectra (KIC\,3231985, 4150611, and 8703413) and for some 
rapidly rotating stars with \vsini\ exceeding $100$\,\kms\ (KIC\,11506607, 3347643, 4681323, 5471091, and 6450107).
The Spearman's and Kendall's rank correlation coefficients between 
\teff({\sc spec})$-$\teff({\sc KIC}) and \teff({\sc spec}) are equal to $0.25$ and $0.17$, respectively, indicating very low correlation. 
The \teff({\sc KIC}) values are underestimated for stars hotter than about $7500$\,K. For stars with 
lower effective temperatures the agreement is better and the outliers can be explained by rapid rotation or 
chemical peculiarities in their spectra, which can influence the determination of atmospheric parameters.
This result is in agreement with the previous spectroscopic determinations of atmospheric parameters given by \citet{molenda2011}, \citet{lehmann2011} and 
\citet{tkachenko2013a}, where the systematic underestimations of the KIC \teff\ values for stars hotter than about $7000$\,K were pointed out.
These deviations probably originate from incorrect values of interstellar reddening.

Recently, \citet{huber2014} (hereafter H2014) published a new catalogue of revised effective temperatures for $196,468$ \textit{Kepler} stars.
The catalogue is based on a compilation of literature values for atmospheric properties (\teff, {\logg} and [Fe/H]) derived from different methods, including
photometry, spectroscopy, asteroseismology, and exoplanet transits. Effective temperature is available from the H2014 catalogue for all targets in our sample.
In Fig.\,\ref{tempcomparison}a we show the differences between \teff({\sc H2014}) and \teff({\sc spec}). As one can see, there is a 
similar scatter in differences [\teff({\sc spec})$-$\teff({\sc H2014})], as was the case for \teff({\sc spec})$-$\teff({\sc KIC}). 
For stars with \teff\,$> 8000$\,K, the differences [\teff({\sc spec})$-$\teff({\sc H2014})] are lower than in the case of \teff({\sc KIC}), 
which means that the \teff({\sc H2014}) values are higher than \teff({\sc KIC}) for such stars. 
Neither \teff({\sc KIC}) nor \teff({\sc H2014}) are systematically underestimated between $7000$ and $8000$\,K.

Figure\,\ref{tempcomparison}\,b compares \teff({\sc spec}) with that determined from the spectral energy distribution, \teff(SED). 
The consistency between these two methods is good: there is no significant trend of \teff({\sc spec})$-$\teff(SED) with effective temperature. 
The Spearman's and Kendall's rank correlation coefficients between \teff({\sc spec})$-$\teff({\sc SED}) and \teff({\sc spec}) are equal to $0.23$ and $0.16$, respectively. 
This shows that the correlation is smaller than between \teff({\sc spec})$-$\teff({\sc KIC}) and \teff({\sc spec}).
The differences exceed $500$\,K only for three rapidly rotating stars: KIC\,5294231, 5786771, and 8915335. 
The \teff({\sc SED}) analysis supports the claimed accuracy of our spectroscopic temperatures.\\
%--

\begin{figure*}
\centering
\includegraphics[width=18cm,angle=0]{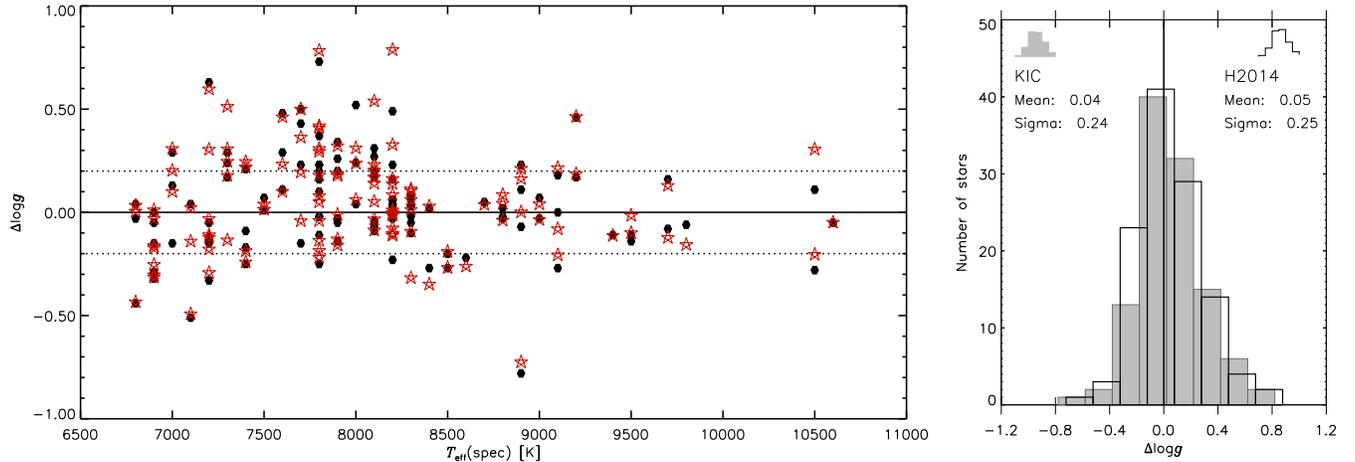}
\caption[]{Left panel: the differences $\Delta$\logg\,$=$\,\logg({\sc spec})$-$\logg(KIC) (filled circles) and $\Delta$\logg\,$=$\,\logg({\sc spec})$-$\logg(H2014) 
(stars) as a function of \teff({\sc spec}). Right panel: the distributions of differences $\Delta$\logg\ and their characteristics.}
\label{loggcomparison}
\end{figure*}

%--

\subsubsection{\it Surface gravities}
Figure\,\ref{loggcomparison} compares the \logg({\sc spec}) values obtained from high-resolution spectroscopy with those taken from the KIC, \logg(KIC), and the H2014 
catalogue, \logg(H2014). For most stars the surface gravities are consistent to within $\pm0.2$\,dex. 
The distribution of differences \logg({\sc spec})$-$\logg(KIC) and \logg({\sc spec})$-$\logg(H2014) does not indicate any trend with surface gravity or effective temperature. 
The higher differences were determined mostly for rapidly rotating stars with \vsini\ $> 100$\,\kms. 
Irrespective of rotational velocity, spectroscopic surface gravity values determined from iron lines are adopted for subsequent analysis of these stars.
The high-resolution spectra are much more sensitive to \logg\ than the photometric indices upon which the KIC is based.\\
%--

\begin{figure}
\centering
\includegraphics[width=8cm,angle=0]{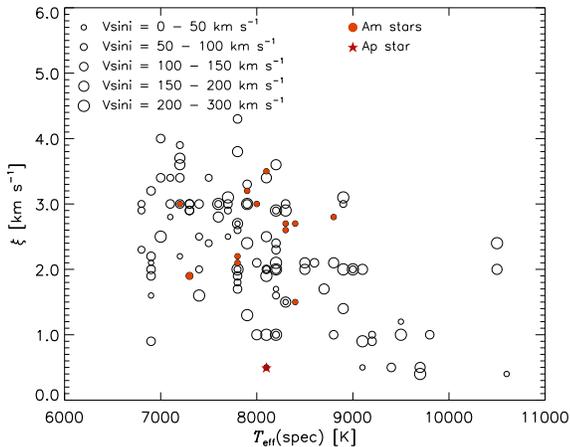}
\caption[]{Microturbulent velocities as a function of effective temperature. Non-CP stars are shown as open circles, Am and Ap stars as filled circles and
star symbol, respectively. Sizes of symbols for non-peculiar stars depend on rotational velocity, see the legend.}
\label{micro-calc}
\end{figure}

\subsubsection{\it Microturbulent velocities}%
In Fig.\,\ref{micro-calc} the derived values of microturbulence are shown as a function of effective temperature, with different
symbols for CP and non-CP stars. The values of \turb\ obtained decrease towards higher effective temperature. 
For stars with \teff\ between $7000$ and $8000$\,K, \turb\ is within the range $2-4$\,\kms, for those with \teff\ between $8000$ and $9000$\,K, \turb\ is approximately
$2$\,\kms, while for even higher values of \teff\ microturbulence decreases to about $0.5-1$\,\kms.
This behaviour is in agreement with the results of \citet{2014psce.conf..193G}, based on the analysis of high-resolution spectroscopic data of 
61 A field stars, 55 A and 58 F in open clusters (Pleiades, Coma Berenices, Hyades and Ursa Major). The results are also in agreement with \citet{2004IAUS..224..131S}. 
The determined microturbulent velocities of CP stars are similar to those of non-CP stars. 
According to \citet{landstreet2009} the microturbulent velocity in Am stars is generally higher than in normal stars. This tendency is not seen in our sample. 
However, in \citet{landstreet2009} it is not possible to directly compare the microturbulence velocities of the Am stars with normal stars of the same temperature, 
because, as \citet{2014PhDT.......131M} notes, the temperature distributions do not overlap.
We conclude that in our sample the distribution of microturblent velocities among the Am stars appears to be the same as that of normal stars.
As shown by \citet{landstreet2009}, the line profiles of slowly rotating Am stars show different asymmetries and anomalies caused by velocity fields. 
These asymmetries affect line shapes and equivalent widths and lead to a higher microturbulence velocities in comparison with chemically normal A stars.
However, since the analysis of \citet{landstreet2009} is based on a sample of stars with low projected rotation velocity (\vsini\ $<12$\,\kms), it cannot be ruled out 
that this phenomenon is diluted for faster rotators. It is beyond the scope of this paper to examine this effect in more detail.

As expected, the microturbulent velocity determined for the Ap star is small, \turb\,$=0.5\pm0.2$\,\kms\ \citep[e.g.][]{folsom2007}.

\begin{figure*}
\centering
\includegraphics[width=18cm,angle=0]{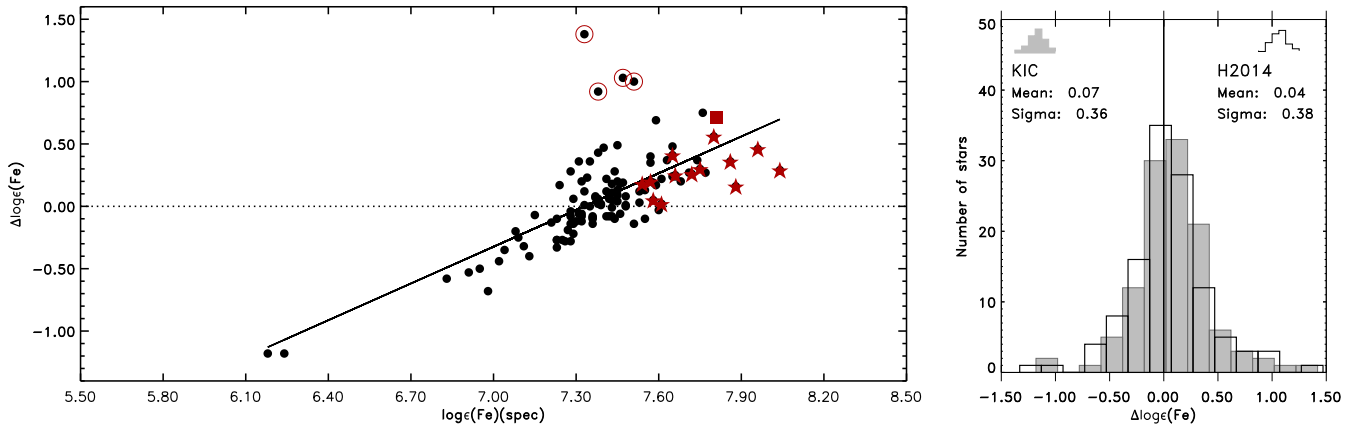}
\caption[]{Left panel: the differences $\Delta{\log\epsilon \rm (Fe)} = \log\epsilon{\rm (Fe)}$({\sc spec})\,$-\log\epsilon{\rm (Fe)}$(KIC) as a function of 
$\log\epsilon{\rm (Fe)}$({\sc spec}). Non-CP and CP stars are plotted by different symbols as explained in the legend.
The four most deviant objects are encircled (see the description in the text). The line is the linear fit to all points.
Right panel: the distribution of differences $\Delta \log\epsilon{\rm (Fe)}$ and their characteristics.}
\label{fedohcomparison}
\end{figure*}

%--

\subsubsection{\it Iron abundances}
Figure~\ref{fedohcomparison} compares the iron abundances derived from high-resolution spectra, $\log\epsilon{\rm (Fe)}$({\sc spec}), 
with those from the KIC, $\log\epsilon{\rm (Fe)}$(KIC). As can be seen, the differences $\log\epsilon{\rm (Fe)}$({\sc spec})$ - \log\epsilon{\rm (Fe)}$(KIC) are 
strongly correlated with the iron abundance. 
The same situation occurs for differences between $\log\epsilon{\rm (Fe)}$({\sc spec}) and $\log\epsilon{\rm (Fe)}$(H2014), taken from
the H2014 catalogue. The Spearman's and Kendall's rank correlation coefficients are equal to $0.74$ and $0.57$, respectively.
Only four stars (KIC\,4150611, 6951642, 7119530, 7299869; the encircled symbols in Fig.~\ref{fedohcomparison})
do not follow this general tendency. KIC\,4150611 is a possible triple system classified 
as F1\,V with weak metal lines suggesting spectral type A9. This classification was not confirmed by detailed analysis 
of chemical abundances, as iron abundance is only $0.16$\,dex lower than the solar abundance from \citet{asplund2009}. 
The other elements do not show significant deviations from the solar values either. This effect can be caused by analysing a triple object as a single one.
Additionally, the rotational velocity of this star is $128\pm5$\,\kms. For a star with \teff\,$=7400$\,K, this causes strong blending of all metal lines. 
It is worth mentioning that the differences between effective temperatures derived from the KIC photometry, SED, and high-resolution spectrum are also high for this star. 
Rapid rotation (\vsini\,$ > 120$\,\kms) is also likely to be responsible for the deviations in chemical abundances for the other three stars. 
Additionally, the analysis of high-resolution spectra resulted in higher iron abundances for KIC\,4768731, 
an  Ap (SrCrEu) star for which a high iron abundance is expected. 
The same trend of $\log\epsilon{\rm (Fe)}$({\sc spec})$ - \log\epsilon{\rm (Fe)}$(KIC) with iron abundance was noticed by \citet{tkachenko2013a}.
We found no correlation of $\log\epsilon{\rm (Fe)}$({\sc spec})$ - \log\epsilon{\rm (Fe)}$(KIC) 
differences with effective temperature or surface gravity. 

%--

\section{Projected rotational velocities}
\label{sec:rotational}

\begin{figure*}
\centering
\includegraphics[width=18cm,angle=0]{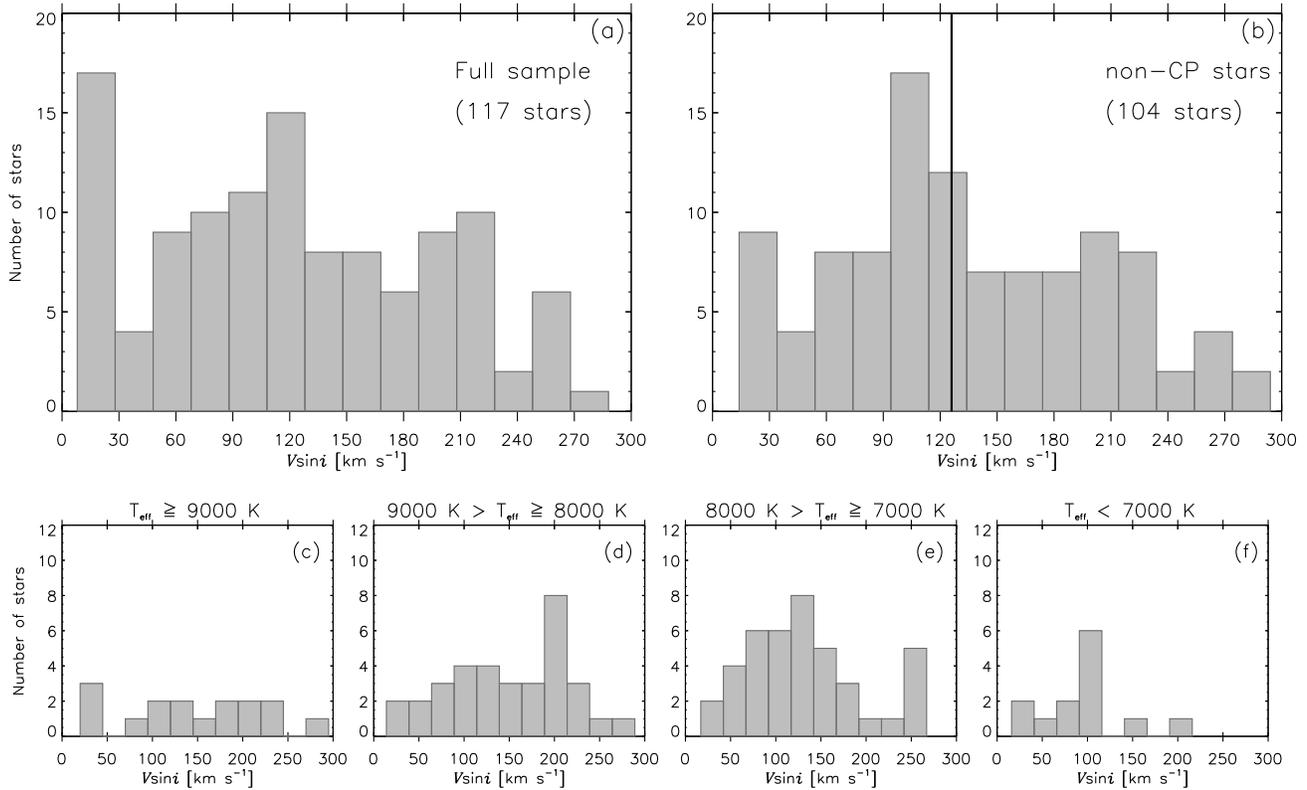}
\caption[]{The distribution of rotational velocities for all analysed stars (panel {\it a}) and for all non-CP stars (panel {\it b}). 
Panels {\it c, d, e}, and {\it f} show distributions of \vsini\ for stars with \teff\,$ \geq 9000$\,K, $9000 > $\teff\,$ \geq 8000$\,K, 
$8000 > $\teff\,$ \geq 7000$\,K, and \teff\,$ < 7000$\,K, respectively.}
\label{vsini}
\end{figure*}

Rotational velocity is one of the fundamental parameters that influences stellar evolution. It has been frequently demonstrated that stellar 
rotation strongly depends on spectral type, with most A stars being rapid rotators \citep[e.g.][]{royer2009, abt2009}.

Figure\,\ref{vsini}\,a shows the distribution of derived values of \vsini. The stars analysed have projected rotational velocities ranging 
from $8$ to about $280$\,\kms. Stars with \vsini\ lower than $40$\,\kms\ are considered here as slowly rotating stars. 
There are 21 such objects in our sample. The obvious members of this group are all either Am or Ap (CrSrEu) stars. 
These objects will be thoroughly discussed in the next section. Another slowly-rotating star, KIC\,3231985, also shows small peculiarities 
its spectrum and was classified as A4\,IV\,Ca\,weak\,(A3). Figure\,\ref{vsini}\,b shows the distribution of \vsini\ for all non-CP stars. 
One can see that the peak for slow rotation has disappeared and the mean value of \vsini\ for this sample is equal to $134$\,\kms. 
The maximum of the distribution is located around $110$\,\kms. This result is consistent with the rotational velocities of A stars given in the 
literature \citep{royer2009}. \citet{abt2000} pointed out that most of the normal A0--F0 main-sequence stars have equatorial rotational velocities 
greater than $120$\,\kms. Panels c-f in Fig.\,\ref{vsini} show distributions of \vsini\ for stars with different ranges of \teff. 
Most of the stars with \teff\,$ < 7000$\,K have projected rotational velocities below $100$\,\kms, consistent with the predictions for the late A and 
early F stars \citep{royer2009}. 

There are nine slowly rotating stars in our sample that were not classified as CP objects. Three of them were classified as 
F stars: KIC\,6519869 (F1\,IV-Vs), KIC\,7661054 (F2.5\,V), and KIC\,4077032 (F2\,IIIs). On average, F stars have smaller rotation
velocities than A stars. The rotation velocity drops rapidly from high values ($> 100$\,\kms) for late A stars to low values ($< 20$\,\kms) for late F 
stars \citep[][]{1986LNP...254..130N, royer2009}.
All slowly rotating non-CP F stars have abundance patterns close to solar, except for some elements. For KIC\,4077032 heavy and rare-earth elements are overabundant, 
while for KIC\,6519869 only abundances of elements derived from weak blends are significantly different from solar.
Four of the slow rotators were classified as mid-A stars: KIC\,10721930 (A5\,IV), KIC\,3231985 (A4\,IV Ca weak (A3)), KIC\,5355850 (A5\,IVs), and KIC\,8386982 (A4\,IV/V:).
For KIC\,10721930 heavy and rare-earth elements are overabundant. For KIC\,3231985 almost all elements are overabundant. This star has been identified as a binary  
\citep[e.g.][]{dommanget2002, fabricius2002}. Two objects were classified as early-A stars: KIC\,12153021 (A2\,IV-V) and KIC\,12736056 (A0.5\,IIIs).

The occurrence of non-CP A stars with low \vsini\ has been  
discussed in many papers \citep[e.g. ][]{takeda2008, abt2009, royer2014}. \citet{abt2009} presented this problem in detail, considereding
binarity and the amount of time necessary for slow rotators to become Ap or Am stars by a diffusion mechanism 
(see also \citet{2006ApJ...645..634T} and \citet{2013EAS....63..199M}). 
The frequency of binaries among A stars turned out to be normal. On the other hand, the time-scale for developing peculiarities is 
rapid for Ap(Si), Ap(HgMn), and Am stars, but slow for Ap(SrCrEu) stars, for which it takes about half of their main-sequence lifetime. 
According to \citet{abt2009}, if a constant formation rate of field A stars is assumed, about half of the potential Ap(SrCrEu) stars will appear 
as normal slow rotators and that is why there are normal slow rotators in the A0-A3 group. 
Therefore, slowly rotating, non-CP stars could be Ap stars that already underwent magnetic braking, but do not yet show chemical peculiarity.
Accordingly, the five slowly rotating stars from our sample (KIC\,10721930, 12153021, 12736056, 5355850, and 8386982) might become 
Ap(CrSrEu) stars in the future.

A more probable reason for the low observed \vsini\ in some A stars is the projection effect. 
The influence of this effect on the spectral lines of Vega has been investigated by \citet{gulliver1994}. For many years Vega was considered 
as a standard star with \vsini\ equal to $24$\,\kms\ \citep[e.g. ][]{royer2007}. 
\citet{gray1985vega} first suggested that Vega is a rapid rotator seen pole-on. \citet{gulliver1994} analysed high-resolution, high signal-to-noise
spectra of Vega and found out two distinct types of profiles. The strong lines exhibit classical rotational profiles with enhanced wings, 
but the weak lines have different, flat-bottomed profiles. 
\citet{gulliver1994} used {\small\sc ATLAS9} model atmospheres and {\small\sc SYNTHE} synthetic spectra to model the spectral
lines and confirmed that Vega is in fact a rapidly rotating, nearly pole-on star with a gradient in temperature and gravity in the photosphere. 
The equatorial rotational velocity of Vega ($V_{\rm eq} = 245 \pm 15$\,\kms) and inclination ($5.1^{\circ} \pm 0.3 $) were derived 
by fitting the flat-bottomed line profiles of \specline{Fe}{i}{4528}\AA\ and \specline{Ti}{ii}{4529}\AA. 
To investigate the effect that low inclination has on the observed spectrum of a rapid rotator, 
high-resolution (R\,$\geq 70000$), and very high signal-to-noise (S/N\,$\sim 1000$) spectra are necessary. 
\citet{takeda2008} found that lines of ionized elements in the spectrum of Vega show classical profiles, while the lines of neutral elements 
display very characteristic shapes: from flat-bottomed to reversed profiles, due to the temperature gradient from the pole to the equator. 

\citet{royer2012} analysed a sample of A stars to find possible rapid rotators seen pole-on. The line profiles in each spectrum were analysed globally, 
using a Least-Square Deconvolution (LSD) method \citep[][]{donati1997, kochukhov2010} to enhance the signal and recover the broadening function. 
To investigate the possibility that the slow rotators in our sample are in fact rapid rotators seen pole-on, we followed the methods presented 
by \citet{gulliver1994} and \citet{takeda2008}. We used the revised LSD code \citep{tkachenko2013} to calculate the LSD profiles for intrinsic line depths 
from $0$ to $0.6$ and from $0.6$ to $1.0$, which enables the distinction between the line profiles of strong and faint lines in the spectra of 
the slow rotators.
All line lists were computed with the individual parameters of \teff, \logg, \vsini, {\turb} and [M/H] determined in this paper. 
The boundary for the separation of line depths was chosen so that the S/N in both LSD profiles, computed from weak and strong lines, were approximately equal. 
The wavelength range used (4890 to 5670\,\AA) included the region between H$\beta$ and the first strong telluric lines. We found no significant difference in profile shapes for any of the investigated stars, which can be attributed to the insufficient quality of our spectra.

\section{Chemical abundances}
\label{sec:abundances}

\begin{figure*}
\centering
\includegraphics[width=18cm,angle=0]{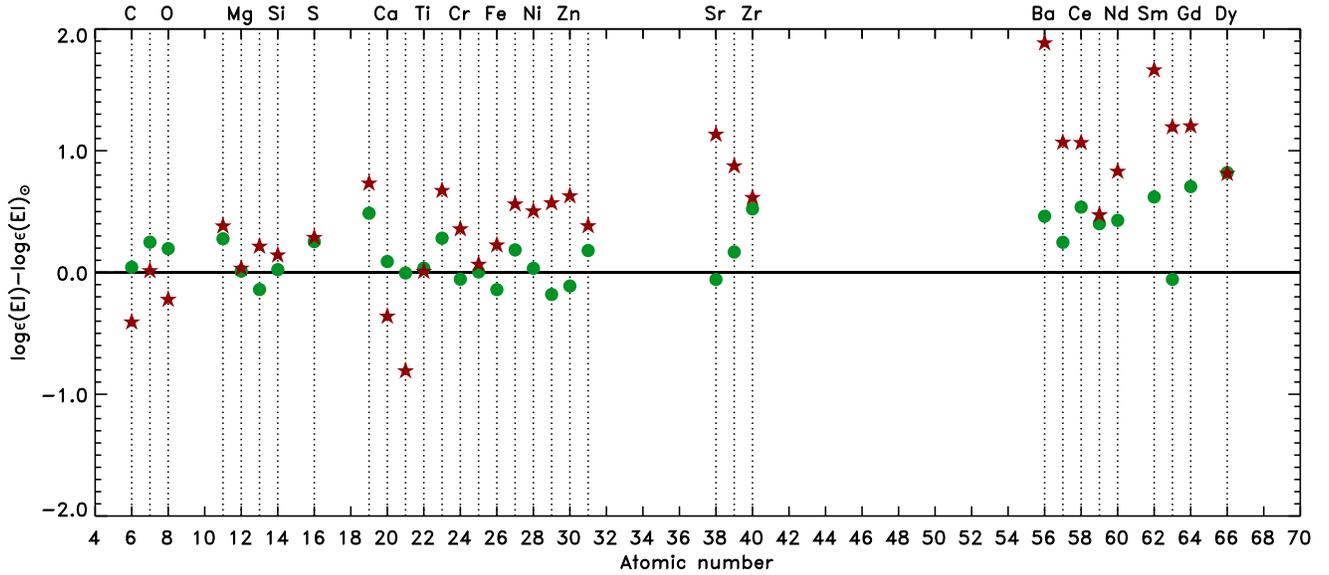}
\caption[]{The difference between the obtained chemical abundances and solar values \citep{asplund2009} as a function of the atomic number. 
The results for non-CP stars are given as filled circles; for CP stars as filled stars.}
\label{abundance-average-fig}
\end{figure*}

\begin{figure*}
\centering
\includegraphics[width=18cm,angle=0]{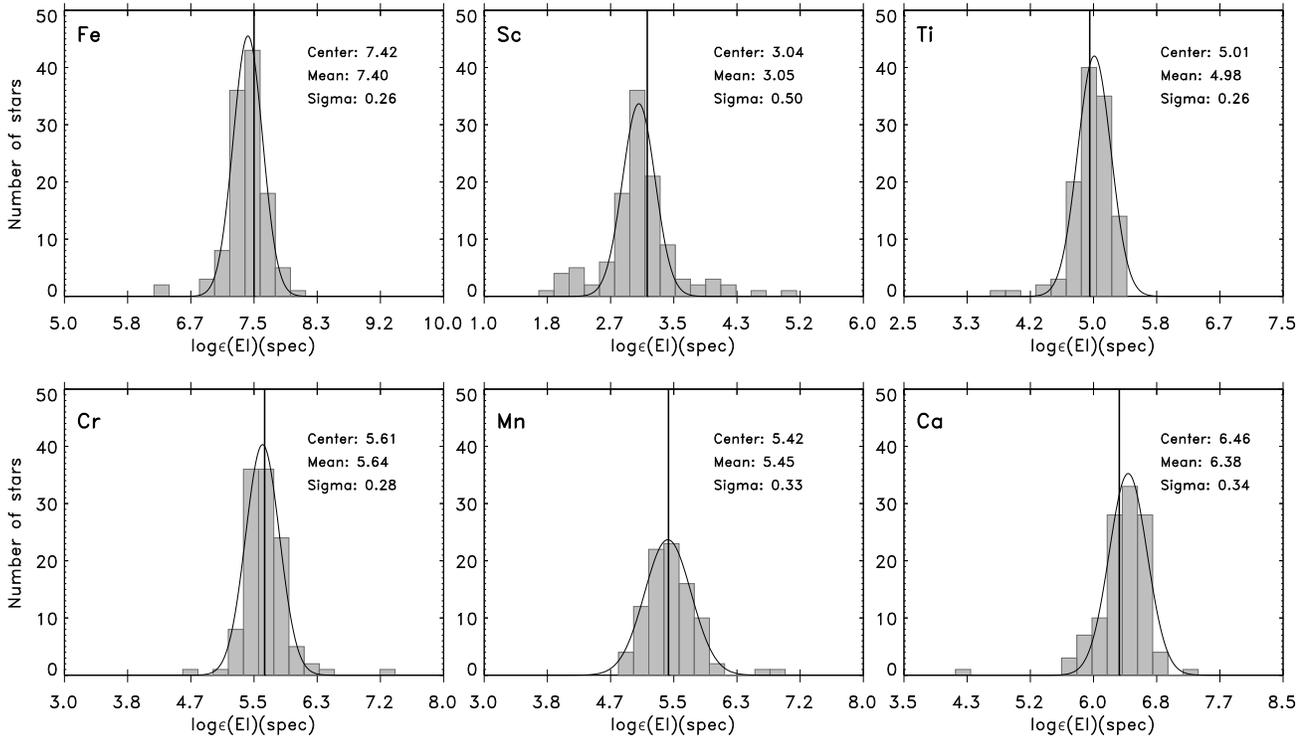}
\caption[]{The distributions of the abundances of the elements: Fe, Sc, Ti, Cr, Mn, and Ca in the analysed sample of stars. 
The solar abundances taken from \citet{asplund2009} are shown as vertical lines.}
\label{elements}
\end{figure*}

\begin{figure*}
\centering
\includegraphics[width=18cm,angle=0]{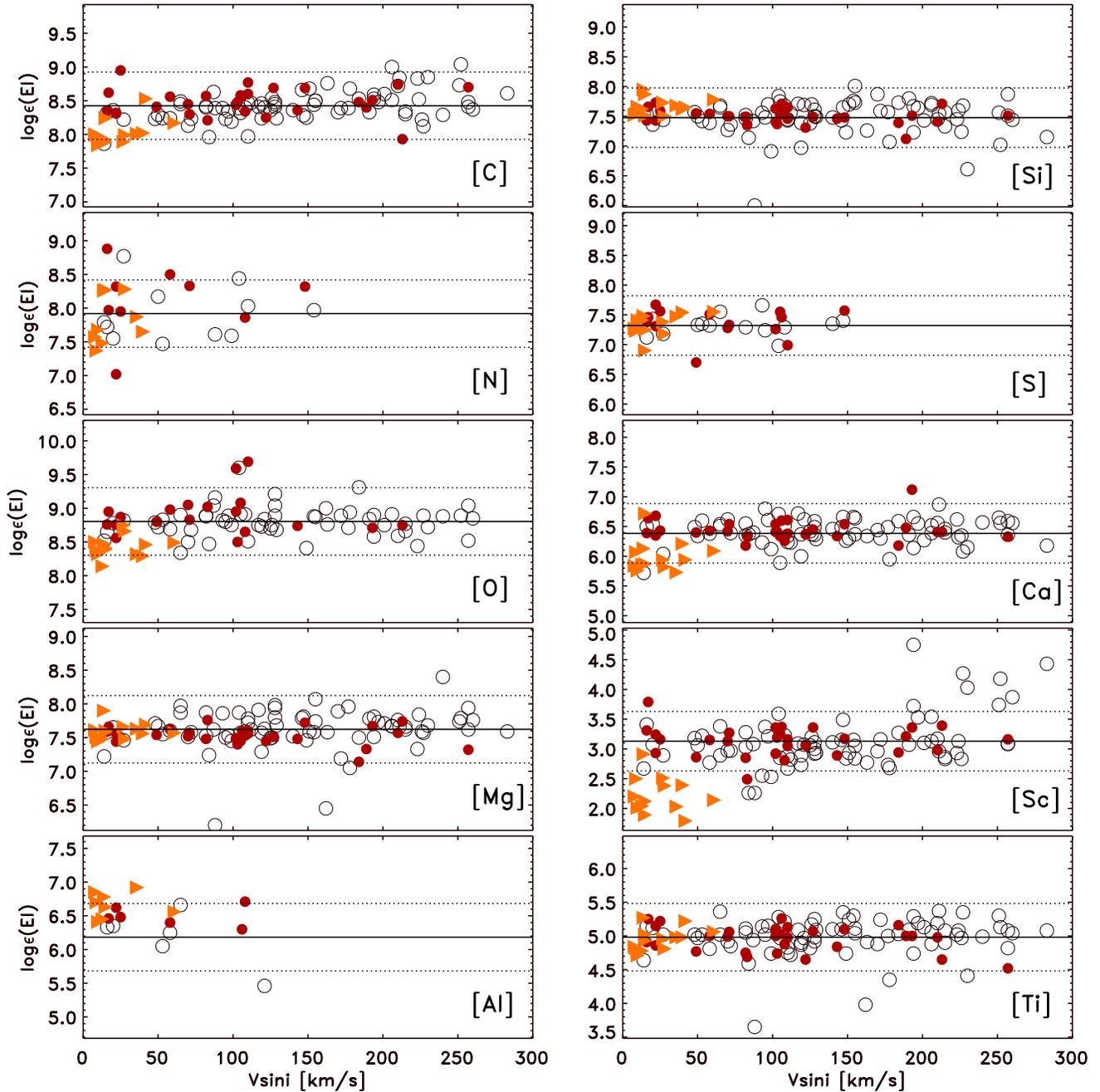}
\caption[]{Abundances of different elements as a function of rotational velocity for CP stars (triangles), non-CP 
stars with \teff\,$>$\,$7200$\,K (filled circles), and non-CP stars with \teff\,$<$\,$7200$\,K (open circles).}
\label{vsini-elem}
\end{figure*}

\begin{figure*}
\centering
\includegraphics[width=18cm,angle=0]{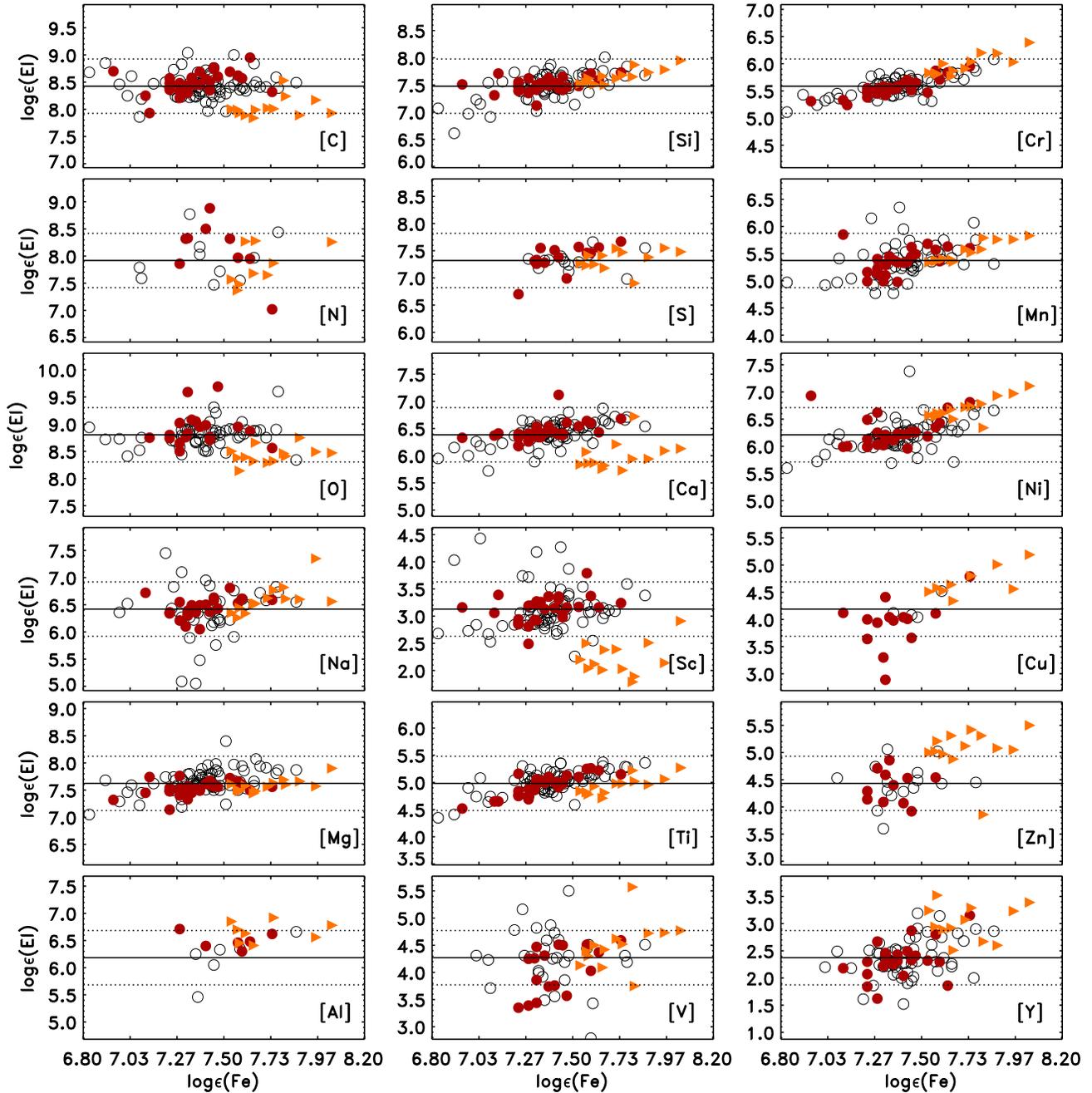}
\caption[]{Abundances of different elements as a function of iron abundance for CP stars (triangles), non-CP 
stars with \teff\,$>$\,$7200$\,K (filled circles), and non-CP stars with \teff\,$<$\,$7200$\,K (open circles).}
\label{fe-elem}
\end{figure*}

\begin{table}
\centering
\caption{Average abundances for the entire sample and the non-CP stars compared to the solar values of \citet{asplund2009}.}
\label{abundance-average-tab}
\begin{tabular}{c|r|ll|cc}
\toprule
El.     & No. of & Normal            & All                       &  Solar  \\
        & all    & stars             & stars                     &  values \\
        & stars  &$\log\epsilon(\rm El)$ &$\log\epsilon(\rm El)$ &  \\
\midrule
     C &$   116   $&$  8.47   \pm  0.23   $&$  8.42   \pm  0.27   $&$  8.43   $\\
     N &$    29   $&$  8.08   \pm  0.45   $&$  8.01   \pm  0.42   $&$  7.83   $\\
     O &$    91   $&$  8.88   \pm  0.28   $&$  8.82   \pm  0.31   $&$  8.69   $\\
    Na &$    76   $&$  6.52   \pm  0.49   $&$  6.54   \pm  0.46   $&$  6.24   $\\
    Mg &$   117   $&$  7.61   \pm  0.28   $&$  7.61   \pm  0.27   $&$  7.60   $\\
    Al &$    20   $&$  6.31   \pm  0.33   $&$  6.47   \pm  0.32   $&$  6.45   $\\
    Si &$   115   $&$  7.53   \pm  0.24   $&$  7.55   \pm  0.24   $&$  7.51   $\\
     P &$     1   $&$  5.32               $&$  5.32               $&$  5.41   $\\
     S &$    40   $&$  7.37   \pm  0.21   $&$  7.37   \pm  0.20   $&$  7.12   $\\
     K &$     4   $&$  5.52   \pm  0.14   $&$  5.58   \pm  0.17   $&$  5.03   $\\
    Ca &$   116   $&$  6.43   \pm  0.31   $&$  6.38   \pm  0.34   $&$  6.34   $\\
    Sc &$   115   $&$  3.14   \pm  0.42   $&$  3.04   \pm  0.50   $&$  3.15   $\\
    Ti &$   117   $&$  4.98   \pm  0.26   $&$  4.98   \pm  0.25   $&$  4.95   $\\
     V &$    61   $&$  4.21   \pm  0.51   $&$  4.29   \pm  0.51   $&$  3.93   $\\
    Cr &$   116   $&$  5.58   \pm  0.20   $&$  5.65   \pm  0.28   $&$  5.64   $\\
    Mn &$    92   $&$  5.43   \pm  0.35   $&$  5.45   \pm  0.33   $&$  5.43   $\\
    Fe &$   117   $&$  7.36   \pm  0.24   $&$  7.40   \pm  0.26   $&$  7.50   $\\
    Co &$    18   $&$  5.18   \pm  0.59   $&$  5.36   \pm  0.55   $&$  4.99   $\\
    Ni &$   115   $&$  6.25   \pm  0.27   $&$  6.31   \pm  0.30   $&$  6.22   $\\
    Cu &$    27   $&$  4.01   \pm  0.43   $&$  4.26   \pm  0.53   $&$  4.19   $\\
    Zn &$    39   $&$  4.45   \pm  0.39   $&$  4.67   \pm  0.49   $&$  4.56   $\\
    Ga &$     5   $&$  3.22   \pm  0.03   $&$  3.35   \pm  0.14   $&$  3.04   $\\
    Sr &$    70   $&$  2.79   \pm  0.75   $&$  2.98   \pm  0.84   $&$  2.87   $\\
     Y &$    93   $&$  2.38   \pm  0.44   $&$  2.48   \pm  0.50   $&$  2.21   $\\
    Zr &$    58   $&$  3.10   \pm  0.36   $&$  3.12   \pm  0.38   $&$  2.58   $\\
    Ba &$   108   $&$  2.64   \pm  0.52   $&$  2.80   \pm  0.70   $&$  2.18   $\\
    La &$    24   $&$  1.35   \pm  0.58   $&$  1.84   \pm  0.93   $&$  1.10   $\\
    Ce &$    17   $&$  2.12   \pm  0.44   $&$  2.46   \pm  0.40   $&$  1.58   $\\
    Pr &$     4   $&$  1.12   \pm  0.65   $&$  1.14   \pm  0.53   $&$  0.72   $\\
    Nd &$    15   $&$  1.85   \pm  0.23   $&$  2.13   \pm  0.35   $&$  1.42   $\\
    Sm &$     2   $&$  1.58               $&$  2.11   \pm  0.74   $&$  0.96   $\\
    Eu &$    10   $&$  0.46   \pm  0.21   $&$  1.34   \pm  0.81   $&$  0.52   $\\
    Gd &$     3   $&$  1.78   \pm  0.08   $&$  1.94   \pm  0.30   $&$  1.07   $\\
    Dy &$     4   $&$  1.92   \pm  0.20   $&$  1.92   \pm  0.16   $&$  1.10   $\\
    Er &$     1   $&$  1.48               $&$  1.48               $&$  0.92   $\\
\bottomrule
\end{tabular}
\end{table}

In Table\,\ref{abundance-average-tab} the average abundances for the stars in our sample are compared with the photospheric solar values of \citet{asplund2009}.
Only the abundances of Mg, Ti, and Fe were derived for all stars. 
We excluded the \ion{O}{i} 7771-5\,\AA\ triplet from our analysis and used other \ion{O}{i} lines (e.g., 5331, 6157, 6158, and 6456\,\AA) that are much less affected by 
non-LTE effects \citep[e.g.][]{przybilla2000}. Lines of heavy elements except for Sr, Y, Ba, and rare-earth elements are very sparse in spectra of A stars. 
Abundances of these elements were investigated only for slowly and moderately rotating stars and in most cases, only one or two blends were available.

In Fig.\,\ref{abundance-average-fig}, the determined average abundances for all stars are compared with the solar abundances of \citet{asplund2009}.
In Fig.\,\ref{elements}, histograms show the distribution of abundances of the iron-peak elements, Ti, Cr, Mn, Fe, Sc, and Ca. 
The average abundances of most of the light and iron-peak elements are close to the solar values. The largest differences occur for the elements
represented by weak blends only. Naturally, the abundances of some elements determined for CP stars 
are significantly different from the solar values. For instance, the highest Cr abundance, $\log \epsilon ({\rm Cr}) = 7.17\pm0.26$, was determined for the Ap\,CrSrEu star KIC4768731, and the lowest 
abundances of Sc were derived for Am stars. 

We found that apart from 14 CP stars, there are two low-metallicity objects in our sample, KIC\,9828226 and KIC\,8351193. 
KIC\,9828226 was classified as an  A1.5\,V star for which the metal line spectrum is weak and matches A0 type. The moderate rotational velocity, $88\pm9$\,\kms,  
allowed us to obtain the abundances of Mg, Si, Ca, Sc, Ti, Cr, Fe, Ni, Sr, C, and N. All these elements 
are underabundant. KIC\,8351193 (HD\,177152) was classified as a B9.5: star. With a rotational velocity of $162$\,\kms\ the spectrum of 
this star is almost featureless. It was possible to determine only the abundances of Mg, Ti, and Fe on the basis of single lines. All these elements 
are underabundant. However, because of the high \vsini, a small error in continuum location significantly changes the results.
\citet{tkachenko2013a} analysed the high-resolution TLS spectrum of this star and obtained \teff\,$= 9980\pm250$\,K and \logg\,$= 3.80\pm0.15$\,dex from
the analysis of Balmer lines. The effective temperature and surface gravity obtained in the present paper, \teff({\sc spec})\,$= 10500\pm250$\,K and 
\logg({\sc spec})\,$=4.1\pm0.1$ are higher than given by \citet{tkachenko2013a}. The high rotational velocities obtained in both papers are consistent 
within the error bars. Although \citet{tkachenko2013a} also found a low metallicity for this star, their iron abundance of $5.15$\,dex 
is much lower than the value obtained in this paper ($6.24$\,dex). The difference is caused mostly by differences in effective temperatures. 
The effective temperatures \teff(SED) and \teff({\sc spec}) of KIC\,8351193 obtained in this paper are consistent with each other and with the values 
$10210$\,K and $11000\pm400$\,K, determined from the SED and Str\"{o}mgren photometry by \citet{2011MNRAS.414..792B}.
On the basis of \textit{Kepler} data, \citet{2011MNRAS.414..792B} classified KIC\,8351193 as a ``rotationally modulated'' star with period of $0.56767$\,d. 
We confirm this classification (see Sect.\,\ref{sec:conclusions}). 

%%%%%%%%%%%%%%%%%%%%%%%%%%%%%%%%%%%%%%%%%%%%%%%%%%%%%%%%%%%%%%%%%%%%%%%%%%%%%%%%%%%%%%%%%%%%

\subsection{Correlation of chemical abundances with other parameters}

The analysis of a large sample of stars allowed us to search for correlations between the abundances of chemical elements and stellar parameters. 
We analysed separately three groups: the CP stars formed one group, and the normal stars were separated into hotter (\teff\,$>$\,$7200$\,K) 
and cooler (\teff\,$<$\,$7200$\,K) populations.
For some elements, large scatter or a small number of stars with determined abundances prevents the correlation analysis. 
This is the case for most of the heavy and rare-earth elements. Comparing samples with hotter and cooler stars, we found a larger scatter of abundances for hotter stars, 
especially for \vsini\ higher than $150$\,\kms. The same effect was observed for the abundances of A and F stars in the Hyades 
\citep{2010A&A...523A..71G}, Pleiades \citep{2008A&A...483..567G} and Coma Berenices open clusters \citep{2008A&A...479..189G}. 

No significant correlations of element abundances with effective temperature and surface gravity were found in the normal stars, with 
absolute values of Spearman's and Kendall's correlation coefficients less than $\sim0.20$. No significant correlations between the 
abundances of most elements and rotational velocity were found either. Typical relations for some elements are presented in Fig.\,\ref{vsini-elem}. 
A small positive correlation was found between the abundances of C and \vsini\ for hotter stars, with correlation coefficients of about 0.3--0.4 according 
to both considered statistics. In an investigation of 23 normal A and F stars, \citet{fossati2008} found no correlation between abundances and rotational 
velocity. A larger sample was examined by \citet{takeda2008}, 
who found negative correlations for C, O, Ca and strong positive correlations for the Fe-peak elements plus Y and Ba, with rotational velocity.

Significant positive correlations with iron abundance in normal stars were found for almost all elements (see Fig.\,\ref{fe-elem}). 
The highest correlation coefficients were obtained for light elements Na, Mg and Si, iron-peak elements Ca, Ti, Cr, Mn, Ni, and heavy elements Sr and Ba, 
all of which have correlation coefficients larger than $0.50$. Less significant positive correlations were obtained for Sr and Y, 
for which both correlation coefficients were smaller than $0.40$. 
\citet{takeda2008} also found strong positive correlations between the abundances of Si, Ti and Ba, and that of Fe.

We classified 21 CP stars in our sample, including 17 Am stars, one Ap CrSrEu, two $\lambda$\,Boo stars and one He-strong B star. 
We checked for correlations between abundances in Am stars and their atmospheric parameters, iron abundances, and rotational velocities.
The abundances of most chemical elements appeared to be independent of the effective temperature. The negative correlation determined for O and the positive correlations found for Cu, Zn, and Zr could be caused by the small sample size.
We found no correlations between abundances and surface gravity. On the other hand, the abundances of 
Na, Si, Ti, V, Cr, Mn, Ni, Cu, and rare-earth element La were all positively correlated with the iron abundance (Fig.\,\ref{fe-elem}). 
The only element found to be negatively correlated with iron was Zr, with a correlation coefficient of $-0.20$, according to Spearman's statistic. 
The abundances of many elements were also found to be strongly correlated with that of iron  for non-CP stars.

The correlation between Na and Fe abundances was discussed by \citet{takeda2011}, who performed a non-LTE analysis of the abundances of alkali elements 
(Li, Na, and K) determined from the \ion{Li}{i} $6708$\,{\AA}, \ion{Na}{i} $5682$/$5688$\,{\AA}, and \ion{K}{i} $7699$\,{\AA} lines  
for 24 slowly rotating A stars (\vsini\,$<$\,$50$\,\kms) in the broad temperature range $7000$\,K $<$ \teff\,$<$ $10000$\,K. Many of these stars showed 
Am peculiarities to different degrees. They found a significant trend of Na abundances with Fe, independent of \teff. This means that Na becomes enriched
along with Fe in accordance with the degree of Am peculiarity. The same trend was derived for our sample of stars with a Spearman's correlation coefficient of $+0.76$. 
The positive correlation of Na abundance with Fe disagrees with the predictions of atomic diffusion theory \citep[e.g.][]{2006ApJ...645..634T}. 
The same is true of Si abundances in Am stars, where we see a positive correlation of $+0.80$, according to Spearman's statistic, that is not theoretically expected.

No significant correlations with \vsini\ were found for the abundances of any element, albeit in a small analysed range of $v\sin i$.
This contradicts the finding of \citet{fossati2008} who analysed Am stars in the Praesepe cluster and reported strong 
correlations between abundances and \vsini\ of Am stars for all peculiar elements, except for Sc and Ti. 
One reason for the contradiction may be that the field stars analysed here have different ages and initial chemical composition.

\subsection{Chemically peculiar stars}

\begin{figure*}
\centering
\includegraphics[width=18cm,angle=0]{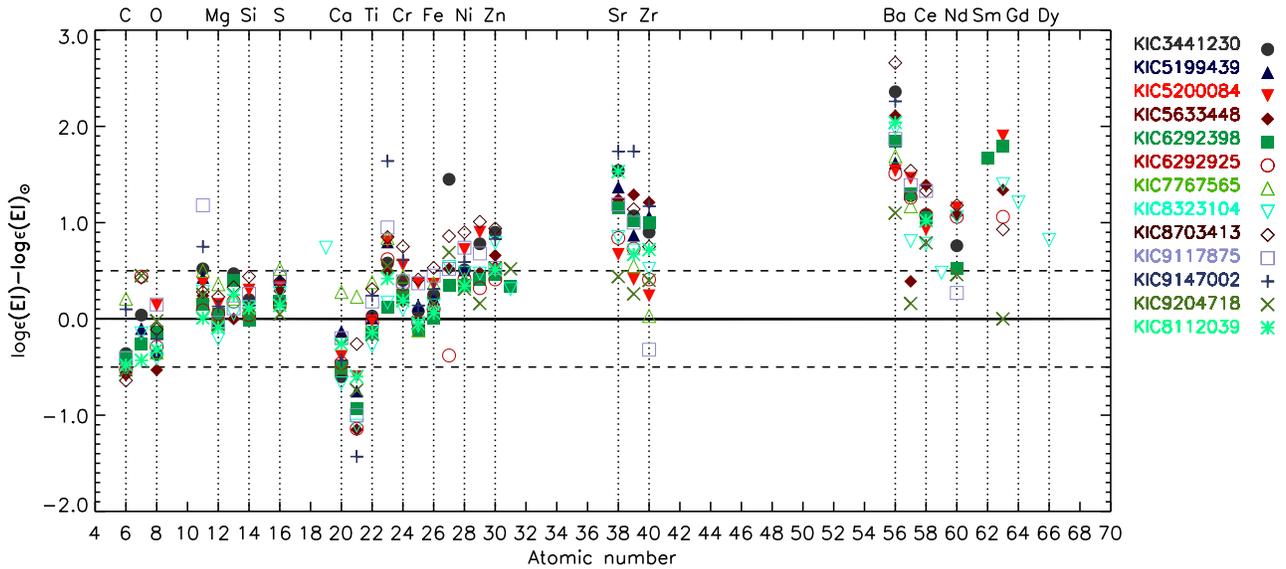}
\caption[]{Chemical abundances of Am stars compared with the solar values \citep{asplund2009} as a function of the atomic number.}
\label{elementsAm}
\end{figure*}

\begin{table}
\centering
\caption{CP stars investigated in this work.}
\label{cpstars}
\begin{tabular}{l|l|l}
\toprule
KIC number &HD number &  SpC (new) \\
\midrule
\multicolumn{3}{c}{New Am stars} \\
\hline
3441230    &          &kA2hA5mF0\,V      Am \\
5199439    &225365    &kA3hA5mA5\,(IV)s  Am: \\
5200084    &225410    &kA3hA6mF1\,(IV)   Am \\
5633448    &225463    &kA3hA4mA7\,V      Am \\
6292398    &          &kA3hA7mA7\,V      Am: \\
6292925    &          &kA2.5hA3mA7\,(IV) Am: \\
7767565    &196995    &kA5hA7mF1\,IV     Am \\
\hline
\multicolumn{3}{c}{Known Am stars} \\
\hline
8323104    &188911    &kA2.5hA6mA7\,(IV) Am \\
8703413    &187254    &kA3hA5mA9\,(IV)s  Am \\
9117875    &190165    &kA3hF0.5mF3\,(III) Am \\
9147002    &180237    &kA3hA5mF3\,V Am \\
9204718    &176843    &kA3hA9mF1\,(IV) Am \\
\hline
\multicolumn{3}{c}{Am stars in SB2 systems} \\
\hline
8112039    &187091    &kA3hA5mA5\,(IV)s Am: \\
4660665    &          &kA3hA5mA5\,IV Am: \\
6865077    &          &kA3hA8mA6\,(IV)  Am \\
8692626    &          &kA2hA4mA6\,(IV)  Am: \\
10616138   &          &kA3hA6mA7\,(IV)s Am: \\
\hline 
\multicolumn{3}{c}{Ap star} \\
\hline
4768731    &225914    & A5\,Vp (SrCrEu) \\
\hline 
\multicolumn{3}{c}{$\lambda$\,Boo stars} \\
\hline
9828226    &          & hA2kA0mB9\,Vb. \\
5724440    &187234    & F0\,Vnn\,weak met\,(A5) \\
\bottomrule
\end{tabular}
\end{table}

There are two main classes of CP stars among the main-sequence Population I stars: the metallic line (Am) and the peculiar 
A stars (Ap). The Am stars manifest overabundances of most iron-peak elements and some heavy elements like Zn, Sr, Zr and Ba, and 
underabundances of Ca and Sc \citep{gray&corbally2009}. These anomalies are explained by radiative diffusion in a non-magnetic star 
 \citep[e.g.][]{ 2005EAS....17...43R, 2006ApJ...645..634T}. 
Their projected rotational velocities are generally smaller than those of normal A stars \citep{abthuston1971}, which permits the 
segregation of elements by diffusion \citep{abtmoyd1973}. The majority of Am stars are members of close binary systems, where their slow rotation results from tidal braking.

The Ap stars form the second group of CP stars with overabundances of some elements, such as Si, Sr, Cr, and rare-earth 
elements \citep[e.g.][]{kurtz2007a, kurtz2007b}. Most of these stars have strong magnetic fields with polar field strengths typically 
of several kG \citep{hubrig2005, kurtz2006}. These stars have much slower rotational velocities than normal A and B stars. 
The distribution of chemical abundances changes both horizontally, with the formation of spots, and vertically, where they are stratified
 \citep[e.g.][]{2005A&A...438..973R, 2014psce.conf..220R, 2015A&A...574A..79K}.
The abundance anomalies and spots are caused by diffusion in the presence of a magnetic field  \citep[e.g.][]{1993ASPC...44..458B, 2003ASPC..305..199T}.

There are other classes of CP stars, such as the $\lambda$\,Bootis stars, 
which are a class of metal-weak population\,I A stars \citep{gray&corbally2009}. They are characterised by a weak \specline{Mg}{II}{4481}\,{\AA} line and solar abundances of 
C, N, O, and S \citep{gray}. Their projected rotational velocity distribution matches that of normal A stars \citep{abtmorrell1995, gray&corbally2009}.
The metal deficiencies of $\lambda$\,Boo can be distinguished from those of  Population II stars on the basis of detailed abundance analyses \citep{2014A&A...567A..67P}. 
This suggests that the abundances of $\lambda$\,Boo stars are restricted to the outer layers 
of their atmospheres and can be explained by accretion of dust-depleted circumstellar material \citep{1990ApJ...363..234V, 1994MNRAS.269..209K}.

There are 20 peculiar A stars in our sample (see Table\,\ref{cpstars}), which are comprised of 17 Am stars, one Ap star 
(KIC\,4768731), and two $\lambda$\,Bootis candidates (KIC\,9828226 and KIC\,5724440). 

Four of the Am stars were previously known from multicolour photometry \citep{mendoza1974} and spectroscopy \citep{floquet}.
These are KIC\,8323104, 8703413, 9117875, and 9204718. KIC\,8323104 has a very low \vsini\, of $10\pm1$\,\kms, which enabled us to derive the abundances of alkali elements 
Na and K. Determinations of Na and K abundances in the atmospheres of Am stars are rare \citep{takeda2011}, but important for establishing the theory 
explaining the chemical abundance peculiarities of A stars. 
The effective temperature of KIC\,9117875, at $7300$\,K, is low in comparison with other Am stars in our sample. The star falls among the $\gamma$\,Dor stars 
in the \teff\,--\logg\ diagram, but it is not a $\gamma$\,Dor star (see Sect.\,\ref{sec:conclusions}). 
The fifth known Am star, KIC\,9147002 was previously classified as an A2p star by \citet{bertaud1960}. 
Its abundance pattern is typical of Am stars, with a significant Sc underabundance and strong overabundances of heavy elements. 
The Am stars KIC\,8703413, 9117875 and 9204718, were also studied by \citet{CatanzaroRipepi}. Their results are in agreement with the values obtained here, 
within the errors. 

Six newly discovered Am stars are KIC\,3441230, 5199439, 5200084, 5633448, 6292398, and 6292925. These have low $v\sin i$ and 
abundance patterns typical of Am stars. A seventh Am star is KIC\,7767565, of spectral type kA5hA7mF1\,IV, whose chemical 
abundances are not typical, because Ca and Sc are overabundant. This star has a moderate \vsini\ of $64\pm3$\,\kms. 

KIC\,8112039 is a highly eccentric binary, also known as KOI-54 \citep{welsh2011, burkart2012}. \citet{welsh2011} analysed 
the high-resolution spectrum taken in quadrature and determined atmospheric parameters of both components, 
finding \teff$_A$ and \teff$_B$ equal to $8500$ and $8800$\,K, and surface gravities \logg\,$= 3.8$ and $4.1$, respectively. Both stars appeared to be metal-rich, with [Fe/H]\,$= 0.4 \pm 0.2$\,dex. 
They also determined low \vsini\ values for the components, at $7.5$ and $4.5$\,\kms. 
The spectrum investigated in this paper was taken in conjunction, so that the lines of both stars overlap. 
From the analysis of our spectrum we derived \teff\,$=8400$\,K, \logg\,$=3.7$\,dex, [Fe/H]\,$= 0.0\pm 0.13$\,dex, and \vsini\,$ = 8\pm1$\,\kms. 

Four other stars were classified as Am stars and belong to SB2 binary systems: KIC\,4660665, KIC\,6865077, KIC\,8692626, and KIC\,10616138. 
These stars, together with KIC\,8112039, will be analysed in a forthcoming paper (Catanzaro et al., in preparation).
 
\begin{figure}
\centering
\includegraphics[width=8cm,angle=0]{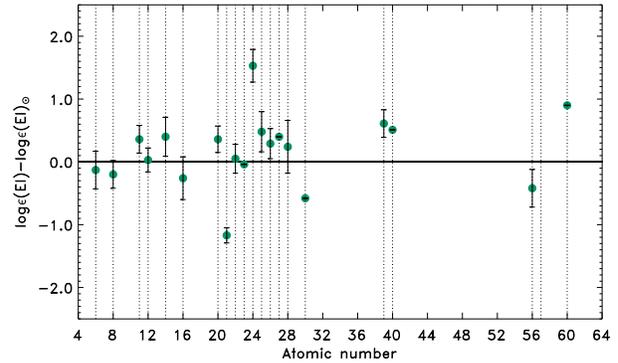}
\caption[]{Chemical abundances of the Ap star KIC\,4768731 compared with the solar values \citep{asplund2009} as a function of the atomic number.}
\label{elementsAp}
\end{figure}

The only Ap star in our sample, KIC\,4768731, has the spectral type A5\,Vp (SrCrEu). According to \citet{mason2001}, the star is the primary of a visual binary 
with a separation of about $12$\,arcsec with $V = 9.19$\,mag (primary) and $12.00$\,mag (secondary). Previously, KIC\,4768731 was classified as an A7 star (see Table\,\ref{journal1}), 
without indication of chemical peculiarity. The chemical pattern of KIC\,4768731 is typical for Ap (SrCrEu) stars, with significant 
overabundances of Cr and Sr (see Fig.\,\ref{elementsAp}). All Fe-peak elements, except for Sc, V, and Zn, are overabundant. Additionally, KIC\,4768731 was discovered as a 
rapidly oscillating (roAp) star (Smalley et al., in preparation). 
Detailed spectral analysis of Ap stars requires taking into account stratification of the elements in their atmospheres. 
This effect has a substantial impact on spectral lines, so atmospheric parameters determined here from the ionisation and excitation equilibria and wings of hydrogen lines
can be incorrect. Results of spectral analysis with stratification of chemical elements for KIC\,4768731 will be presented by Niemczura, Shulyak et al. (in preparation).

We classified two stars as probable $\lambda$\,Boo stars. The spectral type obtained for KIC\,9828226 is hA2kA0mB9\,Vb. Its hydrogen lines are very broad, the metal lines are weak, and 
the magnesium \specline{Mg}{ii}{4481} line is especially weak when compared with the metal line type. The star has \vsini\,$=88\pm9$\,\kms.
KIC\,5724440 was given the spectral type F0\,Vnn\,(weak met (A5)). Its hydrogen lines are broad and shallow -- typical for a rapid rotator -- and are consistent with an F0 spectral type, whereas 
its metal lines are weak and are similar in strength to the A5 standard HD\,23886. The \ion{Ca}{ii}\,K line is broad but shallow, having the depth of the A3\,V standard but the width of the A7\,V standard. 
The moderate resolution spectrum of this star was analysed by \citet{catanzaro2011}, who indicated a problem with the effective temperature determination from Balmer lines. 
This problem was also noticed in this paper. The \specline{Mg}{ii}{4481} in the spectrum of KIC\,5724440 is weaker than in the spectrum of a normal star with the same atmospheric parameters. 
The iron-peak elements and especially Sr are slightly underabundant.
This supports the $\lambda$\,Bootis classification of this star.

\begin{figure}
\centering
\includegraphics[width=8cm,angle=0]{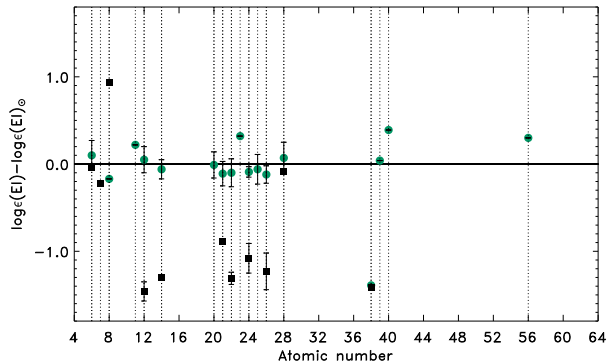}
\caption[]{Obtained chemical abundances of the suspected $\lambda$\,Boo stars KIC\,5724440 (filled circles) and KIC\,9828226 (filled squares) 
compared with the solar values \citep{asplund2009} as a function of the atomic number.}
\label{elementslBoo}
\end{figure}

\section{Discussion and Conclusions}
\label{sec:conclusions}

We analysed high-resolution HERMES/Mercator spectra of 117 A and F stars collected during the spectroscopic campaign of \textit{Kepler} stars.
Spectral classification showed that our sample consists mostly of non-evolved stars of luminosity types V to IV and revealed 20 chemically peculiar (CP) stars 
of Am, Ap (CrSrEu), and $\lambda$\,Bootis types. These peculiarities were confirmed by detailed investigation of the stellar spectra. Nine new CP stars
were found, of which six were Am, one was Ap, and two were $\lambda$\,Boo stars.

%----------

The atmospheric parameters, i.e. \teff, \logg, \turb, abundances of chemical elements, and \vsini\  
for all stars were determined by using spectral synthesis. The atmospheric parameters were compared with the values taken from the KIC and those determined from SED fitting. 
Similar to the results reported by \citet{tkachenko2013a} and \citet{lehmann2011}, 
the KIC values of effective temperatures were found to be lower than the spectroscopic ones for stars with \teff\ $\gtrsim 7500$\,K.
One of the reasons for these differences is probably the use of incorrect interstellar reddening when estimating the KIC temperatures. 
The E(B$-$V) values from the KIC were compared with the spectroscopic values determined from interstellar Na D lines. For \teff\ higher than $7500$\,K 
the differences E(B$-$V)(Na)--E(B$-$V)(KIC) increase. The spectroscopic effective temperatures derived from iron and Balmer lines are in good agreement 
with values obtained from SED fitting. For most stars, the surface gravities determined from iron lines and taken from the KIC are consistent to $\pm0.2$\,dex. 
There is a positive correlation between the differences of iron abundances determined from spectroscopy and taken from KIC and $\log\epsilon{\rm (Fe)}$. 
The same trend was noticed by \citet{tkachenko2013a}. This suggests that the KIC iron abundances of metal-weak stars are systematically overestimated, whereas
iron abundances of metal-strong stars are underestimated. The obtained microturbulence velocities follow the trend presented by \citet{landstreet2009} and are smaller 
towards higher effective temperatures. The obtained values of \vsini\ range from $8$ to about $280$\,\kms\ with the average equal to $134$\,\kms, which is  
typical for A and early F stars.

%----------
\begin{figure*}
\centering
\includegraphics[width=16cm,angle=0]{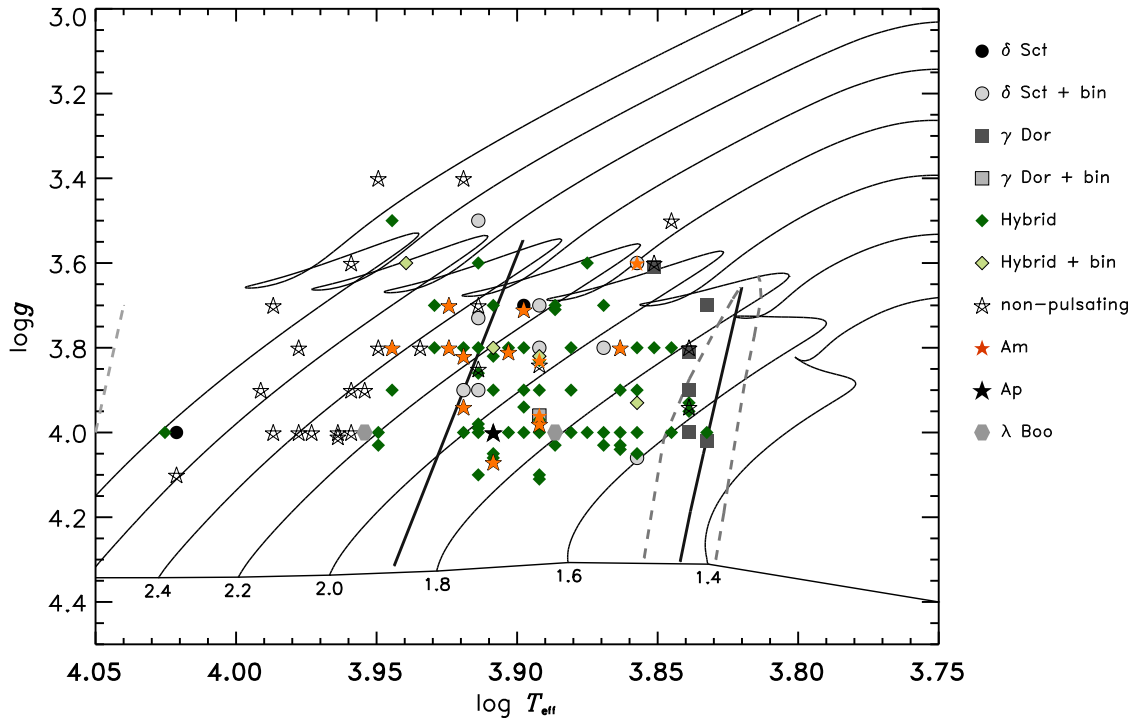}
\caption[]{$\log$\,\teff\ -- \logg\ diagram for the stars analysed in this paper. Stars identified as single $\delta$\,Sct are indicated by filled circle, 
$\delta$\,Sct in a binary systems as open circle, single $\gamma$\,Dor as filled squares, $\gamma$\,Dor in binary systems as light blue squares. Single stars with
hybrid pulsations are plotted as filled diamonds, whereas hybrid stars in binary systems are open diamonds. Non-pulsating stars (including Am objects) 
are shown as stars.}
\label{hrall}
\end{figure*}

%----------
The atmospheric parameters determined in this work are very important as input ingredients for further seismic modelling. We analysed A and F stars, 
so our sample contains pulsating $\delta$\,Sct, $\gamma$\,Dor, and hybrid stars. 
To find out if the obtained atmospheric parameters of the stars are consistent with theoretical instability strips in this part of the H-R diagram, we performed 
an analysis of the \textit{Kepler} data for all stars investigated here. We used the corrected light curves available through the MAST \footnote{https://archive.stsci.edu/kepler/} 
database. We considered all the long cadence data available for a given star. If necessary, short cadence data were used to verify our classification. 
The \textit{Kepler} data from each quarter were shifted to have the same mean level and all quarters were merged. 
For some stars, it was necessary to remove the trends from each quarter separately. 
Then, we visually examined all light curves, the frequency spectra, and detected frequencies to identify candidate $\delta$\,Sct, $\gamma$\,Dor, 
hybrid, and other types of variability. The results of our initial classification are presented in the last column of Table\,\ref{journal1}. 
We classified 15 stars as $\delta$\,Sct, 13 as $\gamma$\,Dor, and 71 as $\delta$\,Sct/$\gamma$\,Dor hybrid stars. In this paper, all objects 
for which we can find frequencies typical for $\delta$\,Sct and $\gamma$\,Dor stars are termed ``hybrid''.
Moreover, we found 14 pulsating stars in binary systems, including eclipsing binaries, and 30 stars which do not show pulsations (these stars
were discussed in detail by \citet{2015MNRAS.447.3948M}. The spectroscopic parameters provided here were critical for assigning 
locations of the non-pulsators with respect to the $\delta$\,Sct instability strip. Here we only mention that light curves of the stars belonging to the latter group
show variability resulting from rotation and spots on the surfaces of these stars and/or from binarity (see Table\,\ref{journal1}). 

All of the stars investigated are shown in the \teff\ $-$ \logg\ diagram (Fig.\,\ref{hrall}). We add evolutionary tracks, calculated with 
Time Dependent Convection \citep[TDC,][]{grigahcene2005}, for stars having [Fe/H]\,$=0.0$ and $\alpha_{\rm MLT} = 2.0$ covering a range of masses 
from $1.4$ to $2.8$\,M$_\odot$. The stars are plotted with different symbols depending on the type of pulsations, type of chemical peculiarity 
and other characteristics. As we can see in Fig.\,\ref{hrall}, non-pulsating stars are situated inside and on the 
blue edge of the $\delta$\,Sct instability strip. Most Am stars are located inside the instability strip. 
Most hybrid stars are located inside or close to the blue edge of the $\delta$\,Sct instability strip. 
The exceptions are KIC\,12736056 and 8489712, whose effective temperatures are too high for hybrid $\delta$\,Sct/$\gamma$\,Dor stars. The same applies for KIC\,4056136,
classified as a $\delta$\,Sct star, KIC\,11180361, showing variability typical for hybrid stars and binarity, and for KIC\,6292398, for which we 
found $\gamma$\,Dor frequencies and evidence for binarity. All these stars need further investigation to verify the possibility of
contamination by another star. The frequency analysis of the \textit{Kepler} data revealed 31 stars with low-frequency peaks which we attribute to binarity.
For these stars lines from only one star are present in the spectrum and we see $\delta$\,Sct and/or $\gamma$\,Dor frequencies in the \textit{Kepler} data. 

%----------

We investigated 13 Am type stars, of which 7 are newly discovered. Most of them show only the variability typical for stars in binary systems or
due to rotation and spots. Only three stars are possible pulsators. KIC\,6292398 has a frequency pattern found in $\gamma$\,Dor stars and binarity, KIC\,7767565 is 
probably also a $\gamma$\,Dor star, and KIC\,9204718 shows frequencies typical of binarity, rotational changes and $\delta$\,Sct pulsations. The \textit{Kepler} photometric data for
these stars need to be analysed in detail. 

%---------- 

We have also investigated a possible relationship between metallicities or chemical peculiarities and the hybrid pulsations. 
The first $\delta$\,Sct/$\gamma$\,Dor hybrid, HD\,8801, was discovered by \citet{henry} and classified as an Am star. This classification was
confirmed by \citet{neuteufel2013}. \citet{hareter} analysed the spectra of two other hybrid stars, 
HD\,114839 and BD\,+18$^{\circ}$\,4914, discovered with MOST photometry by \citet{rowe} and \citet{king}, respectively. The evolved star BD$^{\circ}$\,+18\,4914 has an abundance pattern 
typical for an Am star, but the less evolved star HD\,114839, previously classified in the literature as a mild Am star, does not show any peculiarity. 
For 58 hybrid candidates investigated here the spectral analyses do not indicate any peculiarity.

%________________________________________________________________

\section*{Acknowledgments}
We are grateful to the anonymous referee for their thorough examination of this paper.
EN acknowledges the support from Funda\c{c}\,{a}o para a Ci\^{e}ncia e a Tecnologia (FCT) through the grant 
`Coopera\c{}c\,{a}o Cientica e Teccnologica FCT/Polonia 2011/2012 (Proc. 441.00 Polonia)' funded by FCT/MCTES, Portugal and the support from
1007/S/IAs/14 funds. This work was supported by NCN grant 2011/01/B/ST9/05448. DD would like to acknowledge the NSC grant No. 2011/01/N/ ST9/00400. 
Calculations have been carried out in Wroc{\l}aw Centre for Networking and Supercomputing (http://www.wcss.pl), grant No.\,214.
This research was supported by the Australian Research Council. Funding for the Stellar Astrophysics Centre is provided by the Danish National 
Research Foundation (grant agreement no.: DNRF106). The research is supported by the ASTERISK project (ASTERoseismic Investigations with SONG and Kepler) 
funded by the European Research Council (grant agreement no.: 267864).
KU acknowledges support by the Spanish National Plan of R\&D for 2010, project AYA2010-17803.
The research leading to these results has received funding from the European Research 
Council under the European Community’s Seventh Framework Programme (FP7/2007–2013)/ERC grant agreement No. 227224 (PROSPERITY).
This publication makes use of data products from the Two Micron All Sky Survey, which is a joint project of the University of Massachusetts
and the Infrared Processing and Analysis Center/California Institute of Technology, funded by the National Aeronautics and Space Administration
and the National Science Foundation. MB is F.R.S.-FNRS Post-doctoral Researcher, Belgium.
AP acknowledges support from the NCN grant No. 2011/03/B/ST9/02667.
PIP is a Postdoctoral Fellow of the Fund for Scientific Research (FWO), Flanders, Belgium.
SB is supported by the Foundation for Fundamental Research on Matter (FOM),
which is part of the Netherlands Organisation for Scientific Research (NWO).

\label{lastpage}

%%%%%%%%%%%%%%%%%%%%%%%%%%%%%%%%%%%%%%%%

\setcounter{table}{1}

\small

\label{lastpage}

\begin{table*}
\centering
\caption{Journal of spectroscopic observations and the derived spectral classes. $N$ is the number of available spectra; SpT1 indicates the spectral
classification from the literature, whereas SpT2 means spectral classifications obtained in this work. Superscript $\star$ in the first column 
denotes the stars with additional information below the table. Additional notations for luminosity class: ``b'' -- lower luminosity main-sequence star; 
``a'' or ``a+'' -- higher luminosity main-sequence star; ``IV-V'' -- between IV and V; ``IV/V'' -- either IV or V; ``n'' -- nebulous (i.e. broad-lined); ``s'' -- sharp lines;
``nn'' -- very rapid rotators; ``wk'' met -- weak metal lines; ``met str'' -- strong metal lines. In the last column additional information taken from spectra and 
\textit{Kepler} data are presented. ``SB2'' means double-lined spectroscopic binary; ``rot.'' -- rotational frequency detected; 
``bin.'' -- orbital frequency detected; ``mult.'' -- orbital frequencies of both components detected; ``hybrid'' -- frequencies of $\gamma$\,Dor and $\delta$\,Sct visible in the spectra; 
``contamination'' -- \textit{Kepler} data contaminated by a nearby star; ``?'' -- our classification of variability type is uncertain.} 
\begin{tabular}{llcrlll}
\toprule
KIC          & HERMES        & N   & V   & SpT1     & SpT2        & Notes \\
Number       & observations  &     & [mag]& & [this paper]&       \\
\midrule
 1294756               & 2010 May-September  & 1     & 8.96&  A2$^{1}$ & A3\,IV              & mult. + rot. + $\delta$\,Sct \\ 
 2571868               & 2011 July           & 1     & 8.67&  A0$^{1}$ & A3\,IVn             & hybrid (low. ampl. $\gamma$\,Dor) \\ 
 2694337               & 2010 May-September  & 1     &10.35&           & F0.5\,IVs           & hybrid\\   
 2695344               & 2012 July           & 1     & 9.62&  F2$^{1}$ & F2\,V               & SB2; $\gamma$\,Dor\\
                       & 2012 September      & 1     &     &           &                     & \\
 3230227$^{\star}$     & 2010 May-September  & 1     & 8.94&  A5$^{1}$ & A5\,IV:             & SB2; ecl. bin. + $\delta$\,Sct\\  
 3231985$^{\star}$     & 2011 September      & 1     & 9.22&  A2$^{1}$ & A4\,IV Ca weak (A3) & rot.\\  
 3347643               & 2010 May-September  & 1     & 7.69&  A2$^{1}$ & A3\,Van             & hybrid (low. ampl. $\gamma$\,Dor)\\  
 3441230               & 2012 July           & 1     & 9.79&  A2$^{1}$ & kA2hA5mF0\,V Am     & rot.\\     
 3656913               & 2012 July           & 1     & 9.81&  A0$^{1}$ & kA2hA6mF0\,IV Am    & SB2; bin.\\
 3850810$^{\star}$     & 2010 May-September  & 1     &10.10&           & F1\,Vs wk met (F0)  & hybrid + rot.\\     
 3851151               & 2010 May-September  & 1     & 9.82&  A2$^{1}$ & A3\,V               & hybrid + rot.\\    
 3942392               & 2012 July           & 1     & 9.87&  F2$^{1}$ & F2\,Vs              & rot. + $\gamma$\,Dor \\     
 4035667               & 2010 May-September  & 1     &10.05&           & A3\,IVn             & hybrid + bin. + rot.\\   
 4044353               & 2010 May-September  & 1     & 9.76&  A2$^{1}$ & A5\,IVs             & hybrid\\     
 4048494               & 2011 July           & 1     & 9.55&           & A6\,V               & SB2; bin. + $\delta$\,Sct \\
 4056136               & 2011 September      & 1     & 9.55&  A0$^{1}$ & B9\,IV-Vnn          & rot.; $\delta$\,Sct; contamination\\  
 4077032               & 2011 July           & 1     & 9.71&  F0$^{1}$ & F2\,IIIs            & hybrid (low ampl. $\gamma$\,Dor )\\   
 4150611               & 2010 May-September  & 1     & 8.00&  A2$^{1}$ & F1\,V wk met (A9)   & $\delta$\,Sct + ecl. bin. \\       
 4281581               & 2011 July           & 1     & 9.32&  A2$^{1}$ & A3\,IV-Vs           & hybrid \\      
 4480321$^{\star}$     & 2011 September      & 1     &10.27&           & A9\,V wk met (A5)   & SB2; hybrid + bin. + rot. \\
 4572373$^{\star}$     & 2012 July           & 1     & 9.84&           & A3\,Van (wk met A2) & rot.\\     
                       & 2012 September      & 1     &     &           &                     & \\
 4660665$^{\star}$     & 2011 September      & 1     & 7.86&  A2$^{1}$ & kA3hA5mA5\,IV Am:   & SB2; bin. + rot. + $\gamma$\,Dor \\
 4671225               & 2011 September      & 1     & 9.88&  A7$^{1}$ & A3\,IV-Vn           & hybrid\\       
 4681323$^{\star}$     & 2011 July           & 2     & 9.06&  A0$^{1}$ & A1\,IV-s            & bin. + rot.\\      
 4768731$^{\star}$     & 2011 July           & 1     & 9.17&  A7$^{1}$ & A5\,Vp SrCrEu       & Ap; bin. + rot.\\      
 4831769               & 2011 September      & 1     & 9.55&  A2$^{1}$ & A3\,Va              & bin.\\ 
 4832225               & 2011 September      & 1     & 9.08&  A0$^{1}$ & B9.5\,V             & SB2; bin. + $\gamma$\,Dor\\
 4840675               & 2012 July           & 1     & 9.66&           & A5\,Vnn             & SB2; hybrid (low ampl. $\gamma$\,Dor )\\
 5018590               & 2012 July           & 1     & 9.87&           & F0.5\,V             & $\gamma$\,Dor\\   
                       & 2012 September      & 1     &     &           &                     & \\
 5113797               & 2011 July           & 1     & 9.17&  A3$^{1}$ & A3\,IV-V            & hybrid (low ampl. $\delta$\,Sct)\\       
 5199439               & 2012 July           & 1     & 9.51&  A0$^{1}$ & kA3hA5mA5\,(IV)s Am:& bin. + rot.\\      
                       & 2012 September      & 1     &     &           &                     & \\
 5200084$^{\star}$     & 2011 September      & 1     & 9.16&  A0$^{1}$ & kA3hA6mF1\,(IV) Am  & bin.\\    
 5294231               & 2012 July           & 1     & 9.74&  A2$^{1}$ & A0\,IVn             & rot.\\ 
                       & 2012 September      & 1     &     &           &                     & \\
 5355850               & 2011 September      & 1     & 9.44&  A5$^{1}$ & A5\,IVs             & bin. + hybrid\\  
 5429163               & 2011 July           & 1     & 9.80&  A5\,V$^{2}$& A4\,V             & hybrid? + rot.\\ 
 5471091               & 2012 July           & 1     & 9.97&  F2$^{1}$ & F2\,IV              & bin. + rot.\\ 
                       & 2012 September      & 1     &     &           &                     & \\
 5473171$^{\star}$     & 2010 May-September  & 1     & 8.97&  A2$^{1}$ & A5\,IV/Vn           & hybrid\\    
 5524045               & 2011 September      & 1     & 9.36&  A0$^{1}$ & A0.5\,Va+           & rot.\\   
 5525210               & 2012 July           & 1     & 9.90&           & F1\,V               & SB2; bin. + rot.\\
                       & 2012 September      & 1     &     &           &                     & \\
 5608334               & 2011 July           & 1     & 9.88&           & F2\,Vs wk met (A7)  & hybrid?\\  
 5633448$^{\star}$     & 2011 September      & 1     & 8.96&  A2$^{1}$ & kA3hA4mA7\,V Am     & rot.\\   
 5724440$^{\star}$     & 2011 July           & 1     & 7.89&  A5$^{1}$ & F0\,Vnn wk met (A5) & $\lambda$\,Boo?; hybrid\\   
 5786771$^{\star}$     & 2011 July           & 2     & 9.05&  A2$^{3}$ & A0.5nn:             & bin. + rot.\\
 5954264               & 2010 May-September  & 1     & 8.16&  F0$^{1}$ & F1\,IV-V            & $\gamma$\,Dor\\ 
 5988140$^{\star}$     & 2010 May-September  & 1     & 8.81&  A7.5\,IV-III$^{4}$ & A8\,IIIs  & $\delta$\,Sct + bin.\\ 
\bottomrule
\end{tabular}
\end{table*}

\setcounter{table}{1}

\begin{table*}
\centering
\caption{continuation}
\begin{tabular}{llcrlll}
\toprule
KIC        & HERMES   & N   & V    & SpT1     & SpT2        & Notes \\
Number     & observations&     & [mag]&  & [this paper]&  \\
\midrule
 6032730               & 2011 July           & 1     & 8.61&  A2$^{1}$ & A3\,IVn             & hybrid\\    
 6106152               & 2012 July           & 1     & 8.06&  A0$^{1}$ & A3\,IV wk met (A1)  & bin.\\      
 6128236               & 2011 September      & 1     & 8.79&  F2$^{1}$ & F0\,IV+             & bin. + rot.\\       
 6192566               & 2011 September      & 1     & 9.31&  A2$^{1}$ & A6\,IV-V            & SB2; bin. + rot. ?\\
 6292398               & 2011 September      & 1     & 9.98&           & kA3hA7mA7\,V Am:    & bin. + $\gamma$\,Dor \\
 6292925               & 2011 September      & 1     & 9.66&           & kA2.5hA3mA7\,(IV) Am:& bin.\\    
                       & 2012 July           & 1     &     &           &                     & \\
 6352430$^{\star}$     & 2011 September      & 1     & 7.91&  B8$^{1}$ & B8\,Vs He str (B6)  & SB2; $\gamma$\,Dor\\
 6380579               & 2011 September      & 1     & 9.99&           & F2\,Vs              & $\gamma$\,Dor\\   
 6381306           & 2011 September      & 1     & 8.63&  A0$^{1}$ &                       & SB2; hybrid\\   
 6450107           & 2011 September      & 1     & 7.53&  A0$^{1}$ & A1\,IV-s              & rot.\\    
 6509175$^{\star}$ & 2011 September      & 1     & 9.95&  A2$^{1}$ & A6\,IVn:              & hybrid\\    
 6519869$^{\star}$ & 2011 September      & 1     &10.38&           & F1\,IV-Vs             & $\gamma$\,Dor\\     
 6587551           & 2010 May-September  & 1     & 9.72&  A0$^{1}$ & A3\,IV                & hybrid\\       
 6756386           & 2010 May-September  & 1     & 8.62&  A2$^{1}$ & A3\,IVn               & hybrid\\      
 6761539$^{\star}$ & 2011 September      & 1     &10.27&           & F0\,V wk met (A5)     & hybrid\\       
 6844024$^{\star}$ & 2012 September      & 1     & 9.92&           & F1\,Vn: wk met (A6)   & bin. + hybrid\\      
 6865077$^{\star}$ & 2011 July           & 1     & 9.76&           & kA3hA8mA6\,(IV)       & SB2; hybrid\\
 6937758$^{\star}$ & 2010 May-September  & 1     & 9.72&  A2$^{1}$ & A5\,V                 & hybrid (low ampl. $\gamma$\,Dor)\\         
 6951642           & 2011 September      & 1     & 9.70&  A5$^{1}$ & F1\,V                 & hybrid; contamination?\\     
 6967360           & 2011 September      & 1     & 9.42&  A2$^{1}$ & A1\,Vann              & hybrid\\ 
 7060333$^{\star}$ & 2011 September      & 1     & 9.09&  A2$^{1}$ & A5\,IV                & bin. + $\delta$\,Sct\\      
 7119530$^{\star}$ & 2010 May-September  & 3     & 8.44&  A3$^{1}$ & A3\,IVn               & hybrid\\      
 7299869           & 2012 July           & 1     & 9.70&  A2$^{1}$ & A3\,IV-n              & hybrid\\      
                   & 2012 September      & 1     &     &           &                       & \\
 7345479           & 2011 September      & 1     & 7.90&  A0$^{1}$ & A2\,Vnn               & bin. + rot.\\      
 7530366           & 2011 September      & 1     & 8.31&  A0$^{1}$ & A0.5\,IVnn            & bin. + rot.\\      
 7533694           & 2011 September      & 1     &10.29&           & A2.5\,IVn             & hybrid\\    
 7583939$^{\star}$ & 2011 July           & 1     & 9.64&  A0$^{1}$ & A2\,IV/V              & hybrid (low amp. $\gamma$\,Dor)\\ 
 7661054           & 2012 July           & 1     & 9.91&           & F2.5\,V               & $\gamma$\,Dor\\      
                   & 2012 September      & 1     &     &           &                       & \\
 7756853$^{\star}$ & 2010 May-September  & 1     & 8.86&  A0$^{1}$ & A3\,Vs                & SB2; hybrid \\
 7767565$^{\star}$ & 2011 July           & 1     & 9.29&  A5$^{1}$ & kA5hA7mF1\,IV Am      & hybrid (low ampl. $\delta$\,Sct) \\     
 7770282           & 2011 July           & 1     & 9.81&  F0$^{1}$ & A6\,IV                & hybrid (low ampl. $\delta$\,Sct) \\      
 7842286$^{\star}$ & 2011 July           & 1     & 9.93&  A2$^{1}$ & A3\,IV-n met str (A4) & hybrid (low ampl. $\gamma$\,Dor) \\    
 7959867           & 2010 May-September  & 1     & 9.73&  A2$^{1}$ & A4\,IV                & hybrid\\      
 8027456           & 2011 September      & 1     & 9.67&  A0$^{1}$ & A1\,V                 & hybrid\\    
 8112039           & 2011 September      & 1     & 8.39&  A0$^{1}$ & kA3hA5mA5\,(IV)s Am:  & mult. \\  
 8323104$^{\star}$ & 2012 July           & 1     & 9.62&  Am...$^{5}$& kA2.5hA6mA7\,(IV) Am& bin. + rot.\\  
 8351193$^{\star}$ & 2011 July           & 1     & 7.57&  B9$^{1}$  & B9.5:                & rot.\\     
                   & 2012 July           & 1     &     &            &                      & \\
 8355130           & 2010 May-September  & 1     & --  &  F0\,III$^{6}$& F2\,V             & $\gamma$\,Dor\\ 
 8367661           & 2011 September      & 1     & 8.56&  A0$^{1}$   & A2\,IVn             & rot.\\    
 8386982$^{\star}$ & 2012 July           & 1     & 9.77&  A0$^{1}$   & A4\,IV/V:           & bin. + rot.\\ 
 8489712           & 2011 July           & 1     & 8.60&  A0$^{1}$   & A2\,IVs             & hybrid (low ampl. $\delta$\,Sct) + rot./bin.\\     
 8692626$^{\star}$ & 2012 July           & 1     & 8.29&  B9$^{1}$   & kA2hA4mA6\,(IV) Am: & SB2; bin.\\
 8703413           & 2012 July           & 1     & 8.71&  Am...$^{6}$& kA3hA5mA9\,(IV)s Am & bin. + rot.\\     
 8750029           & 2010 May-September  & 1     & 9.62&  A5$^{1}$   & F0\,Vn              & bin. + $\delta$\,Sct\\       
 8915335           & 2011 September      & 1     & 9.59&  A2$^{1}$   & A2.5\,IV-Vn         & hybrid\\     
 8933391$^{\star}$ & 2010 May-September  & 1     & 8.85&  F0$^{1}$   & A9\,IVs             & hybrid (low ampl. $\gamma$\,Dor)\\
 8975515$^{\star}$ & 2010 May-September  & 1     & 9.48&  A2$^{1}$   & A6\,V:              & SB2; hybrid + bin.?\\
\bottomrule
\end{tabular}
\end{table*}

\setcounter{table}{1}

\begin{table*}
\centering
\caption{continuation}
\begin{tabular}{llcrlll}
\toprule
KIC              & HERMES   & N   & V   & SpT1     & SpT2         & Notes \\
Number           & observations&     & [mag]& & [this paper] & \\
\midrule
 9117875$^{\star}$ & 2011 July           & 1     & 7.53&  Am...$^{6}$& kA3hF0.5mF3\,(III) Am & bin. + rot.\\     
 9147002$^{\star}$ & 2012 July           & 1     & 9.83&  A2p$^{7}$  & kA3hA5mF3\,(IV) Am  & rot.\\ 
 9204718           & 2011 September      & 1     & 8.74&  Am...$^{8}$& kA3hA9mF1\,V Am     & $\delta$\,Sct + bin. + rot.\\  
 9246481           & 2010 May-September  & 1     & 9.04&  A0$^{1}$   & A5\,IV              & hybrid (low ampl. $\gamma$\,Dor)\\
 9286638           & 2010 May-September  & 1     & 7.32&  F0$^{1}$   & F2\,V wk met (F0)   & bin.+rot.\\       
 9291618$^{\star}$ & 2011 September      & 1     & 9.66&  A5$^{1}$   & A9\,Vn wk met (A6)p & hybrid (low ampl. $\gamma$\,Dor)\\     
 9300946$^{\star}$ & 2012 July           & 1     &10.00&  F5\,V$^{9}$& F2\,Vs              & $\gamma$\,Dor\\      
                   & 2012 September      & 2     &     &             &                     & \\
 9351622           & 2011 July           & 1     & 9.08&  F0$^{1}$   & A8\,V               & hybrid\\     
 9419182           & 2011 September      & 1     & 9.21&  F2$^{1}$   & F2\,V               & hybrid (low ampl. $\delta$\,Sct)\\    
 9509296           & 2012 July           & 1     & 9.88&             & F0\,V wk met (A7)   & hybrid\\    
                   & 2012 September      & 1     &     &             &                     & \\
 9552758           & 2012 July           & 1     & 9.94&  A2$^{10}$  & A3\,IVn             & hybrid (low ampl. $\gamma$\,Dor)\\     
                   & 2012 August         & 1     &     &             &                     & \\
 9699848           & 2011 September      & 1     & 9.12&  A2$^{1}$   & A2.5\,Vn            & rot.\\  
 9764965           & 2010 May-September  & 1     & 8.84&  A5m$^{6}$  & F0\,Vs              & hybrid\\     
 9775454           & 2010 May-September  & 1     & 8.19&  F1\,IV$^{11}$& F1\,Vs            & hybrid\\     
 9828226$^{\star}$ & 2012 August         & 1     & 9.92&  A0$^{1}$   & hA2kA0mB9\,Vb       & $\lambda$\,Boo; hybrid ?\\   
 9970568           & 2011 July           & 1     & 9.58&  A2$^{1}$   & A4\,IVn             & hybrid\\   
 10026614$^{\star}$& 2012 July           & 1     & 9.59&             & F4\,V:              & SB2; bin. + rot.? \\
 10263800          & 2011 September      & 1     & 9.61&  A0$^{1}$   & A0.5\,Vn            & hybrid + rot.\\    
 10264728           & 2012 July           & 1     & 9.90& A2$^{1}$   & --             & hybrid\\     
                    & 2012 September      & 1     &     &            &                & \\
 10355055           & 2011 July           & 1     & 9.34& A2$^{1}$   & A3\,IVn        & hybrid (low ampl. $\gamma$\,Dor)\\     
 10533616           & 2011 July           & 1     & 9.57& A5$^{1}$   & A4\,IVn        & $\delta$\,Sct \\
 10549371           & 2011 July           & 1     & 9.43& A5$^{1}$   & F0\,Vs         & hybrid (low ampl. $\gamma$\,Dor)\\ 
 10555142$^{\star}$ & 2011 September      & 1     & 6.70& K0\,III$^{12}$& F1\,Vn wk met (A5)& hybrid + rot. ?\\    
 10590857           & 2010 May-September  & 1     & 9.99& F0$^{1}$   & F0\,V          & hybrid (low ampl. $\gamma$\,Dor)\\    
 10616138$^{\star}$ & 2011 September      & 1     & 9.59& A2$^{1}$   & kA3hA6mA7\,(IV)s Am: & SB2; hybrid + rot. ?  \\
 10721930$^{\star}$ & 2012 July           & 1     & 9.61& A2$^{1}$   & A5\,IV:        & rot. \\     
 10977859           & 2010 May-September  & 1     & 8.76& A2$^{1}$   & A5\,IVs        & bin. + $\delta$\,Sct\\   
 11013201           & 2010 May-September  & 1     & 9.31& A2$^{1}$   & A6\,IV         & bin. + $\delta$\,Sct\\    
 11090405           & 2011 July           & 1     & 9.53& A5$^{1}$   & A9\,Vn         & hybrid\\    
 11180361$^{\star}$ & 2011 July           & 1     & 7.68& A2$^{1}$   & A3\,IV-n       & ecl. bin. + rot. + hybrid\\ 
 11189959           & 2011 July           & 2     & 8.15& A0$^{1}$   & A1\,Va+n       & rot. + bin.\\
 11193046           & 2011 July           & 1     & 9.48& A2$^{1}$   & A3\,IV-V       & hybrid\\
 11402951$^{\star}$ & 2012 July           & 1     & 8.12& A5$^{1}$   & F0\,IV+        & hybrid (low ampl. $\gamma$\,Dor)\\    
 11497012           & 2010 May-September  & 1     & 9.66& F0$^{1}$   & F0\,Vs         & hybrid (low ampl. $\gamma$\,Dor)\\     
 11506607           & 2011 September      & 1     & 9.61& A2$^{1}$   & A3\,Va+        & rot. + bin.\\      
 11509728           & 2012 July           & 1     & 9.75& A2$^{1}$   & A6\,IVn        & $\delta$\,Sct + rot.\\      
                    & 2012 September      & 2     &     &            &                & \\
 11572666$^{\star}$ & 2012 September      & 1     & 9.84&            & kA4hF1:mA\,Vn  & SB2; hybrid\\
 11602449$^{\star}$ & 2010 May-September  & 1     & 9.84&            & A6\,IV/V       & hybrid + rot.\\     
 11661993           & 2011 July           & 1     & 9.33&            & A9\,V          & SB2; hybrid (low ampl. $\gamma$\,Dor)\\
 11821140           & 2011 September      & 1     &10.01& F0$^{1}$   & A5\,IV         & hybrid\\     
 12020590$^{\star}$ & 2011 September      & 1     & 9.91&            & A3\,IVn: met str (A5) & hybrid\\    
 12153021$^{\star}$ & 2011 July           & 1     & 8.65& A2$^{1}$   & A2\,IV/V       & bin.\\     
 12736056           & 2011 September      & 1     & 9.16& A0$^{1}$   & A0.5\,IIIs     & hybrid + rot.?\\   
 12784394           & 2010 May-September  & 1     & 9.67& A5$^{1}$   & A4\,V          & SB2; $\delta$\,Sct + rot. ?\\
\bottomrule
\end{tabular}
\begin{description}
\item[ ]{KIC\,3230227:} Wings of Balmer lines are unusable but core matches A5\,IV. Metal line type matches A5\,V very well.
\item[ ]{KIC\,3231985:} Slight Ca weakness in both the K line and $\lambda$4226 line.
\item[ ]{KIC\,3850810:} Either considerably metal weak or poorly normalised, but probably a bit of both.
\item[ ]{KIC\,4480321:} \ion{Ca}{ii} K line is broad but shallow. H lines are around A9, having very shallow cores compared to their wing strengths, but the metal 
lines are not that late.
\item[ ]{KIC\,4572373:} Metal lines and \ion{Ca}{ii} K line are slightly weaker than A3 in spite of rotational, but hydrogen lines match A3\,Va.
\item[ ]{KIC\,4660665:} Has the appearance of a very marginal Am star (e.g. slightly enhanced \specline{Sr}{ii}{4077}, Fe/Ti\,{\sc ii}\,4172--9).
\item[ ]{KIC\,4681323:} Very close to the A1\,IVs standard except for a slightly fainter luminosity class.
\item[ ]{KIC\,4768731:} The \ion{Ca}{ii} K line is only as deep as A3, but is comparable to H8 in the core width. The \specline{Ca}{i}{4226} line also has an unusual 
profile, being shallow and broad. \specline{Sr}{ii}{4077} and $\lambda$4216 are exceptionally strong. Hydrogen lines match A5\,V well. The $\lambda$4128-31 doublet, 
usually attributable to Si, is enhanced, which given the Sr overabundance probably implicates a \ion{Eu}{ii} overabundance. This is confirmed with an 
\specline{Eu}{ii}{4205} enhancement. All three of $\lambda\lambda$4172, 4111 and 3866 are enhanced, confirming a Cr enhancement. Clearly an Ap\,SrCrEu star.
\end{description}
\end{table*}

\setcounter{table}{1}

\begin{table*}
\centering
\caption{continuation}
\begin{description}
\item[ ]{KIC\,5200084:} No obvious weakness in \specline{Ca}{i}{4226}, but \specline{Sr}{ii}{4216} is enhanced. \specline{Sr}{ii}{4077} is clearly enhanced. 
The redward third of \specline{Mn}{i}{4030} is weakened, as is characteristic of Am stars. Slight ALE.
\item[ ]{KIC\,5461344:} He lines and \ion{Ca}{ii} K are intermediate between B8 and B9\,V, and the hydrogen line type is loosely consistent with that, 
but the metal line spectrum resembles a late-A or F giant. Composite.
\item[ ]{KIC\,5633448:} Typical Am star: Ca weak, Sr strong; however, not especially sharp lined.
\item[ ]{KIC\,5724440:} Broad but shallow hydrogen lines (nearest match would be $\sim$F0). Metallic line spectrum is weak, at around A5, even accounting for very rapid 
rotation. \ion{Ca}{ii} K line is broad but shallow, having the depth of the A3\,V standard but the breadth of the A7\,V standard. $\lambda$\,Boo candidate.
\item[ ]{KIC\,5786771:} Rapid rotator. \ion{Ca}{ii} K line best matches A0.5 and is broad. No obvious helium lines. Metals weaker than A1. Shallow hydrogen cores 
do not fit at any type.
\item[ ]{KIC\,5988140:} Metal lines slightly stronger and hydrogen lines slightly narrower than the A7\,III standard 78\,Tau.
\item[ ]{KIC\,6352430:} He lines are consistent with B6\,V; \specline{Mg}{ii}{4481} is much too strong for that. H line type is B8\,V. \ion{Ca}{ii} K line is slightly 
stronger than B8. Si lines agree with B8.
\item[ ]{KIC\,6509175:} Although there are four spectra, normalisation is an issue and no precisely matching standard was found.
\item[ ]{KIC\,6519869:} Hydrogen lines are intermediate between F0 and F2 (thus F1). Normalisation has rendered \ion{Ca}{ii} K line unusable. 
The $\lambda$4172-9 blend indicates a luminosity class of Va, but the $\lambda$4395-4400 and Ti forest at $\lambda$4500 favour a luminosity class of IV.
\item[ ]{KIC\,6761539:} Hydrogen lines match F0, but both \ion{Ca}{ii} K line and metallic-line spectrum match A5 well.
\item[ ]{KIC\,6844024:} Hydrogen lines are shallow; the cores are as deep as F2 V but about 30\:per\:cent wider than F2 V at the core-wing boundary. 
A hydrogen type of F1\,V is given as a compromise. Metal lines, including \ion{Ca}{ii} K, are much weaker, at about A6.
\item[ ]{KIC\,6865077:} Very similar to KIC\,4660665.
\item[ ]{KIC\,6937758:} Problem with normalisation of Balmer lines.
\item[ ]{KIC\,7060333:} \ion{Ca}{ii} K line is rather broad for its A5 depth.
\item[ ]{KIC\,7119530:} Matches the A3\,IVn standard well, except for the poor normalisation around the hydrogen lines. Even then, the core and core-wing boundary 
profile of the hydrogen lines still match the A3\,IVn standard.
\item[ ]{KIC\,7583939:} Lies somewhere between the A1\,Va and A3\,IV standards, but no standard available at A2.
\item[ ]{KIC\,7756853:} Temperature type is A2, but luminosity class requires more standards. Metal lines match the A3\,V standard. 
The \ion{Ca}{ii} K line is unusually shallow and sharp (i.e. weak).
\item[ ]{KIC\,7767565:} No ALE, simply luminosity class IV.
\item[ ]{KIC\,7842286:} Mild but definite enhancement of metals.
\item[ ]{KIC\,8323104:} ALE.
\item[ ]{KIC\,8351193:} Almost featureless spectrum. Metal lines are absent, with the exception of a very weak \specline{Mg}{ii}{4481} (B7) and weak \ion{Ca}{ii} K line (B8). 
Helium lines at $\lambda\lambda$4026 and 4471 are slightly stronger than A0. Hydrogen wings are only slightly narrower than A0 (indicating B9.5, like the helium lines), 
but are too shallow in the core.
\item[ ]{KIC\,8386982:} Metal line type is about A3 or A4s (assuming luminosity class IV/V), but hydrogen line type could not be matched, possibly due 
to a normalisation issue.
\item[ ]{KIC\,8692626:} Classical Am signature, with enhanced Sr and weak Ca. The redward third of \specline{Mn}{i}{4030} is weak. 
Luminosity class based on Fe/\specline{Ti}{ii}{4172-9} blend, as is typical for Am stars, but hydrogen lines match the A3\,Va standard.
Exhibits the ALE in the same way as KIC\,10616138.
\item[ ]{KIC\,8933391:} Problems with normalisation.
\item[ ]{KIC\,8975515:} Problems with normalisation of Balmer lines, leading to an uncertain luminosity class. \ion{Ca}{ii} K and neutral metals agree at A6.
\item[ ]{KIC\,9117875:} Extreme Am star. Sr lines stronger still than the F3 standard. Large ALE.
\item[ ]{KIC\,9147002:} Extreme Am. Slight ALE.
\item[ ]{KIC\,9291618:} H$\gamma$ wings are as broad as A7, but the core is shallow like F2. Overall hydrogen line appearance is $\sim$A9. \ion{Ca}{ii} K line is around A6. 
Metal lines also consistent with A6\,Vn.
\item[ ]{KIC\,9300946:} A superb match to the F2\,Vs standard, 78 UMa.
\item[ ]{KIC\,9828226:} Tremendously broad H lines. Mg II 4481 is weaker than the metal line type, that is, even weaker than B9. Other metals are barely discernible. 
Thus a solid $\lambda$\,Boo candidate.
\end{description}
\end{table*}

\setcounter{table}{1}

\begin{table*}
\centering
\caption{continuation}
\begin{description}
\item[ ]{KIC\,10026614:} Hydrogen line type is about F4, which agrees with the strength of the G-band, but the metal lines are closer to F3.
\item[ ]{KIC\,10616138:} Anomalous Luminosity Effect (ALE). Typical morphology for an Am star. The $\lambda\lambda$4395--4400 lines are particularly weak compared 
to \specline{Fe}{i}{4383} and the metal line spectrum.
\item[ ]{KIC\,10555142:} Hydrogen lines do not match any standard well, but fit best between F0 and F1. Metals are noticeably broadened but nevertheless weak.
\item[ ]{KIC\,10721930:} Luminosity criteria are in disagreement. The width of the hydrogen lines favours a dwarf classification, but the \ion{Fe}{ii}/\ion{Ti}{ii} blends 
favour a giant. This illustrates the difficulty of luminosity classification at A5. The \ion{Ca}{ii} K line type is A6. Compromise type given is A5\,IV, but the star does not 
match the A5\,IV standard well. A metal-rich, early A ($\sim$A3) star was considered, but neutral metals appear no earlier than A3. Oddly, \specline{Mn}{i}4030 is enhanced, 
but is usually expected to show a mild negative luminosity effect.
\item[ ]{KIC\,11180361:} Very close to the A3\,IVn standard except for a slightly fainter luminosity class.
\item[ ]{KIC\,11402951:} Definitely between IV and III in luminosity class, but possibly slightly metal strong, too.
\item[ ]{KIC\,11572666:} Weak metals even when accounting for rapid rotation. Hydrogen lines have a peculiar profile that is not exactly intermediate between F0 and F2, 
rather, they have the cores of an F2 star and wings of an F0 star.
\item[ ]{KIC\,11602449:} Problem with normalisation. Metallic and \ion{Ca}{ii} K line types lie between A5 and A7. The temperature type of the hydrogen lines is consistent with 
this, but problems with normalisation prevents a certain luminosity classification, especially at this temperature type, where luminosity-sensitive features are lacking.
\item[ ]{KIC\,12020590:} The (rotationally broadened) metal lines and \ion{Ca}{ii} K line indicate a temperature type of A5, but hydrogen cores are too shallow for A5, 
matching A3\,IV much better.
\item[ ]{KIC\,12153021:} Temperature type is clearly A2, but the luminosity class is hard to determine without A2 standards.
\end{description}
\begin{description}
 \item[ ]{$^1$ } \citet{1975AGK3..C......0H}
 \item[ ]{$^2$ } \citet{1969AJ.....74.1082M}
 \item[ ]{$^3$ } \citet{2010A&A...517A...3C}
 \item[ ]{$^4$ } \citet{2012AN....333..975L}
 \item[ ]{$^5$ } \citet{2009A&A...498..961R}
 \item[ ]{$^6$ } \citet{1952ApJ...116..592M}
 \item[ ]{$^7$ } \citet{bertaud1960}
 \item[ ]{$^8$ } \citet{floquet}
 \item[ ]{$^9$ } \citet{2010PASP..122.1437P}
 \item[ ]{$^{10}$} \citet{fabricius2002}
 \item[ ]{$^{11}$} \citet{1950ApJ...112...48M}
 \item[ ]{$^{12}$} \citet{1973AJ.....78..401W}
 \end{description}
\end{table*}

\begin{table*}
\centering
\caption{Atmospheric parameters of investigated \textit{Kepler} A- and F-type stars. Photometric values (KIC), those obtained from spectral energy distributions (Na \& SED)
and from analysis of metal and Balmer lines are presented. The $1\sigma$ uncertainties are given for effective temperatures from SED and for 
\vsini\ and $\log\epsilon{\rm (Fe)}$ values from metal lines.}
\begin{scriptsize}
\begin{tabular}{c|cccc|cc|rccrc}
\toprule
KIC          &\multicolumn{4}{c}{\hrulefill \,KIC \,\hrulefill}
             &\multicolumn{2}{c}{\hrulefill \,Na\,D \& SED\,\hrulefill}
             &\multicolumn{5}{c}{\hrulefill \,Metal \& Balmer lines\,\hrulefill}\\
Number       & \teff      & \logg     & $E(B-V)$   & $\log\epsilon{\rm (Fe)}$   & $E(B-V)$   &  \teff    &  \multicolumn{1}{c}{\teff}   & \logg    &  \turb    &   \multicolumn{1}{c}{\vsini} & $\log\epsilon{\rm (Fe)}$\\
             & [K]        &           & [mag]      &                            & [mag]      &     [K]   & \multicolumn{1}{c}{[K]}      &          & [\kms]    &    \multicolumn{1}{c}{[\kms]}&  \\
\midrule
 1294756  &$  8411 $&$   3.92 $&$   0.08 $&$   7.38 $&$   0.03 $&$    8410 \pm   280 $&$  8300 \pm   200 $&$    3.9 \pm    0.2 $&$    1.5 \pm    0.3 $&$   170 \pm     9 $&$   7.39 \pm   0.10 $\\
 2571868  &$  7929 $&$   3.56 $&$   0.09 $&$   7.29 $&$   0.02 $&$    8120 \pm   260 $&$  8200 \pm   200 $&$    3.6 \pm    0.2 $&$    2.0 \pm    0.4 $&$   210 \pm     7 $&$   7.41 \pm   0.13 $\\
 2694337  &$  7285 $&$   3.99 $&$   0.08 $&$   7.39 $&$   0.06 $&$    7640 \pm   190 $&$  7500 \pm   100 $&$    4.0 \pm    0.1 $&$    3.4 \pm    0.1 $&$    97 \pm     2 $&$   7.61 \pm   0.07 $\\
 3230227  &$  7972 $&$   3.89 $&$   0.07 $&$   7.35 $&$   0.02 $&$    8150 \pm   220 $&$  8200 \pm   100 $&$    3.9 \pm    0.1 $&$    2.9 \pm    0.1 $&$    50 \pm     4 $&$   7.38 \pm   0.08 $\\
 3231985  &$  8763 $&$   3.87 $&$   0.10 $&$   7.37 $&$   0.03 $&$    8800 \pm   340 $&$  9100 \pm   100 $&$    3.6 \pm    0.1 $&$    0.5 \pm    0.1 $&$    22 \pm     1 $&$   7.74 \pm   0.07 $\\
\bottomrule
\end{tabular}
\end{scriptsize}
\end{table*}

\setcounter{table}{2}

\begin{table*}
\centering
\caption{continuation}
\begin{scriptsize}
\begin{tabular}{c|cccc|cc|rccrc}
\toprule
KIC          &\multicolumn{4}{c}{\hrulefill \,KIC \,\hrulefill}
             &\multicolumn{2}{c}{\hrulefill \,Na\,D \& SED\,\hrulefill}
             &\multicolumn{5}{c}{\hrulefill \,Metal \& Balmer lines\,\hrulefill}\\
Number       & \teff      & \logg     & $E(B-V)$   & $\log\epsilon{\rm (Fe)}$   & $E(B-V)$   &  \teff    &  \multicolumn{1}{c}{\teff}   & \logg    &  \turb    &   \multicolumn{1}{c}{\vsini} & $\log\epsilon{\rm (Fe)}$\\
             & [K]        &           & [mag]      &                            & [mag]      &     [K]   & \multicolumn{1}{c}{[K]}      &          & [\kms]    &    \multicolumn{1}{c}{[\kms]}&  \\
\midrule
 8027456  &$  8732 $&$   3.77 $&$   0.12 $&$   7.56 $&$   0.04 $&$    8730 \pm   320 $&$  8900 \pm   200 $&$    4.0 \pm    0.2 $&$    1.4 \pm    0.3 $&$   194 \pm     6 $&$   7.28 \pm   0.13 $\\
 8112039  &$  8156 $&$   3.97 $&$   0.06 $&$   7.37 $&$   0.03 $&$    8230 \pm   230 $&$  8400 \pm   100 $&$    3.7 \pm    0.1 $&$    1.5 \pm    0.1 $&$     9 \pm     1 $&$   7.57 \pm   0.13 $\\
 %8211500  &$  7403 $&$   3.91 $&$   0.04 $&$   7.63 $&$   0.01 $&$    8040 \pm   290 $&$  7800 \pm   100 $&$    3.8 \pm    0.1 $&$    2.0 \pm    0.1 $&$    93 \pm     8 $&$   7.60 \pm   0.11 $\\
 8323104  &$  7587 $&$   3.92 $&$   0.08 $&$   7.24 $&$   0.04 $&$    7700 \pm   180 $&$  7800 \pm   100 $&$    3.9 \pm    0.1 $&$    2.1 \pm    0.1 $&$    10 \pm     1 $&$   7.65 \pm   0.13 $\\
 8351193  &$  8467 $&$   3.99 $&$   0.04 $&$   7.36 $&$   0.01 $&$   10760 \pm   320 $&$ 10500 \pm   200 $&$    4.1 \pm    0.2 $&$    2.0 \pm    0.3 $&$   162 \pm     0 $&$   6.18 \pm   0.00 $\\
 8355130  &$  7043 $&$   4.11 $&$   0.06 $&$   7.56 $&$   0.02 $&$    7030 \pm   150 $&$  7100 \pm   100 $&$    3.6 \pm    0.1 $&$    2.8 \pm    0.1 $&$    49 \pm     2 $&$   7.23 \pm   0.13 $\\
\bottomrule

\end{tabular}
\end{scriptsize}
\end{table*}

\setcounter{table}{3}

\begin{landscape}

\begin{table}
\begin{scriptsize}
\caption{Abundances of chemical elements for a sample of analysed stars. Median value is given; standard deviations were calculated only if the number of analysed
parts is greater than three. In other cases the average value calculated from standard deviations of other elements are given. Number of analysed parts is given in brackets.}
\begin{tabular}{lllllllllll}
\midrule
   & KIC\,10263800 &     KIC\,10264728 &     KIC\,10355055 &     KIC\,10533616 &     KIC\,10549371 &     KIC\,10555142 &     KIC\,10590857 &     KIC\,10721930 &     KIC\,10977859 &     KIC\,11013201\\
    C & 8.85$\pm$0.14 (1) &   8.48$\pm$0.12 (2) &   8.90$\pm$0.12 (2) &   8.38$\pm$0.07 (7) &  8.31$\pm$0.14 (14) &   8.78$\pm$0.16 (5) &   8.36$\pm$0.19 (5) &  8.30$\pm$0.17 (23) &   8.50$\pm$0.19 (8) &  8.48$\pm$0.17 (2) \\
    N               & $-$ &                 $-$ &                 $-$ &                 $-$ &   8.33$\pm$0.14 (1) &                 $-$ &                 $-$ &   7.78$\pm$0.17 (2) &                 $-$ &                $-$ \\
    O & 8.74$\pm$0.14 (2) &   9.04$\pm$0.12 (1) &   8.44$\pm$0.12 (1) &   8.92$\pm$0.14 (2) &   8.83$\pm$0.14 (1) &                 $-$ &   8.91$\pm$0.17 (1) &   8.75$\pm$0.17 (2) &   8.92$\pm$0.22 (2) &  8.75$\pm$0.17 (1) \\
   Ne               & $-$ &                 $-$ &                 $-$ &                 $-$ &                 $-$ &                 $-$ &                 $-$ &                 $-$ &                 $-$ &                $-$ \\
   Na               & $-$ &   6.60$\pm$0.12 (1) &   7.10$\pm$0.12 (1) &   6.89$\pm$0.14 (2) &   6.46$\pm$0.14 (2) &   6.63$\pm$0.18 (1) &   6.22$\pm$0.17 (1) &   6.21$\pm$0.05 (3) &   5.91$\pm$0.22 (1) &  6.63$\pm$0.17 (2) \\
   Mg & 7.68$\pm$0.14 (1) &   8.01$\pm$0.12 (4) &   7.28$\pm$0.19 (3) &   7.98$\pm$0.10 (6) &   7.56$\pm$0.13 (8) &   7.64$\pm$0.18 (2) &   7.57$\pm$0.10 (4) &  7.55$\pm$0.18 (10) &   8.03$\pm$0.21 (6) &  7.95$\pm$0.18 (3) \\
   Al               & $-$ &                 $-$ &                 $-$ &                 $-$ &                 $-$ &                 $-$ &                 $-$ &   6.33$\pm$0.17 (1) &                 $-$ &                $-$ \\
\bottomrule
\end{tabular}
\end{scriptsize}
\end{table}
\end{landscape}

\newpage

\setcounter{table}{3}
\begin{landscape}

\begin{table}
\begin{scriptsize}
\caption{continuation}
\begin{tabular}{lllllllllll}
\midrule
   & KIC\,11090405 &     KIC\,11180361 &     KIC\,11189959 &     KIC\,11193046 &     KIC\,11402951 &     KIC\,11497012 &     KIC\,11506607 &     KIC\,11509728 &     KIC\,11602449 &     KIC\,11821140\\
    C & 8.37$\pm$0.23 (3) &   8.59$\pm$0.17 (3) &   8.43$\pm$0.11 (2) &   8.79$\pm$0.26 (4) &   8.52$\pm$0.40 (5) &   8.22$\pm$0.28 (9) &   8.16$\pm$0.25 (5) &   8.42$\pm$0.08 (5) &   8.32$\pm$0.16 (5) &  8.28$\pm$0.12 (3) \\
    N               & $-$ &                 $-$ &                 $-$ &                 $-$ &   8.44$\pm$0.19 (1) &                 $-$ &                 $-$ &                 $-$ &                 $-$ &                $-$ \\
    O               & $-$ &   8.93$\pm$0.18 (2) &   8.74$\pm$0.01 (3) &                 $-$ &   9.61$\pm$0.19 (2) &   8.83$\pm$0.16 (2) &   8.78$\pm$0.22 (2) &   9.58$\pm$0.08 (2) &                 $-$ &  8.68$\pm$0.12 (1) \\
   Ne               & $-$ &                 $-$ &                 $-$ &                 $-$ &                 $-$ &                 $-$ &                 $-$ &                 $-$ &                 $-$ &                $-$ \\
   Na & 6.99$\pm$0.14 (2) &   6.83$\pm$0.18 (1) &                 $-$ &                 $-$ &   6.73$\pm$0.19 (2) &   6.38$\pm$0.16 (2) &                 $-$ &   6.55$\pm$0.08 (1) &   6.35$\pm$0.15 (1) &  5.05$\pm$0.12 (1) \\
   Mg & 7.92$\pm$0.18 (5) &   7.95$\pm$0.30 (5) &   7.51$\pm$0.06 (3) &   7.69$\pm$0.20 (3) &   7.94$\pm$0.22 (8) &   7.52$\pm$0.10 (4) &   7.63$\pm$0.12 (7) &   7.98$\pm$0.10 (5) &   7.58$\pm$0.15 (2) &  7.64$\pm$0.10 (4) \\
   Al               & $-$ &                 $-$ &                 $-$ &                 $-$ &                 $-$ &                 $-$ &                 $-$ &                 $-$ &                 $-$ &                $-$ \\
\bottomrule
\end{tabular}
\end{scriptsize}
\end{table}
\end{landscape}

\newpage

\setcounter{table}{3}
\begin{landscape}

\begin{table}
\begin{scriptsize}
\caption{continuation}
\begin{tabular}{lllllllllll}
\midrule
   & KIC\,12020590 &     KIC\,12153021 &     KIC\,12736056 &      KIC\,1294756 &      KIC\,2571868 &      KIC\,2694337 &      KIC\,3230227 &      KIC\,3231985 &      KIC\,3347643 &      KIC\,3441230\\
    C & 8.24$\pm$0.13 (9) &   8.40$\pm$0.10 (9) &   8.91$\pm$0.20 (3) &   8.36$\pm$0.11 (2) &   8.79$\pm$0.17 (4) &   8.34$\pm$0.04 (5) &   8.29$\pm$0.05 (7) &  8.29$\pm$0.12 (19) &   8.41$\pm$0.07 (5) &  8.07$\pm$0.35 (5) \\
    N               & $-$ &   8.45$\pm$0.12 (2) &   8.04$\pm$0.23 (4) &                 $-$ &                 $-$ &                 $-$ &   8.17$\pm$0.10 (1) &   7.02$\pm$0.14 (1) &                 $-$ &  7.87$\pm$0.16 (1) \\
    O & 8.42$\pm$0.15 (2) &   8.81$\pm$0.16 (9) &   8.88$\pm$0.20 (7) &                 $-$ &   8.86$\pm$0.19 (2) &   8.89$\pm$0.09 (1) &   8.76$\pm$0.10 (2) &   8.58$\pm$0.14 (2) &   8.73$\pm$0.13 (2) &  8.32$\pm$0.16 (1) \\
   Ne               & $-$ &                 $-$ &                 $-$ &                 $-$ &                 $-$ &                 $-$ &                 $-$ &                 $-$ &                 $-$ &                $-$ \\
   Na & 6.52$\pm$0.15 (1) &   8.77$\pm$0.12 (2) &                 $-$ &                 $-$ &   6.97$\pm$0.19 (2) &   6.50$\pm$0.09 (1) &   6.39$\pm$0.10 (2) &   6.59$\pm$0.14 (2) &                 $-$ &  6.76$\pm$0.09 (3) \\
   Mg & 7.51$\pm$0.17 (4) &   7.62$\pm$0.15 (7) &  7.47$\pm$0.15 (13) &   7.88$\pm$0.10 (4) &   7.74$\pm$0.15 (3) &   7.53$\pm$0.03 (3) &   7.79$\pm$0.23 (4) &   7.59$\pm$0.19 (8) &   7.96$\pm$0.08 (4) &  7.64$\pm$0.17 (7) \\
   Al               & $-$ &   6.35$\pm$0.12 (1) &   6.48$\pm$0.17 (1) &                 $-$ &                 $-$ &                 $-$ &                 $-$ &   6.62$\pm$0.14 (1) &                 $-$ &  6.92$\pm$0.16 (1) \\
\bottomrule
\end{tabular}
\end{scriptsize}
\end{table}
\end{landscape}

\newpage

\setcounter{table}{3}
\begin{landscape}

\begin{table}
\begin{scriptsize}
\caption{continuation}
\begin{tabular}{lllllllllll}
\midrule
    & KIC\,3850810 &      KIC\,3851151 &      KIC\,3942392 &      KIC\,4035667 &      KIC\,4044353 &      KIC\,4056136 &      KIC\,4077032 &      KIC\,4150611 &      KIC\,4281581 &      KIC\,4572373\\
    C & 8.52$\pm$0.21 (6) &   8.58$\pm$0.10 (2) &   8.60$\pm$0.04 (4) &   8.34$\pm$0.05 (3) &   8.47$\pm$0.05 (5) &   8.12$\pm$0.09 (1) &  8.63$\pm$0.14 (24) &   8.51$\pm$0.19 (2) &   8.33$\pm$0.03 (4) &  8.60$\pm$0.15 (2) \\
    N               & $-$ &                 $-$ &                 $-$ &                 $-$ &   8.03$\pm$0.12 (1) &                 $-$ &   7.97$\pm$0.19 (1) &                 $-$ &                 $-$ &                $-$ \\
    O & 8.50$\pm$0.13 (1) &   8.68$\pm$0.13 (3) &                 $-$ &   8.84$\pm$0.15 (1) &                 $-$ &   8.99$\pm$0.09 (2) &   8.96$\pm$0.19 (2) &   9.04$\pm$0.19 (1) &   8.91$\pm$0.10 (1) &  9.78$\pm$0.15 (2) \\
   Ne               & $-$ &                 $-$ &                 $-$ &                 $-$ &                 $-$ &                 $-$ &                 $-$ &                 $-$ &                 $-$ &                $-$ \\
   Na & 6.83$\pm$0.13 (2) &   6.36$\pm$0.10 (1) &   6.38$\pm$0.13 (2) &                 $-$ &                 $-$ &                 $-$ &   6.60$\pm$0.19 (6) &   6.27$\pm$0.19 (1) &                 $-$ &                $-$ \\
   Mg & 7.38$\pm$0.10 (4) &   7.45$\pm$0.10 (2) &   7.51$\pm$0.07 (5) &   7.78$\pm$0.15 (1) &   7.81$\pm$0.12 (2) &   7.62$\pm$0.09 (2) &  7.64$\pm$0.16 (10) &   7.69$\pm$0.22 (5) &   7.60$\pm$0.15 (7) &  7.41$\pm$0.20 (5) \\
   Al               & $-$ &                 $-$ &                 $-$ &                 $-$ &                 $-$ &                 $-$ &   6.46$\pm$0.19 (1) &                 $-$ &                 $-$ &                $-$ \\
\bottomrule
\end{tabular}
\end{scriptsize}
\end{table}
\end{landscape}

\newpage

\setcounter{table}{3}
\begin{landscape}

\begin{table}
\begin{scriptsize}
\caption{continuation}
\begin{tabular}{lllllllllll}
\midrule
    & KIC\,4671225 &      KIC\,4681323 &      KIC\,4768731 &      KIC\,4831769 &      KIC\,5018590 &      KIC\,5113797 &      KIC\,5199439 &      KIC\,5200084 &      KIC\,5294231 &      KIC\,5355850\\
    C & 8.52$\pm$0.08 (3) &   8.25$\pm$0.15 (2) &  8.30$\pm$0.30 (13) &   7.98$\pm$0.16 (2) &   8.64$\pm$0.28 (4) &   8.47$\pm$0.10 (1) &   7.93$\pm$0.22 (7) &  7.87$\pm$0.14 (16) &   7.93$\pm$0.11 (1) & 8.23$\pm$0.09 (22) \\
    N               & $-$ &   7.85$\pm$0.15 (2) &                 $-$ &                 $-$ &                 $-$ &                 $-$ &   7.73$\pm$0.12 (2) &                 $-$ &                 $-$ &  8.77$\pm$0.11 (1) \\
    O               & $-$ &   8.77$\pm$0.04 (3) &   8.49$\pm$0.22 (5) &   8.48$\pm$0.19 (3) &   9.65$\pm$0.15 (2) &                 $-$ &   8.37$\pm$0.12 (2) &   8.83$\pm$0.14 (2) &   8.75$\pm$0.11 (2) &  8.86$\pm$0.11 (2) \\
   Ne               & $-$ &                 $-$ &                 $-$ &                 $-$ &                 $-$ &                 $-$ &                 $-$ &                 $-$ &                 $-$ &                $-$ \\
   Na & 7.45$\pm$0.12 (1) &                 $-$ &   6.60$\pm$0.22 (5) &                 $-$ &   6.55$\pm$0.15 (2) &                 $-$ &   6.67$\pm$0.12 (2) &   6.61$\pm$0.14 (2) &                 $-$ &  5.90$\pm$0.11 (3) \\
   Mg & 7.73$\pm$0.03 (3) &   7.53$\pm$0.20 (7) &  7.63$\pm$0.19 (10) &   7.24$\pm$0.25 (3) &   7.45$\pm$0.14 (9) &   7.68$\pm$0.10 (2) &   7.58$\pm$0.15 (8) &   7.75$\pm$0.13 (8) &   7.78$\pm$0.05 (4) &  7.39$\pm$0.15 (9) \\
   Al               & $-$ &                 $-$ &                 $-$ &                 $-$ &                 $-$ &                 $-$ &                 $-$ &                 $-$ &                 $-$ &                $-$ \\
\bottomrule
\end{tabular}
\end{scriptsize}
\end{table}
\end{landscape}

\newpage

\setcounter{table}{3}
\begin{landscape}

\begin{table}
\begin{scriptsize}
\caption{continuation}
\begin{tabular}{lllllllllll}
\midrule
    & KIC\,5429163 &      KIC\,5471091 &      KIC\,5473171 &      KIC\,5524045 &      KIC\,5608334 &      KIC\,5633448 &      KIC\,5724440 &      KIC\,5786771 &      KIC\,5954264 &      KIC\,5988140\\
    C & 8.76$\pm$0.09 (1) &   8.49$\pm$0.05 (4) &   8.42$\pm$0.05 (4) &   8.30$\pm$0.12 (1) &   8.92$\pm$0.24 (4) &  7.85$\pm$0.19 (20) &   8.53$\pm$0.17 (4) &   8.70$\pm$0.15 (1) &   8.60$\pm$0.07 (6) & 8.23$\pm$0.13 (10) \\
    N               & $-$ &                 $-$ &                 $-$ &                 $-$ &                 $-$ &   7.60$\pm$0.14 (2) &                 $-$ &                 $-$ &                 $-$ &  7.47$\pm$0.10 (1) \\
    O & 8.76$\pm$0.09 (1) &   8.95$\pm$0.13 (1) &   8.89$\pm$0.11 (1) &   8.80$\pm$0.12 (2) &                 $-$ &   8.16$\pm$0.04 (4) &   8.52$\pm$0.13 (1) &                 $-$ &   9.08$\pm$0.10 (1) &  8.90$\pm$0.10 (2) \\
   Ne               & $-$ &                 $-$ &                 $-$ &                 $-$ &                 $-$ &                 $-$ &                 $-$ &                 $-$ &                 $-$ &                $-$ \\
   Na               & $-$ &   6.49$\pm$0.13 (1) &   6.27$\pm$0.11 (1) &                 $-$ &   6.60$\pm$0.09 (3) &   6.48$\pm$0.10 (3) &   6.46$\pm$0.13 (1) &                 $-$ &   6.52$\pm$0.10 (2) &  6.42$\pm$0.10 (2) \\
   Mg & 7.59$\pm$0.09 (2) &   7.55$\pm$0.14 (6) &   7.19$\pm$0.01 (3) &   7.66$\pm$0.05 (3) &   7.59$\pm$0.23 (7) &  7.55$\pm$0.21 (10) &   7.65$\pm$0.15 (3) &   7.30$\pm$0.12 (3) &   7.53$\pm$0.16 (6) &  7.48$\pm$0.15 (3) \\
   Al               & $-$ &                 $-$ &                 $-$ &                 $-$ &                 $-$ &   6.45$\pm$0.14 (1) &                 $-$ &                 $-$ &                 $-$ &  6.05$\pm$0.10 (1) \\
\bottomrule
\end{tabular}
\end{scriptsize}
\end{table}
\end{landscape}

\newpage

\setcounter{table}{3}
\begin{landscape}

\begin{table}
\begin{scriptsize}
\caption{continuation}
\begin{tabular}{lllllllllll}
\midrule
    & KIC\,6032730 &      KIC\,6106152 &      KIC\,6128236 &      KIC\,6292398 &      KIC\,6292925 &      KIC\,6380579 &      KIC\,6450107 &      KIC\,6509175 &      KIC\,6519869 &      KIC\,6587551\\
    C & 8.35$\pm$0.11 (3) &   8.42$\pm$0.15 (3) &   8.61$\pm$0.08 (6) &  8.03$\pm$0.27 (10) &  7.90$\pm$0.13 (22) &   8.23$\pm$0.15 (6) &   8.52$\pm$0.12 (2) &   8.74$\pm$0.12 (2) &  8.35$\pm$0.14 (20) &  8.57$\pm$0.08 (2) \\
    N               & $-$ &   7.86$\pm$0.10 (1) &                 $-$ &   7.57$\pm$0.17 (1) &   8.27$\pm$0.12 (1) &                 $-$ &                 $-$ &                 $-$ &   8.32$\pm$0.17 (1) &                $-$ \\
    O & 8.85$\pm$0.15 (1) &   8.67$\pm$0.10 (2) &                 $-$ &   8.55$\pm$0.19 (5) &   8.40$\pm$0.11 (5) &   9.02$\pm$0.22 (1) &   8.85$\pm$0.09 (6) &                 $-$ &   8.76$\pm$0.17 (1) &  8.87$\pm$0.08 (1) \\
   Ne               & $-$ &                 $-$ &                 $-$ &                 $-$ &                 $-$ &                 $-$ &                 $-$ &                 $-$ &                 $-$ &                $-$ \\
   Na & 6.63$\pm$0.15 (2) &                 $-$ &   7.00$\pm$0.13 (2) &   6.39$\pm$0.17 (2) &   6.33$\pm$0.02 (3) &   6.95$\pm$0.22 (2) &   6.35$\pm$0.12 (1) &                 $-$ &   6.10$\pm$0.05 (6) &                $-$ \\
   Mg & 7.83$\pm$0.15 (2) &   7.60$\pm$0.13 (6) &   7.52$\pm$0.18 (7) &   7.65$\pm$0.21 (9) &  7.65$\pm$0.15 (10) &   7.84$\pm$0.18 (7) &   7.62$\pm$0.09 (5) &   7.61$\pm$0.12 (3) &  7.43$\pm$0.17 (10) &  8.12$\pm$0.08 (2) \\
   Al               & $-$ &   6.71$\pm$0.10 (1) &   6.30$\pm$0.13 (1) &   6.85$\pm$0.17 (1) &   6.63$\pm$0.12 (1) &                 $-$ &   5.46$\pm$0.12 (1) &                 $-$ &                 $-$ &                $-$ \\
\bottomrule
\end{tabular}
\end{scriptsize}
\end{table}
\end{landscape}

\newpage

\setcounter{table}{3}
\begin{landscape}

\begin{table}
\begin{scriptsize}
\caption{continuation}
\begin{tabular}{lllllllllll}
\midrule
    & KIC\,6756386 &      KIC\,6761539 &      KIC\,6844024 &      KIC\,6937758 &      KIC\,6951642 &      KIC\,6967360 &      KIC\,7060333 &      KIC\,7119530 &      KIC\,7299869 &      KIC\,7345479\\
    C & 8.51$\pm$0.04 (3) &   8.16$\pm$0.18 (3) &   8.40$\pm$0.14 (2) &   8.33$\pm$0.08 (3) &   8.75$\pm$0.12 (3) &   8.66$\pm$0.20 (2) &   8.12$\pm$0.12 (7) &   8.38$\pm$0.10 (2) &   8.54$\pm$0.16 (2) &  8.34$\pm$0.11 (1) \\
    N               & $-$ &                 $-$ &                 $-$ &                 $-$ &                 $-$ &                 $-$ &                 $-$ &                 $-$ &                 $-$ &                $-$ \\
    O & 8.91$\pm$0.07 (1) &                 $-$ &                 $-$ &   8.70$\pm$0.11 (1) &                 $-$ &                 $-$ &   8.50$\pm$0.12 (1) &   8.88$\pm$0.10 (1) &                 $-$ &  8.79$\pm$0.11 (2) \\
   Ne               & $-$ &                 $-$ &                 $-$ &                 $-$ &                 $-$ &                 $-$ &                 $-$ &                 $-$ &                 $-$ &                $-$ \\
   Na               & $-$ &   6.72$\pm$0.15 (1) &   6.36$\pm$0.14 (2) &   6.43$\pm$0.11 (1) &   6.08$\pm$0.12 (2) &                 $-$ &                 $-$ &                 $-$ &   6.28$\pm$0.16 (2) &                $-$ \\
   Mg & 7.70$\pm$0.07 (2) &   7.47$\pm$0.04 (3) &   7.39$\pm$0.13 (4) &   7.55$\pm$0.11 (2) &   7.52$\pm$0.14 (6) &   7.62$\pm$0.20 (2) &   7.54$\pm$0.13 (5) &   8.40$\pm$0.10 (1) &   7.85$\pm$0.10 (6) &  7.68$\pm$0.17 (3) \\
   Al               & $-$ &                 $-$ &                 $-$ &                 $-$ &                 $-$ &                 $-$ &                 $-$ &                 $-$ &                 $-$ &                $-$ \\
\bottomrule
\end{tabular}
\end{scriptsize}
\end{table}
\end{landscape}

\newpage

\setcounter{table}{3}
\begin{landscape}

\begin{table}
\begin{scriptsize}
\caption{continuation}
\begin{tabular}{lllllllllll}
\midrule
    & KIC\,7530366 &      KIC\,7533694 &      KIC\,7583939 &      KIC\,7661054 &      KIC\,7767565 &      KIC\,7770282 &      KIC\,7842286 &      KIC\,7959867 &      KIC\,8027456 &      KIC\,8112039\\
    C & 8.51$\pm$0.13 (1) &   8.96$\pm$0.12 (2) &   8.74$\pm$0.09 (2) &  8.37$\pm$0.12 (25) &   8.64$\pm$0.27 (4) &   8.21$\pm$0.11 (7) &   8.72$\pm$0.13 (2) &   8.46$\pm$0.03 (4) &   8.61$\pm$0.11 (2) & 7.95$\pm$0.17 (19) \\
    N               & $-$ &                 $-$ &                 $-$ &   8.88$\pm$0.13 (1) &                 $-$ &                 $-$ &                 $-$ &   7.97$\pm$0.09 (1) &                 $-$ &  7.40$\pm$0.12 (2) \\
    O & 8.74$\pm$0.13 (2) &                 $-$ &   8.94$\pm$0.09 (1) &   8.82$\pm$0.13 (2) &   8.34$\pm$0.16 (1) &   8.83$\pm$0.13 (2) &                 $-$ &   8.94$\pm$0.09 (2) &   8.66$\pm$0.11 (2) &  8.37$\pm$0.05 (5) \\
   Ne               & $-$ &                 $-$ &                 $-$ &                 $-$ &                 $-$ &                 $-$ &                 $-$ &                 $-$ &                 $-$ &                $-$ \\
   Na               & $-$ &   6.83$\pm$0.12 (1) &                 $-$ &   6.38$\pm$0.09 (9) &   6.74$\pm$0.16 (2) &   5.09$\pm$0.13 (1) &   6.85$\pm$0.13 (1) &                 $-$ &                 $-$ &  6.25$\pm$0.11 (4) \\
   Mg & 7.70$\pm$0.07 (3) &   7.77$\pm$0.12 (2) &   7.13$\pm$0.09 (2) &  7.58$\pm$0.13 (13) &   7.96$\pm$0.13 (6) &   7.73$\pm$0.14 (6) &   7.77$\pm$0.06 (5) &   7.44$\pm$0.27 (3) &   7.43$\pm$0.15 (5) &  7.51$\pm$0.27 (8) \\
   Al               & $-$ &                 $-$ &                 $-$ &                 $-$ &   6.66$\pm$0.16 (1) &                 $-$ &                 $-$ &                 $-$ &                 $-$ &  6.70$\pm$0.12 (1) \\
\bottomrule
\end{tabular}
\end{scriptsize}
\end{table}
\end{landscape}

\newpage

\setcounter{table}{3}

\begin{landscape}
\begin{table}
\begin{scriptsize}
\caption{continuation}
\begin{tabular}{lllllllllll}
\midrule
    & KIC\,8323104 &      KIC\,8351193 &      KIC\,8355130 &      KIC\,8367661 &      KIC\,8386982 &      KIC\,8489712 &      KIC\,8703413 &      KIC\,8750029 &      KIC\,8915335 &      KIC\,8933391\\
    C& 7.91$\pm$0.29 (18) &                 $-$ &   8.39$\pm$0.06 (7) &   8.60$\pm$0.10 (2) &   7.86$\pm$0.12 (9) &   8.57$\pm$0.16 (3) &  7.79$\pm$0.29 (17) &   8.52$\pm$0.10 (2) &   8.74$\pm$0.17 (2) &  8.43$\pm$0.12 (6) \\
    N & 7.68$\pm$0.17 (1) &                 $-$ &                 $-$ &                 $-$ &   7.79$\pm$0.12 (1) &                 $-$ &   8.26$\pm$0.14 (1) &                 $-$ &                 $-$ &                $-$ \\
    O & 8.28$\pm$0.19 (4) &   9.41$\pm$0.00 (2) &   8.80$\pm$0.11 (1) &   8.77$\pm$0.10 (2) &   8.46$\pm$0.15 (3) &   8.76$\pm$0.10 (3) &   8.58$\pm$0.14 (2) &                 $-$ &                 $-$ &  8.95$\pm$0.11 (2) \\
   Ne               & $-$ &                 $-$ &                 $-$ &                 $-$ &                 $-$ &                 $-$ &                 $-$ &                 $-$ &                 $-$ &                $-$ \\
   Na & 6.52$\pm$0.20 (5) &                 $-$ &   6.36$\pm$0.11 (1) &                 $-$ &                 $-$ &                 $-$ &   6.52$\pm$0.13 (3) &                 $-$ &                 $-$ &  6.38$\pm$0.11 (1) \\
   Mg & 7.39$\pm$0.14 (9) &   6.45$\pm$0.00 (1) &   7.52$\pm$0.15 (3) &   7.72$\pm$0.11 (5) &   7.29$\pm$0.16 (7) &   7.58$\pm$0.18 (7) &  7.83$\pm$0.24 (10) &   7.17$\pm$0.10 (2) &   7.74$\pm$0.15 (3) &  7.52$\pm$0.06 (3) \\
   Al & 6.50$\pm$0.17 (2) &                 $-$ &                 $-$ &                 $-$ &                 $-$ &                 $-$ &   6.78$\pm$0.14 (1) &                 $-$ &                 $-$ &                $-$ \\
\bottomrule
\end{tabular}
\end{scriptsize}
\end{table}
\end{landscape}

\newpage

\setcounter{table}{3}
\begin{landscape}

\begin{table}
\begin{scriptsize}
\caption{continuation}
\begin{tabular}{lllllllllll}
\midrule
    & KIC\,9117875 &      KIC\,9147002 &      KIC\,9204718 &      KIC\,9246481 &      KIC\,9286638 &      KIC\,9291618 &      KIC\,9300946 &      KIC\,9351622 &      KIC\,9419182 &      KIC\,9509296\\
    C & 8.02$\pm$0.30 (5) &   8.53$\pm$0.24 (3) &  7.90$\pm$0.28 (21) &  8.30$\pm$0.20 (10) &   8.68$\pm$0.10 (7) &   9.03$\pm$0.26 (2) &   8.57$\pm$0.04 (6) &   8.42$\pm$0.06 (8) &   8.66$\pm$0.11 (2) &  8.28$\pm$0.10 (5) \\
    N               & $-$ &                 $-$ &   8.28$\pm$0.17 (1) &                 $-$ &   8.32$\pm$0.13 (1) &                 $-$ &   8.50$\pm$0.11 (1) &                 $-$ &                 $-$ &                $-$ \\
    O & 8.84$\pm$0.20 (2) &   8.48$\pm$0.13 (2) &   8.67$\pm$0.17 (2) &   8.72$\pm$0.10 (3) &                 $-$ &                 $-$ &   9.03$\pm$0.11 (2) &   8.89$\pm$0.11 (2) &   9.69$\pm$0.11 (1) &                $-$ \\
   Ne               & $-$ &                 $-$ &                 $-$ &                 $-$ &                 $-$ &                 $-$ &                 $-$ &                 $-$ &                 $-$ &                $-$ \\
   Na & 7.42$\pm$0.12 (4) &   6.99$\pm$0.13 (2) &   6.49$\pm$0.17 (9) &   6.60$\pm$0.14 (2) &   6.81$\pm$0.13 (2) &   7.09$\pm$0.26 (2) &   6.49$\pm$0.02 (3) &   5.82$\pm$0.11 (2) &   6.47$\pm$0.11 (2) &  6.28$\pm$0.11 (2) \\
   Mg & 7.56$\pm$0.13 (5) &   7.73$\pm$0.10 (5) &  7.54$\pm$0.15 (15) &   7.56$\pm$0.14 (6) &   7.74$\pm$0.08 (4) &   7.86$\pm$0.50 (4) &   7.63$\pm$0.05 (8) &   7.90$\pm$0.07 (5) &   7.53$\pm$0.17 (5) &  7.57$\pm$0.08 (5) \\
   Al & 6.56$\pm$0.20 (1) &                 $-$ &                 $-$ &   6.25$\pm$0.14 (1) &                 $-$ &                 $-$ &   6.40$\pm$0.11 (1) &                 $-$ &                 $-$ &                $-$ \\
\bottomrule
\end{tabular}
\end{scriptsize}
\end{table}
\end{landscape}

\newpage

\setcounter{table}{3}
\begin{landscape}

\begin{table}
\begin{scriptsize}
\caption{continuation}
\begin{tabular}{lllllll}
\midrule
    & KIC\,9552758 &      KIC\,9699848 &      KIC\,9764965 &      KIC\,9775454 &      KIC\,9828226 &      KIC\,9970568                                                                                \\
    C & 8.56$\pm$0.17 (2) &   7.97$\pm$0.10 (1) &   8.68$\pm$0.14 (6) &   8.47$\pm$0.10 (8) &   8.39$\pm$0.14 (1) &   9.04$\pm$0.12 (1)                                                                                        \\
    N               & $-$ &                 $-$ &                 $-$ &                 $-$ &   7.61$\pm$0.14 (1) &                 $-$                                                                                        \\
    O               & $-$ &   8.53$\pm$0.10 (2) &   9.06$\pm$0.12 (2) &   9.17$\pm$0.09 (2) &   9.62$\pm$0.14 (2) &   8.89$\pm$0.12 (1)                                                                                        \\
   Ne               & $-$ &                 $-$ &                 $-$ &                 $-$ &                 $-$ &                 $-$                                                                                        \\
   Na               & $-$ &                 $-$ &   6.33$\pm$0.12 (1) &   6.43$\pm$0.09 (2) &                 $-$ &                 $-$                                                                                        \\
   Mg & 7.62$\pm$0.09 (4) &   7.73$\pm$0.15 (8) &   7.63$\pm$0.03 (3) &   7.52$\pm$0.04 (4) &   6.14$\pm$0.11 (3) &   7.84$\pm$0.12 (4)                                                                                        \\
   Al               & $-$ &                 $-$ &                 $-$ &                 $-$ &                 $-$ &                 $-$                                                                                        \\
\bottomrule
\end{tabular}
\end{scriptsize}
\end{table}

\end{landscape}

\end{document}